\documentclass[a4paper, amsfonts, amssymb, amsmath, reprint, showkeys, nofootinbib, twoside]{revtex4-1}
\usepackage[english]{babel}
\usepackage[utf8]{inputenc}
\usepackage[colorinlistoftodos, color=green!40, prependcaption]{todonotes}
\usepackage{amsthm}
\usepackage{mathtools}
\usepackage{physics}
\usepackage{xcolor}
\usepackage{graphicx}
\usepackage[left=23mm,right=13mm,top=35mm,columnsep=15pt]{geometry} 
\usepackage{adjustbox}
\usepackage{placeins}
\usepackage[T1]{fontenc}
\usepackage{lipsum}
\usepackage{csquotes}
\usepackage{siunitx}
\usepackage{import}
\usepackage[pdftex, pdftitle={Article}, pdfauthor={Author}]{hyperref} 
\newcommand{\cuprite}{Cu$_\mathrm{2}$O}

\newcommand{\aMW}{\ensuremath{\Delta\alpha}}
\newcommand{\abMW}{\ensuremath{\Delta\bar{\alpha}}}

\begin{document}
\title{
Microwave-optical coupling via Rydberg excitons in cuprous oxide
}

\author{Liam A. P. Gallagher, Joshua P. Rogers,  Jon D. Pritchett, Rajan A. Mistry, Danielle Pizzey, Charles S. Adams, Matthew P. A. Jones}
    \email[Correspondence email address: ]{m.p.a.jones@durham.ac.uk}
    \affiliation{Department of Physics, Durham University, Durham DH1 3LE, United Kingdom}
\author{Peter Gr\"{u}nwald}
\affiliation{Center for Complex Quantum Systems, Department of Physics and Astronomy, Aarhus University,
Ny Munkegade 120, 8000 Aarhus C, Denmark}
\author{Valentin Walther}
\affiliation{ITAMP, Harvard-Smithsonian Center for Astrophysics, Cambridge, Massachusetts 02138, USA}
\author{Chris Hodges, Wolfgang Langbein, Stephen A. Lynch}
    \affiliation{School of Physics and Astronomy, Cardiff University, Cardiff CF24 3AA, United Kingdom}
\date{\today} 

\begin{abstract}
We report exciton-mediated coupling between microwave and optical fields in cuprous oxide (\cuprite) at low temperatures. Rydberg excitonic states with principal quantum number up to $n=12$ were observed at $4$~\si{\kelvin} using both one-photon (absorption) and two-photon (second harmonic generation) spectroscopy. Near resonance with an excitonic state, the addition of a microwave field significantly changed the absorption lineshape, and added sidebands at the microwave frequency to the coherent second harmonic. Both effects showed a complex dependence on $n$ and angular momentum, $l$. All of these features are in semi-quantitative agreement with a model based on intraband electric dipole transitions between Rydberg exciton states. With a simple microwave antenna we already reach a regime where the microwave coupling (Rabi frequency) is comparable to the nonradiatively broadened linewidth of the Rydberg excitons. The results provide a new way to manipulate excitonic states, and open up the possibility of a cryogenic microwave to optical transducer based on Rydberg excitons. 
\end{abstract}

\maketitle

\section{Introduction}
Improved coupling  between microwave and optical frequencies would enhance classical telecommunications as well as finding applications in distributed quantum networks and quantum communication. Solid-state quantum bits (qubits) that operate at microwave frequencies have been demonstrated using superconducting circuits~\cite{Devoret2013,Kelly2015,Hofheinz2008} and quantum dots~\cite{Watson2018}. These architectures offer a high degree of control over quantum states and qubit coupling. However, the effects of thermal noise present problems for transporting the microwave quantum information over large distances~\cite{Kurpiers2018,Zhong2019}. Conversely, optical quantum communication has been demonstrated over global length scales~\cite{Liao2017,Boaron2018,Yu2020}. To connect these two regimes, a hybrid quantum system allowing efficient conversion between optical and microwave frequencies is required~\cite{Xiang2013,Clerk2020}. Microwave to optical conversion has been demonstrated to varying degrees using mechanical oscillators~\cite{Stannigel2010,Forsch2020}, nonlinear crystals~\cite{Fan2018} and Rydberg atoms~\cite{Fan_2015,Gard2017,Han2018,Petrosyan_2019,chopinaud2021}. 

Rydberg atoms show promise due to the large electric dipole moment associated with transitions between Rydberg states of opposite parity, which scales with principal quantum number, $n$, as $n^2$. However compatibility between the laser cooling technology required for Rydberg atoms and the millikelvin dilution refrigerator environment of superconducting qubits is an ongoing challenge~\cite{Hattermann2017,morgan2019,Petrosyan2009,kaiser2021}. An alternative Rydberg platform which is more compatible with other solid state devices is offered by excitonic states in semiconductors. Excitons in \cuprite~ are a solid state analogue of hydrogen atoms~\cite{hayashi1952,Gross1956,Agekyan1977,Washington1977,Kavoulakis1997}. Rydberg exciton states with principal quantum number $n>25$ have been observed~\cite{Kazimierczuk2014,versteegh2021} with the advantage over Rydberg atoms that these states exist in the solid state and are straightforward to observe in a dilution refrigerator environment required for superconducting qubits ~\cite{Snoke2002,Heckotter2020,versteegh2021}.

Electric dipole transitions between excitonic states with low principal quantum number have  been widely studied using far infra-red and terahertz spectroscopy~\cite{Frohlich1985,FROHLICH1987,Frohlich1990,Frohlich1990-2,Jorger2003,BASSANI2004,Kuwata2004,Kubouchi2005,Tayagaki2005,Huber2006,Yoshiaka2007,Huber2008,Yoshiaka2010,Yoshioka2012}. The energy separations in a Rydberg series scale with $n$ as $n^{-3}$. Combined with the reduced Rydberg constant of excitonic states, this scaling means that for the ``yellow'' series in \cuprite, electric dipole transitions accessible to microwave frequencies on the order of a few tens of \si{\giga\hertz} occur for states as low as $n=8$.  These strong electric dipole transitions are responsible for the long-range van der Waals interactions and Rydberg blockade observed in \cuprite~\cite{Kazimierczuk2014,Heckotter2018,Walther2018}, with potential applications in creating quantum states of light~\cite{Grunwald2016,Walther2018nl,Khazali2017}. A novel tuneable maser has also been proposed based on these transitions~\cite{Ziemkiewicz2018}.

In this paper, we observe a coupling between the electric field produced by a simple planar microwave circuit and the optical properties of cuprous oxide. The effect of the microwave field is studied using both one-photon absorption spectroscopy and second harmonic generation (SHG) spectroscopy~\cite{MATSUMOTO1996,Ketterson2001,Naka2013,Mund2018,Mund2019,Farenbruch2020,Rommel2020}. In contrast to atomic Rydberg states~\cite{Dyubko1995,Ryabtsev2001}, non-radiative broadening of the excitonic energy levels gives rise to a continuous microwave spectrum. Our observations are in good agreement with a model based on intraband electric dipole transitions between excitonic states of opposite parity. These results provide a new tool for manipulating Rydberg states of excitons, and the first step to building a microwave to optical transducer based on Rydberg excitons. 
\begin{figure}
    \centering
    \includegraphics[width=\linewidth]{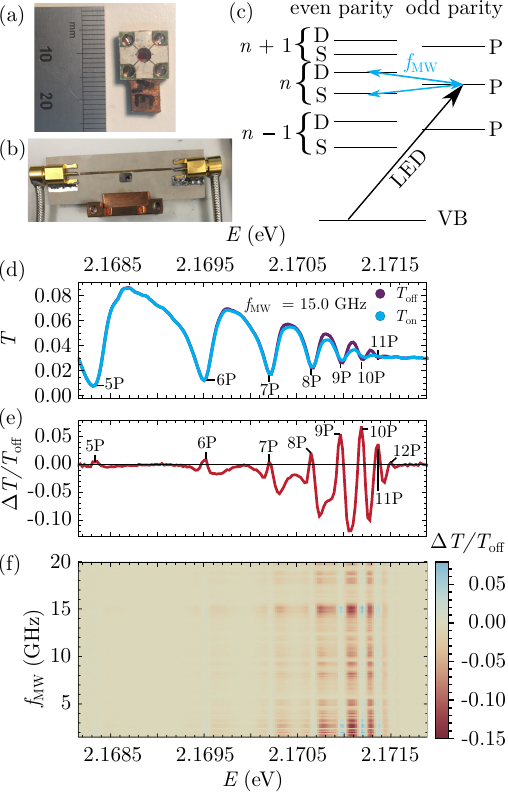}
    \caption{Effect of a microwave field on the one-photon absorption spectrum of \cuprite. (a) Antenna A1 and (b) Antenna A2 used to deliver microwave fields to the sample. (c) Energy level diagram for the one-photon experiment. LED light probes odd parity (P) states. A microwave field of frequency $f_\mathrm{MW}$ introduces coupling between the even and odd parity states. (d) Broadband transmission spectrum with (blue) and without (purple) a microwave field at frequency $f_\mathrm{MW}=15.0~\si{\giga\hertz}$ using antenna A2. Exciton states from 5P to 11P are visible. The effect of the microwave field is more prominent at higher $n$. (e) Relative change in transmission, $\Delta T/T_{\rm off}$, due to the microwave field at $f_\mathrm{MW}=15.0~\si{\giga\hertz}$ using antenna A2. The intensity is increased at P resonances and decreased at the energies of the even parity states, indicating a mixing between the even and odd parity states. (f) Heatmap of  $\Delta T/T_{\rm off}$ as a function of excitation energy, $E$, and microwave frequency, $f_\mathrm{MW}$. Note that (e) is a cross section of this heatmap at  $f_\mathrm{MW}=15~\si{\giga\hertz}$. Fine structure in $f_\mathrm{MW}$ dimension is attributed to the frequency response of the antenna (see Appendix~\ref{app:antenna}).}
    \label{fig:fig1}
\end{figure}
\section{Experiment} \label{sec:experiment}

Cuprous oxide is a direct band gap semiconductor (bandgap energy of 2.172~eV). Spin-orbit coupling leads to a splitting of the valence band. In this paper we study optical transitions between the upper level of the valence band ($\Gamma_7^+$ symmetry) and excitonic states associated with the lowest level of the conduction band ($\Gamma_6^+$ symmetry), referred to as the yellow exciton series (570 -- 610~\si{\nano\meter}). Both energy levels have the same parity, and so excitonic states with odd parity (P, F orbital symmetry) are accessible by single photon electric dipole transitions, while even parity states (S, D) are accessible via two-photon excitation (and electric quadrupole processes). Here we provide details of both one- and two- photon spectroscopy in the presence of microwave fields.

The experiments were performed on a naturally formed \cuprite\ gemstone from the Tsumeb mine in Namibia. The crystal was oriented such that the (111) crystal plane was parallel with the surface and mechanically polished on both sides to a thickness of $\sim 50$~\si{\micro\meter}. A $2\times 3$~\si{\milli\meter} rectangle of this slice was mounted on a \O~5~\si{\milli\meter} CaF$_2$ window in a copper mount, shown in Figure~\ref{fig:fig1}~(a) and (b). A small quantity of glue was applied to one corner of the sample in order to maintain adequate thermal contact with the CaF$_2$ window and copper mount. Details of sample mounting and preparation are available in~\cite{Lynch2021}. The sample was cooled in a low-vibration closed-cycle helium refrigerator to $\sim 4$~\si{\kelvin}. 

Microwave fields were applied using one of the two antennae shown in Fig.~\ref{fig:fig1}(a) and (b). Antenna A1 (Fig.~\ref{fig:fig1}(a)) is a simple printed circuit board with four pads. Two adjacent pads were connected to the microwave generator and two were grounded. Antenna A2 (Fig.~\ref{fig:fig1}(b)) is a stripline design with an input and output port, with the latter terminated externally at 50~$\Omega$. The relevant antenna was connected to a commercial microwave synthesizer delivering frequencies of up to $f_\mathrm{MW} = 20~\si{\giga\hertz}$ and powers $P_\mathrm{MW}$ up to 25~\si{\milli\watt}. In both cases the sample was placed in the near field of the antenna, at the centre of the pads for A1 and between the conductors for the stripline design. The frequency response of each antenna was found to be strongly affected by the presence of metallic components such as the sample mount and lens holders. Using electromagnetic design software, A1 was found to only create an appreciable electric field at the sample within  bands of microwave frequencies, $f_\mathrm{MW}$, around 16 and 19~\si{\giga\hertz}, while the response of A2 was more broadband with superimposed narrow resonances. The maximum achieved screened electric field inside the sample in the simulations at $P_\mathrm{IN}=25~\si{\milli\watt}$ was calculated to be 360~\si{\volt\per\meter} and  1200~\si{\volt\per\meter} for antennae A1 and A2 respectively using a dielectric constant of $\epsilon_r = 7.5$ for \cuprite. Details of these calculations are provided in Appendix~\ref{app:antenna}.

\subsection{Microwave control of optical transmission}\label{sec:LED}

One-photon absorption spectroscopy was performed using a broadband LED as a light source. An energy level diagram of the one-photon experiment is shown in Fig.~\ref{fig:fig1}(c). Broadband LED light (width of 14~nm centred at 580~nm) excites odd parity P states in the yellow series of excitonic energy levels. The spectrum of the light transmitted by the sample was measured using a grating monochromator with a resolution of 70~\si{\micro}eV. The resulting transmission spectrum is shown in Fig.~\ref{fig:fig1}(d). In this energy range, the absorption is dominated by the background associated with phonon-assisted transitions involving the lowest-lying 1S exciton~\cite{Baumeister1961,Schone2017}, with superimposed resonances associated with $n$P excitonic states. Our data shows excitons from $n=5$ to $n=11$, with the observation of higher $n$ states limited by the spectral resolution of the monochromator.

The effect of the application of a microwave electric field at $f_\mathrm{MW}=15.0~\si{\giga\hertz}$ using antenna A2 is shown in Fig.~\ref{fig:fig1}(d). While there is no discernible change for the lowest excitonic states, the region with $n>7$ is substantially modified. These changes are highlighted by plotting the fractional change in transmission ($\Delta T /T_\mathrm{off}~=~(T_\mathrm{on}~-~T_\mathrm{off})/T_\mathrm{off}$) as shown in Fig.~\ref{fig:fig1}(e). Here, we can see that the microwave field changes the transmitted intensity by more than 10\% at certain energies. The microwave frequency dependence is illustrated in Fig.~\ref{fig:fig1}(f), which shows the change in transmission as a function of the microwave frequency $f_\mathrm{MW}$ and excitation energy. A strong response is observed over a broad range of microwave frequencies from $1-20$~GHz, modulated by a complex structure of resonances that are independent of the excitation energy that we attribute to the antenna response (see Appendix~\ref{app:antenna}).

The changes to the transmission spectrum seen in Fig.~\ref{fig:fig1} can be understood in terms of the mixing of opposite parity states (Fig.~\ref{fig:fig1}(c)). State mixing leads to an increase in absorption on the S and D states (which acquire some P character) and a decrease in absorption on the P states. As the exciton states are broad (full width half maximum of 14 GHz at 8P) relative to their separation (8P to 8S is 23 GHz) the microwave response is broadband, with many transitions contributing at each value of the microwave frequency $f_\mathrm{MW}$. This is in contrast to atomic Rydberg states where the atomic linewidth is considerably smaller than the separation between states, leading to sharp resonances at discrete microwave frequencies. This alteration in the microwave response is a consequence of the crystalline environment, since the increased width of the excitonic Rydberg states is due to non-radiative decay via phonons. 
 
\subsection{Microwave modulated second harmonic generation}\label{sec:SHG}
To probe these effects in more detail, we switch to second harmonic generation (SHG) spectroscopy. Second harmonic generation in \cuprite~has been studied by several authors, with a comprehensive discussion of the selection rules provided in~\cite{Mund2018,Mund2019,Farenbruch2020,Rommel2020}. SHG spectroscopy offers several advantages. The second harmonic is coherently generated with an emission spectrum determined by the excitation laser, and is easily separated from the excitation light. In addition, the second harmonic generation spectrum does not exhibit the large phonon-assisted background observed in one-photon transmission spectroscopy~\cite{Baumeister1961,Kazimierczuk2014,Schone2017}. Together these advantages enable us to observe the modulation of an optical carrier by the microwave field.

\begin{figure*}
    \centering
    \includegraphics[width=\linewidth]{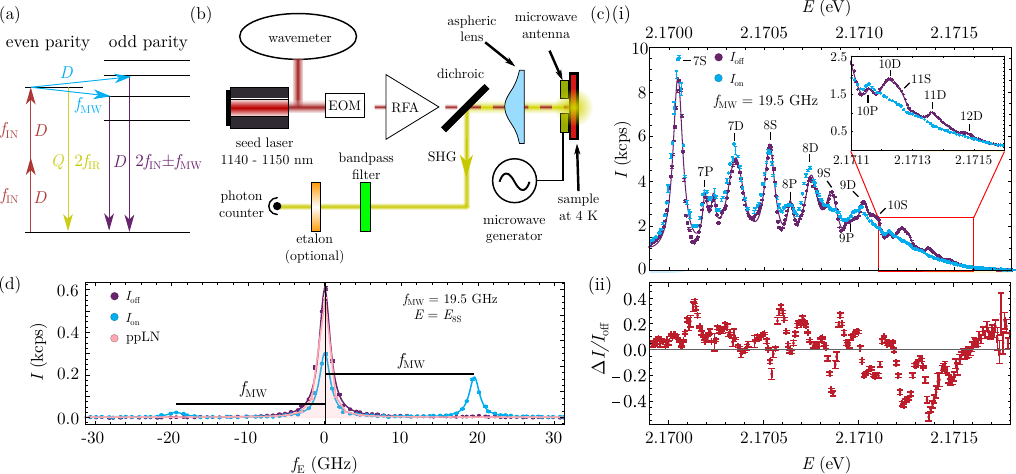}
    \caption{Second harmonic generation spectroscopy of \cuprite\ with a microwave field. (a)~Energy level diagram of the SHG experiment. Labels $D$ and $Q$ indicate whether the step occurs through a dipole or quadrupole process. Two-photon excitation at frequency $f_{\mathrm{IN}}$ excites an even parity (S or D) exciton. Emission can occur from this state through a quadrupole process at frequency $2f_\mathrm{IN}$.  The addition of a microwave field of frequency $f_\mathrm{MW}$ couples the even and odd parity exciton states through electric dipole transitions leading to a four-wave mixing type process and new emission pathways at $2f_\mathrm{IN}\pm f_\mathrm{MW}$. (b)~Experiment block diagram. The seed laser light is sliced into pulses by an electro-optic modulator (EOM) and amplified by a Raman fiber amplifer (RFA), before being focussed onto the sample. Backscattered light is collected and detected using a photon counter. A scanning etalon may be inserted to provide additional filtering. (c)(i)~Emitted second harmonic intensity, $I$, as a function of two-photon excitation energy, $E$, with ($I_\mathrm{on}$; blue) and without ($I_\mathrm{off}$; purple) a microwave field at 19.5~\si{\giga\hertz}. Resonance from $n=7$ to $n=12$ are visible. Solid purple line shows fit to $I_\mathrm{off}$. (ii) Fractional change in intensity, $\Delta I/I_\mathrm{off}$, of the excitation spectrum with $f_\mathrm{MW} = 19.5~\si{\giga\hertz}$. The microwave field alters the SHG spectrum throughout the range of two-photon excitation spectrum. (d)~Spectrally resolved emitted second harmonic intensity, $I$, at  $E=E_\mathrm{8S}$, with (blue) and without (purple) a microwave field at $f_\mathrm{MW}=19.5$~\si{\giga\hertz}
    as function of etalon detuning, $f_{\rm E}$. Red shaded area shows light doubled through a ppLN crystal for comparison. The microwave field causes the appearance of sidebands on the second harmonic.}
    \label{fig:fig2}
\end{figure*}

An energy level diagram of the SHG experiment is shown in Fig.~\ref{fig:fig2}(a). A two-photon excitation of frequency $f_{\mathrm{IN}}$ excites an even parity exciton through two dipole processes. The even parity exciton coherently emits light at twice the input frequency, $2f_{\mathrm{IN}}$. Note that the emission from an even parity state is dipole-forbidden due to parity, and so can only occur as an electric quadrupole process. The microwave field introduces a coupling between the even and odd parity exciton states through electric dipole transitions. This leads to the possibility of a four-wave mixing process occurring and the appearance of two additional frequency components appearing in the spectrum of the emitted light, at a frequency $2f_\mathrm{IN}\pm f_\mathrm{MW}$.

The experimental setup is shown in Figure~\ref{fig:fig2}~(b). The excitation light was generated by an external cavity diode laser that is tunable from 1140 to 1150~\si{\nano\meter} (linewidth of $\sim 10$~\si{\nano\electronvolt}). The frequency of the seed laser was stabilized to a precision wavemeter ($\pm 60~\si{\mega\hertz}$) using a computer-controlled servo loop. The continuous wave (CW) seed laser was amplitude modulated by a fiber-coupled electro-optic modulator (EOM) to create square pulses with duration $\tau=50$~\si{\nano\second} and period $T=200$~\si{\nano\second}. The light was amplified by a commercial Raman fiber amplifier (RFA). The average power reaching the sample was monitored by a pickoff adjacent to the cryostat window and was typically set to be 50~\si{\milli\watt}. An acousto-optic modulator after the amplifier was used to stabilize average power to within 1\%. The excitation light was subsequently focused onto the sample using an aspheric lens with numerical aperture 0.6 to give a 1/e$^2$ waist of approximately 0.5~\si{\micro\meter} inside the sample. 

The same aspheric lens was used to collect the light emitted by the sample in a backscattering geometry (in the bulk of the material SHG is generated in the forward direction).  A 785~\si{\nano\meter} long-pass dichroic mirror followed by two 1000~nm short-pass filters were used to remove residual excitation light. In addition, a bandpass filter centred at 580~\si{\nano\meter} was used to separate the coherently generated second harmonic from photoluminescence (PL) at the energy (wavelength) of the 1S exciton state at 610~\si{\nano\meter}~\cite{Naka2018}. The backscattered second harmonic was coupled into a multimode optical fibre and sent to a photon counter for detection. For some experiments, a planar fused silica Fabry-P\'{e}rot etalon was inserted in the beam path before the detection fiber. The etalon was tuned by varying its temperature. The spectral response of the etalon was calibrated by using a periodically poled lithium niobate (ppLN) crystal to coherently generate the second harmonic of the laser light, yielding a finesse of $44.5 \pm 0.7$ and a free spectral range (FSR) of $60.1 \pm 0.2~\si{\giga\hertz}$.

An excitation spectrum was taken by scanning the laser in 0.5~\si{\giga\hertz} steps. At each step we recorded the wavelength measured by the wavemeter and the SHG intensity  averaged over 4 seconds.  Example results for the spectral region covering  $n=7$ to 12 are shown in Fig.~\ref{fig:fig2}~(c)(i), plotted against the two-photon excitation (TPE) energy $E = 2 h f_{\mathrm{IN}}$. As expected under two-photon excitation, the even parity (S and D) states are prominent. Odd parity excitons are also present between the S and D peaks. The P states have previously been observed in SHG and are attributed to a quadrupole excitation  process~\cite{Mund2018}. 



To observe the effect of the microwaves on the SHG spectrum, microwaves were applied in 0.5 second pulses with a 50\% duty cycle, enabling concurrent measurement of the spectrum both with and without the microwave field. For the experiments involving SHG, antenna A1 (Fig.~\ref{fig:fig1}(a)) was used. The effect of a microwave field ($f_\mathrm{MW}=19.5$~\si{\giga\hertz}) on the SHG spectrum is shown in Fig.~\ref{fig:fig2}(c). The spectrum is modified throughout the range of TPE energy, with some excitonic resonances enhanced, and others suppressed. Fig.~\ref{fig:fig2}(c)(ii) plots the fractional change in intensity, $\Delta I/ I_\mathrm{off} = (I_\mathrm{on} - I_\mathrm{off})/I_\mathrm{off}$. We note that in some regions the fractional change in signal is larger than 40\%. As was the case for the experiments performed in section~\ref{sec:LED} the dependence of the signal on $f_\mathrm{MW}$ was dominated by the response of the antenna (see Appendix~\ref{app:antenna}).

To investigate the effect of the microwaves further, we performed high-resolution spectroscopy of the SHG light using the temperature-tuned etalon shown in Fig.~\ref{fig:fig2}(b). An emission spectrum obtained by scanning the etalon with the TPE energy fixed as $E=E_\mathrm{8S}$ is shown in Fig.~\ref{fig:fig2}(d). With the microwaves off, we observe a single frequency component (the SHG carrier) with a lineshape that is in excellent agreement with that obtained using the ppLN crystal, confirming that this is indeed the coherently generated second harmonic. The addition of the microwave field leads to the appearance of strong sidebands at $f_\mathrm{E}=\pm f_\mathrm{MW}$, accompanied by significant depletion of the carrier. We have not observed higher-order sidebands. The sidebands in Fig.~\ref{fig:fig2}(d) are not of equal strength, the sideband at $+19.5~\si{\giga\hertz}$ (blue sideband) is significantly larger than the one at $-19.5~\si{\giga\hertz}$ (red sideband). The relative amplitude of the sidebands and the carrier, and the sign and magnitude of the asymmetry between the sidebands is strongly dependent on $E$ and $f_\mathrm{MW}$. We note that the total count rate in the SHG spectrum with microwaves (Fig.~\ref{fig:fig2}(c)) represents the sum of these three components.


\section{Theory}\label{sec:theory}

In this section we show that both the changes to the absorption spectrum seen in Fig.~\ref{fig:fig1} and the generation of sidebands in the SHG process shown in Fig.~\ref{fig:fig2} can be explained in terms of the electric dipole transitions between excitonic states of opposite parity. Using the exciton-polariton description of light-matter interactions, we  derive the non-linear susceptibility for one- and two- photon excitation processes (see Appendix \ref{app:theory}), taking into account all dipole--allowed microwave couplings. At low microwave intensities and considering individual excited states, we recover the results of pioneeering studies of the $\mathrm{2P} \to \mathrm{1S}$ electric dipole transition in \cuprite~\cite{Frohlich1985,Saikan1985,Frohlich1990,Frohlich1990-2}. In this limit, the effect can be understood in terms of an AC Stark shift of the excitonic energy levels due to the microwave electric field~\cite{Frohlich1985}.

In general, the light-matter coupling is a tensor depending on the crystallographic orientation and the polarizations of the optical and microwave fields. However, due to stress-induced birefringence in the CaF$_2$ windows, and the complex polarization behaviour of the antenna structures, we could not study polarization effects. Therefore in the following we neglect polarization and consider only an effective scalar coupling. 

\subsection{Microwave modulation of optical transmission}\label{sec:LED_theory}

First we consider the one-photon absorption experiments presented in section~\ref{sec:LED}. Neglecting reflection, the transmission, $T$, through a material can be modelled by the Beer-Lambert law~\cite{f2f_lightmatter} as $T=\exp(-\alpha L)$, where $\alpha$ is the absorption coefficient and $L$ is the thickness of the material. The absorption coefficient is related to the imaginary part of the susceptibility, $\chi$, by $\alpha = k~\mathrm{Im}(\chi)$, where $k$ is the wavenumber of the light. One-photon absorption in \cuprite\ can be described by a linear susceptibility, $\chi^{(1)}$, which has contributions from both $\ket{n,\mathrm{P}}$ states and the phonon background. The contribution to the susceptibility from the $\ket{n,\mathrm{P}}$ state is given by 
\begin{equation}
\chi^{(1)}_{n\mathrm{P}} = \frac{1}{2 \epsilon_0 \hbar \eta} \frac{ \left|D^{\mathrm{VB}\rightarrow n\mathrm{P}}\right|^2}{\delta_{n\mathrm{P}}-i\Gamma_{n\mathrm{P}}}.
\label{eq:chi1}
\end{equation}
Here, $\delta_{n\mathrm{P}} = (E_{n\mathrm{P}}-E)/\hbar$ is the detuning, $\left|D^{\mathrm{VB}\rightarrow n\mathrm{P}}\right|^2$ is the dipole moment per unit volume for the transition between the valance band (VB) and the $\ket{n,\mathrm{P}}$ state, $E_{n\mathrm{P}}$ and $\Gamma_{n\mathrm{P}}$ are the energy and width of the $\ket{n,\mathrm{P}}$ state, $\eta$ is the refractive index of the material and $E$ is the excitation photon energy.

To model the change in absorption due to the microwave field, we introduce a coupling between the even and odd parity exciton states through electric dipole transitions, which gives rise to a third-order cross-Kerr nonlinearity~\cite{Frohlich1985}. The contribution to the nonlinear susceptibility from coupling the $\ket{n,\mathrm{P}}$ and $\ket{n',l'}$ states is given by
\begin{widetext}
\begin{equation}
\chi^{(3)}_{n\mathrm{P}n'l' } = \frac{1}{2\epsilon_0 \hbar^3\eta} \frac{\left|D^{\mathrm{VB}\rightarrow n\mathrm{P}}\right|^2 \left|d^{n\mathrm{P}\rightarrow n'l'}\right|^2 }{(\delta_{n\mathrm{P}}-i\Gamma_{n\mathrm{P}})^2(\delta_{n'l'}^\pm-i\Gamma_{n'l'})}=\chi^{(1)}_{n\mathrm{P}}\frac{\left|d^{n\mathrm{P}\rightarrow n'l'}\right|^2 }{\hbar^2(\delta_{n\mathrm{P}}-i\Gamma_{n\mathrm{P}})(\delta_{n'l'}^\pm-i\Gamma_{n'l'})}.
\label{eq:chi3LED}
\end{equation}
\end{widetext}
Here, $d^{n\mathrm{P}\rightarrow n'l'}= \bra{n',l'}e\cdot\mathbf{r}\ket{n,\mathrm{P}}$ is the dipole matrix element and $\delta^\pm_{n'l'}= (E_{n'l'} - E)/\hbar \mp 2\pi f_\mathrm{MW}$ is the detuning from the $\ket{n',l'}$ state. Both microwave absorption and emission, corresponding to plus or minus, respectively, need to be retained as the linewidths of the states involved are comparable to the microwave frequency meaning the rotating-wave approximation cannot be made. Summing over the P states, and all corresponding dipole-coupled states $\ket{n^\prime, l^\prime}$, we find the change in absorption coefficient due to the microwaves 
\begin{equation}
    \aMW = k~\mathrm{Im}\left(\sum_{n,n',l',\pm} \chi^{(3)}_{n\mathrm{P}n'l'}\right) \mathcal{E}_\mathrm{MW}^2,
    \label{eq:delta a}
\end{equation}
where $\mathcal{E}_\mathrm{MW}$ is the effective microwave electric field inside the sample.

To relate the absorption coefficient to the measured transmission, we must take into account the spectral response of the monochromator. The optical transmission, $T$, measured in Fig.~\ref{fig:fig1}(d), can be written as a convolution $T = (S*e^{-\alpha L})$ over the photon energy $E$, where $S$ is the normalised zero-centred response function of the spectrometer. The absorption coefficient can be written as $\alpha =  \alpha_0 + \aMW$, where $\alpha_0$ is the absorption coefficient in the absence of microwaves. As is apparent in Fig.~\ref{fig:fig1}(d) $\alpha_0 \gg  \aMW$. In this limit, the measured fractional change in transmission can be approximated as
\begin{equation}\label{eq:delta T}
\frac{\Delta T}{T_\mathrm{off}} \approx - (S * \aMW)L\equiv - \abMW L,
\end{equation}
where we introduced the convoluted quantity \abMW.

The parameters in equations~(\ref{eq:chi3LED})--(\ref{eq:delta T}) can be determined experimentally or calculated from theory. $S(E)$ was measured using frequency-doubled laser light. $\chi^{(1)}(E) $ can be obtained from fitting the absorption spectrum in the absence of microwaves (Fig.~\ref{fig:fig1}(d)). The energies and widths of the states can similarly be obtained from fitting the one-photon and two-photon (Fig.~\ref{fig:fig2}(c)(i)) spectra. The matrix elements, $d^{n\mathrm{P}\rightarrow n'l'}$, were calculated from theoretical exciton wavefunctions~\cite{Schone2016-2,Walther2018}.

\begin{figure}
    \centering
    \includegraphics[width=\linewidth]{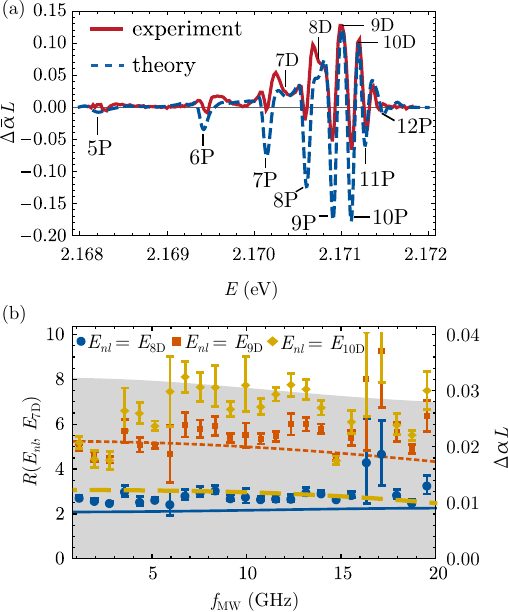}
    \caption{Comparison of the predicted and  measured change in one-photon absorption due to the microwave field. (a) Change in absorption due to the microwave field, $\abMW L$, as a function of excitation energy, $E$, at microwave frequency $f_\mathrm{MW}=15.0~\si{\giga\hertz}$. The range of $E$ spans from the $n=5$ state up to the band edge. Experimental data is shown as solid red line, and theoretical predictions as dashed blue line for $\mathcal{E}_\mathrm{MW}=400$~\si{\volt\per\meter}. (b) Predicted change in absorption as a function of microwave frequency, $f_\mathrm{MW}$ (shaded background, right axis) at excitation energy $E=E_\mathrm{7D}$. Also shown is the measured (points) and predicted (lines) ratio of the change in absorption at $E=E_{nl}$ to $E_\mathrm{7D})$ as a function of microwave frequency (left axis). Three different values of $E_{nl}$ are shown, $E_{nl}=E_\mathrm{8D}$ (blue circles, solid line) $E_{nl}=E_\mathrm{9D}$ (orange squares, dotted line) and $E_{nl}=E_\mathrm{10D}$ (gold diamonds, dashed line) states.}
    \label{fig:LED_theory}
\end{figure}

Fig.~\ref{fig:LED_theory}(a) shows the predicted $\abMW L$ as a function of $E$ at $f_\mathrm{MW}=15.0~\si{\giga\hertz}$. Here the effective (unpolarized) electric field strength $\mathcal{E}_\mathrm{MW}$ is used as a fitting parameter and found to be $400\pm100$~\si{\volt\per\meter}. This value is in reasonable agreement with the calculated field inside the sample for antenna A2 of 1200~\si{\volt\per\meter}, given that polarization effects and experimental insertion losses were not taken into account. More generally, there is good qualitative agreement between the data and the model, which reproduces all of the observed spectral features.  The main discrepancy is that the model overestimates the reduction in absorption seen at the P states. 

We note that this model is a perturbative approach  which assumes a low microwave field strength. This assumption is valid when the Rabi frequency ${\Omega_{n\mathrm{P}n'l'} = d^{n\mathrm{P}\rightarrow n'l'}\mathcal{E}_\mathrm{MW}/\hbar}$ is considerably smaller than $\Gamma$. For the 8S$\to$8P transition with an effective field strength of 400~\si{\volt\per\meter} the effective Rabi frequency is $\Omega_\mathrm{8P8S} = 2\pi \times 9~\si{\giga\hertz}$ and the ratio of the Rabi frequency to the linewidth is $\Omega_\mathrm{8P8S}/\Gamma_\mathrm{8P}\approx 0.4$. This ratio increases with $n$; for the 10P$\to$10D transition we obtain $ \Omega_\mathrm{10P10D}/\Gamma_\mathrm{10D}\approx 0.9$. These values show that it is possible to achieve a coupling strong enough to match the large nonradiative contribution to $\Gamma$ even with the simple antenna designs used in this work. In the limit $\Omega  \sim \Gamma$  higher order terms in the nonlinear susceptibility will start to become significant, which may account for some of the discrepancies seen between the theory and experiment in Fig.~\ref{fig:LED_theory}. 

The microwave frequency dependence predicted by the model at $E=E_\mathrm{7D}$ is shown as the shaded background in Fig.~\ref{fig:LED_theory}(b). As expected, the response is broadband due to the linewidth and large number of states that contribute. To compare to experiment, we remove the effect of the antenna resonances by taking the ratio of the microwave response at different excitation energies, $R(E_1,E_2) = \abMW(E_1)/\abMW(E_2)$. In Fig.~\ref{fig:LED_theory}(b) we fix $E_2 = E_\mathrm{7D}$, and plot $R$ for $E_1 = E_\mathrm{8,9,10 D}$. In all cases $R$ is nearly constant with microwave frequency $f_\mathrm{MW}$ as predicted by the model. Indeed for the $\mathrm{8D}$ and $\mathrm{9D}$ states we observed quantitative agreement between theory and experiment with no free parameters. For $\mathrm{10D}$ $R$ is underestimated by the model, which we attribute to the likely breakdown of the perturbative approach discussed above.

\subsection{Second harmonic generation in the presence of a microwave field}

The model described in the previous section can be extended to the SHG experiments presented in~\ref{sec:SHG}. Here we show that such an extension results in a semi-quantitative description of the appearance of sidebands and reduction of the carrier intensity observed in Fig.~\ref{fig:fig2}(d) that is in good agreement with the experimental data. Details of the derivation are provided in Appendix~\ref{app:theory}. 

First, let us describe the SHG process in the absence of the microwave field.  The excitation is achieved through two dipole transitions and the emission by a quadrupole transition~\cite{Mund2018}. The SHG intensity in the absence of a microwave field, $I_\mathrm{off}$, is given by
\begin{align}
\begin{split}
     I_\mathrm{off} & = A I_\mathrm{IN}^2  \left| \sum_{n,l=\mathrm{S,D}} \frac{1}{\epsilon_0 \hbar} \frac{Q^{nl\rightarrow\mathrm{VB}}M^{\mathrm{VB}\rightarrow nl}}{\delta_{nl}-i\Gamma_{nl}}\right|^2,\\
     & =AI_\mathrm{IN}^2   \left| \sum_{n,l=\mathrm{S,D}} \chi^{(2)}_{nl}\right|^2.\label{eq:SHG_Ioff}
\end{split} 
\end{align} 

Here $\chi_{nl}^{(2)}$ is the contribution from the $\ket{n,l}$ state to the second order nonlinear susceptibility, $\left|Q^{nl\to\mathrm{VB}}\right|^2$ is the quadrupole moment per unit volume, $M$ is an effective matrix element describing the two step excitation process, $\delta_{nl} = (E_{nl} - E)/\hbar$ is the detuning from the $\ket{n,l}$ state with associated excitation energy $E=2 h f_{\mathrm{IN}}$, $I_\mathrm{IN}$ is the intensity of the excitation laser and $A$ is a proportionality constant which depends on the phase matching condition, the length of the crystal, the frequency of the light and the refractive index of \cuprite. We note that neglecting polaritonic effects~\cite{FROHLICH2005,Farenbruch2021} SHG in \cuprite\ is not well phase matched and so is an inefficient process. A similar expression to equation~\ref{eq:SHG_Ioff} can be constructed for the contribution to $I_\mathrm{off}$ from the odd parity states~\cite{Mund2018,Farenbruch2020,Rommel2020}. However for simplicity we neglect this process in the following, and assume that only even-parity states are excited in the absence of microwaves. 

Now let us consider the case where a microwave field is applied. The first two steps of the four-wave mixing process are the same as the SHG process and can be described by the same effective matrix element, $M$. The microwave field introduces an electric dipole coupling to odd-parity Rydberg states $\ket{n',l'}$ which results in an additional electric dipole-allowed emission process. This can viewed as a four-wave mixing process that leads to the creation of the sidebands on the second harmonic in Fig.~\ref{fig:fig2}(d). Again, we note that the large linewidth of the exciton resonances compared to their separation means that the conventional near-resonant rotating-wave approximation cannot be made. The resulting expression for the intensity $I_\mathrm{SB}$ of the sidebands is

\begin{widetext}

\begin{equation}\label{eq:I_sdbnd}
    I_\mathrm{SB}^\pm =  A I^2_\mathrm{IN}  \left| \sum_{n,l=\mathrm{S,D}}\sum_{n'l'}\frac{1}{\epsilon_0 \hbar^2} \frac{D^{n'l'\rightarrow\mathrm{VB}}\Omega_{nln'l'}M^{\mathrm{VB}\rightarrow nl}}{(\delta_{nl}-i\Gamma_{nl})(\delta_{n'l'}^\pm-i\Gamma_{n'l'})-i\Omega_{nln'l'}^2\left(1+\frac{\delta_{n'l'}^\pm-i\Gamma_{n'l'}}{\delta_{n'l'}^\mp-i\Gamma_{n'l'}}\right)}\right|^2,
\end{equation}

\end{widetext}
where the $\pm$ corresponds to the blue and red sidebands respectively.
 
As can be seen in Fig.~\ref{fig:fig2}(d) the intensity of the carrier peak is also altered by the presence of the microwave field. The intensity of the carrier peak when the microwave field is on, $I_\mathrm{CAR}$, is given by
\begin{widetext}
\begin{equation}\label{eq:SHG_Ion}
    I_\mathrm{CAR} = A I_\mathrm{IN}^2  \left| \sum_{n,l=S,D}\sum_{n'l'}\frac{1}{\epsilon_0 \hbar} \frac{Q^{nl\rightarrow\mathrm{VB}}M^{\mathrm{VB}\to nl}}{(\delta_{nl}-i\Gamma_{nl}) +\Omega_{nln'l'}^2\left(\frac{1}{\delta_{n'l'}^--i\Gamma_{n'l'}}+\frac{1}{\delta_{n'l'}^+-i\Gamma_{n'l'}}\right)}\right|^2.
\end{equation}
\end{widetext}
The change in carrier intensity, $I_\mathrm{CAR}$ due to the microwave field is given by 
\begin{equation}\label{eq:SHG_deltaI}
    \Delta I_\mathrm{CAR}= I_\mathrm{CAR} - I_\mathrm{off}.
\end{equation}
Detailed derivations of equations~\ref{eq:SHG_Ioff}, \ref{eq:I_sdbnd} and~\ref{eq:SHG_Ion} are given in Appendix~\ref{app:theory}. 

To compare the model to experiment, we fit each feature (carrier, sidebands) in the experimentally measured etalon scans (an example is shown in Fig.~\ref{fig:fig2}(d)) with the Lorentzian etalon response function. The depletion in the carrier was measured by fitting $I_\mathrm{on} - I_\mathrm{off}$ in the experimentally measured etalon scans.  On the theory side, the product $|M^{\mathrm{VB}\rightarrow nl}|^2 I_\mathrm{IN}$ is obtained by fitting the SHG peak amplitudes in Fig.~\ref{fig:fig2}(c)(i). Here equation~\ref{eq:SHG_Ioff} should provide the appropriate fit function, with $M^{\mathrm{VB}\rightarrow nl} I_\mathrm{IN}$, $\delta_{nl}$ and  $\Gamma_{nl}$ as fit parameters. However, fits using a sum of complex poles are not uniquely defined~\cite{Busson2009}, and therefore we approximate equation~\ref{eq:SHG_Ioff} as a sum of independent Lorentzians for each resonance~\cite{Rogers2021}. All other parameters within the summation in equations~\ref{eq:I_sdbnd} and~\ref{eq:SHG_Ion} are measured or calculated in the same way as in section~\ref{sec:LED_theory}. Thus the only remaining free parameter is an overall amplitude scaling equivalent to the parameter $A$ in equations~\ref{eq:I_sdbnd} and~\ref{eq:SHG_Ion}.

\begin{figure}
    \centering
    \includegraphics[width=\linewidth]{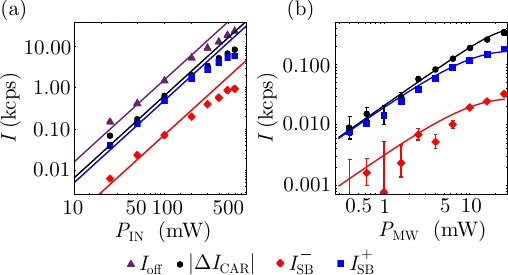}
    \caption{Laser and microwave power dependencies of sideband and carrier amplitudes in the SHG experiment. Here the two-photon excitation energy is $E=E_\mathrm{8S}$ and the microwave frequency is $f_\mathrm{MW}=19.5~\si{\giga\hertz}$. (a) Laser power, $P_\mathrm{IN}$, dependency of SHG (triangles), magnitude of carrier depletion (circles), blue sideband (squares) and red sideband (diamonds) at microwave power $P_\mathrm{MW}= 25\si{\milli\watt}$. Diagonal lines show the predicted quadratic dependencies on laser power. (b) Microwave power dependencies of the magnitude of carrier depletion (circles), blue sideband (squares) and red sideband (diamonds) at $P_\mathrm{IN}=50~\si{\milli\watt}$. Solid lines show predicted power dependencies from equations~\ref{eq:I_sdbnd} and~\ref{eq:SHG_Ion}.}
    \label{fig:power_dep}
\end{figure}

In Fig.~\ref{fig:power_dep} we compare the experimentally measured variation of the carrier and sideband amplitudes with  laser power ($P_\mathrm{IN}$) and microwave power ($P_\mathrm{MW}$)  to that predicted by  equations~\ref{eq:SHG_Ioff},~\ref{eq:I_sdbnd} and~\ref{eq:SHG_deltaI}. As shown in Fig.~\ref{fig:power_dep}(a), all four features show a quadratic dependence on $P_\mathrm{IN}$ in agreement with the model, before deviating at about 200~\si{\milli\watt}. The deviation appears to occur at the same value of $P_\mathrm{IN}$ for all of the features. We attribute this deviation to effects such as localised heating which are not included in the model. For microwave power, the model predicts a linear dependence at low power, which saturates as the power-dependent second term on the denominator of equations~\ref{eq:I_sdbnd} and~\ref{eq:SHG_deltaI} becomes significant (i.e. when $\Omega/\Gamma \sim 1$). As shown in Fig.~\ref{fig:power_dep}(b), this predicted behaviour is in excellent agreement with the experimental data. Here we have introduced a free parameter $B$, shared between all three features, which relates the power produced by the microwave generator to the effective field at the sample by $ P_\mathrm{IN}= B |\mathcal{E}_\mathrm{MW}|^2$. This parameter accounts for the efficiency of the antenna, the dielectric screening, and any losses in the feedthroughs to the cryostation. For a single value of $B$, a good fit is achieved for all three features. At the maximum input power ($P_\mathrm{MW} = 25~\si{\milli\watt}$) and using the fitted value for $B$, the effective field in the sample (produced by antenna A1) was found to be $200 \pm 50~\si{\volt\per\meter}$ which gives $\Omega_{\mathrm{8S}\mathrm{8P}}/\Gamma_\mathrm{8S}\approx 0.3$. The value of $\mathcal{E}_\mathrm{MW}=200~\si{\volt\per\meter}$ is in reasonable agreement with the calculated field strength inside the sample of 360~\si{\volt\per\meter} for antenna A1.

\begin{figure*}
    \centering
    \includegraphics[width=\linewidth]{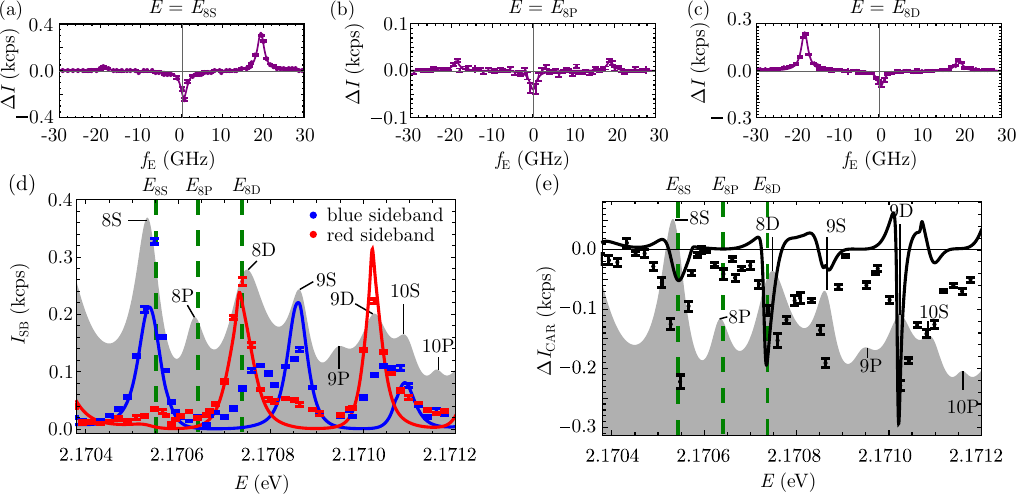}
    \caption{Dependence of sideband and carrier intensities on two-photon excitation energy, $E$. (a), (b) and (c) show the spectrally resolved change in second harmonic intensity due to a microwave field at $f_\mathrm{MW}=19.0~\si{\giga\hertz}$ for three different two-photon excitation energies $E_\mathrm{8S}$, $E_\mathrm{8P}$ and $E_\mathrm{8D}$ respectively. An asymmetry is observed, with the blue sideband larger when two-photon resonant with a S state and the red sideband larger when two-photon resonant with a D state. (d) Theory curve (solid) from equation~\ref{eq:I_sdbnd} and experimental data (points) showing the intensity of the blue and red sidebands at microwave frequency $f_\mathrm{MW} = 19~\si{\giga\hertz}$ and microwave power $P_\mathrm{MW} = 3~\si{\milli\watt}$ as a function of two-photon excitation energy. (e) Theory curve (solid) from equation~\ref{eq:SHG_deltaI} and experimental data (points) showing change in carrier intensity at $f_\mathrm{MW} = 19~\si{\giga\hertz}$ and $P_\mathrm{MW} = 3~\si{\milli\watt}$ as a function of two-photon excitation energy. The shaded background in (d) and (e) is the fit to the SHG excitation spectrum in Fig.~\ref{fig:fig2}(c) to show the range of exciton states explored. Dashed vertical lines indicate the three two-photon excitation energies of the etalon scans in parts (a) to (c).}
    \label{fig:sidebands_I}
\end{figure*}

The data in Fig.~\ref{fig:power_dep}(b) highlights a key feature of the observed emission spectra, which is that there is a strong asymmetry in the strength of the sidebands. As shown in Fig.~\ref{fig:sidebands_I}(a) to (c), this asymmetry is dependent on the excitation energy $E$. When the excitation energy is resonant with an S exciton (Fig.~\ref{fig:sidebands_I}(a)) the blue sideband is larger and when it is resonant with a D exciton (Fig.~\ref{fig:sidebands_I}(c)) the red sideband is larger. When two-photon resonant with a P state (Fig.~\ref{fig:sidebands_I}(b)) the sidebands are of similar strengths. This observation is a direct consequence of the fact that the nearest odd parity state to the state $n$S state is the $n$P state at higher energy, whereas for the $n$D state it is at lower energy. This strong, energy dependent asymmetry is apparent in equations~\ref{eq:I_sdbnd}, where it arises from the presence of both positive and negative components in the $(\delta_{n'l'}^\pm-i\Gamma_{n'l'})$ term.

The predicted amplitude of the sidebands as a function of two-photon excitation energy, $E$, is shown in Fig.~\ref{fig:sidebands_I}(d) at $P_\mathrm{MW}=3~\si{\milli\watt}$. The only fit parameter is the amplitude $A$ which is constrained to be the same for both the red and blue curves.  A similar plot for the carrier depletion $\Delta I_\mathrm{CAR}$ is shown in Fig.~\ref{fig:sidebands_I}(e) (note $A$ is different for the black curve).  Overall, the excitation energy dependence of both the sideband asymmetry and the carrier depletion is well described by the model across the full range of Rydberg states shown. Given the complexity of the experiment this overall agreement demonstrates that the model provides a solid basis for understanding the observed microwave-exciton coupling. Nevertheless, there are some regions of $E$ where the agreement is less good. As expected, one of these is in the vicinity of the P states, which were not included in our model for the SHG process. There also appear to be additional features in the experimental data close to the D peaks. One possible explanation is that we have not included all of the relevant states. In the theoretical model only D states of $\Gamma_5^+$ symmetry were considered. The $\Gamma_5^+$ D states are optically active in TPE due to a mixing with the S states via the exchange interaction~\cite{Heckotter2017-2,Uihlein1981}. However, there are additional D states ($\Gamma_1^+$ and $\Gamma_3^+$ symmetry) which are not active in TPE but have been observed in the presence of external fields~\cite{Uihlein1981,Rommel2020,Farenbruch2020,Semina2018,Heckotter2017-2,Heckotter2021}. Coupling between Rydberg states and these additional D states may explain some of the discrepancies between theory and experiment. By using higher quality samples with less strain and polarization-sensitive measurements, we hope to explore the origin of these additional features in future work. We also note that the model predicts values of $E$ where $\Delta I_\mathrm{CAR}>0$ which was not observed in the experiment.

Lastly, we note that the data in Figs.~\ref{fig:power_dep} and Figs.~\ref{fig:sidebands_I} was measured at a fixed  microwave frequency $f_{\mathrm{MW}}$. The measured and predicted variation with microwave frequency is similar to that observed using one-photon spectroscopy and is described in Appendix~\ref{app:SHGMW}.

\section{Discussion}

Our experiments demonstrate that Rydberg excitons couple strongly to microwave electric fields. An ``atomic physics'' view of the process, based on electric dipole transitions between excitonic states of opposite parity, provides a convincing explanation of the observed microwave-optical coupling in both one and two photon experiments. So far, the dominant effect of the crystal lattice is the non-radiative broadening of the excitonic states due to phononic decay channels, which leads to a broadband microwave frequency response. The final state for these phononic decay channels is the 1S exciton, which then decays radiatively via a quadrupole process.
Using two-photon excitation to the Rydberg states we have observed that the microwave field can modify the intensity of photoluminescence from the 1S exciton. A detailed study of this effect and its relation to exciton-phonon coupling is ongoing. Another effect of the lattice is the dielectric screening in the material. The dielectric screening of the material reduces the exciton binding energy, which leads to a higher dipole moment. However, this effect is cancelled by the dielectric screening of the microwave electric field, leading to similar Rabi frequencies to atomic systems at the same principal quantum number $n$. 

The one-photon transmission measurements show that the microwave field has a sizeable effect on the optical properties. Even with our inefficient microwave antennae, we achieved a coupling parameter $\Omega/\Gamma$ of 0.9. Reaching the strong-driving limit ($\Omega/\Gamma >1$) should be straightforward with improvements to our antenna design, using for example a copper coplanar resonator~\cite{Hogan2012}. In this limit, we anticipate that the physics will change significantly. New effects could include microwave-induced dipole-dipole interactions~\cite{Park2011,Tanasittikosol2011,Sevincli2014}, which could potentially be stronger and longer range ($\propto R^{-3}$)~\cite{Barredeo2015} than the van der Waals interactions so far observed in \cuprite~($\propto R^{-6}$). Multi-photon processes, including possible ionization, will also play a significant role, and a new model will be required. As the exciton linewidths are comparable to the separation between the exciton states, reaching the strong driving limit also implies a new regime where the Rabi frequency is comparable to the transition frequency between adjacent dipole-coupled states~\cite{Tommey2020}.

The SHG measurements clearly demonstrate the modulation of an optical carrier by the microwave field - an important feature of an optical to microwave interface. Within the resolution of our experiment (set by the etalon) this effect appears to be coherent; an obvious future direction is to perform more detailed measurements of the coherence using {e.g.} homodyne detection of the optical beat signal. Together with Hanbury Brown Twiss-type measurements, these experiments could also provide information on the quantum statistics of the generated light, which may be modified by interactions~\cite{Walther2018nl}. We note that the extremely high resolution of this microwave/two-photon method provides a new tool for studying the physics of Rydberg excitons more generally, potentially including states that are difficult to reach optically, such as high-lying paraexciton levels~\cite{Farenbruch2020-2}.

\begin{figure}
    \centering
    \includegraphics[width=\linewidth]{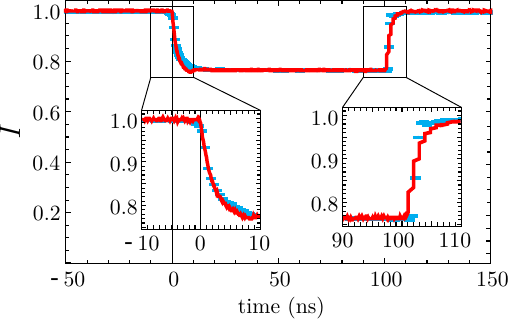}
    \caption{Time dependence of the second harmonic microwave response compared with turn on and off times of microwave pulse. Normalised SHG intensity (not spectrally resolved) as a function of time relative to the microwave pulse trigger. Data taken with $E = E_\mathrm{8S}$,  $f_\mathrm{MW} = 19.5~\si{\giga\hertz}$ and $P_\mathrm{MW} = 25~\si{\milli\watt}$. The microwave field causes a decrease in the intensity of the light detected (blue data). The red curve shows the microwave pulse shape as measured on an oscilloscope. Insets show zoom of the turn on and turn off times. }
    \label{fig:time_resolved}
\end{figure}

Lastly we consider the prospects for a fully quantum interface, for example with superconducting quantum circuits. Our estimates suggest that in comparison to Rydberg atoms, reaching the single-particle strong coupling regime  of cavity QED  (defined as $g^2 /(\kappa \Gamma) \gg 1$, where $g$ is the microwave vacuum Rabi frequency and $\kappa$ is the photon loss rate) is difficult due to the large non-radiative contribution to $\Gamma$. However, a strong collective enhancement of the coupling can be obtained by exciting many excitons within the cavity mode, and we believe that the collective strong coupling regime~\cite{Weisbuch1992,Bernardot1992} is within reach. We note also that the large microwave bandwidth implies extremely fast switching times - Fig.~\ref{fig:time_resolved} shows that we achieved nanosecond switching times in the SHG, limited only by our microwave generator. This is in contrast to atomic systems where the response is inherently narrowband, and may be useful for some applications. In terms of the optical side, read-out via one-photon spectroscopy suffers from a strong non-resonant phonon background which dominates the absorption. To overcome this, the microwave field could be combined with recent proposals to suppress this background using electromagnetically induced transparency~\cite{Grunwald2020}. For readout via SHG, the main issue is the low SHG efficiency in a centro-symmetric material and the absence of phase-matching. Here significant improvements could be made by using external fields such as strain or static electromagnetic fields to break the symmetry, as well using optical waveguides and resonator structures to maximise the local pump intensity.

\section{Summary and Conclusion}
In summary, we have studied the coupling between Rydberg excitons and microwave electric fields using one-photon and two-photon spectroscopy techniques. Even with inefficient microwave coupling a significant effect is observed, and the coherent modulation of an optical carrier was achieved. In the short term, our work provides a new tool for exciting and studying Rydberg exciton states. We expect to reach the strong driving regime, where this control will extend to many-body physics and quantum states of light. Looking further ahead, there is the potential to engineer an optical-to-microwave interface at the quantum level, with potential applications in quantum computing.

The data presented in this paper is available for download at \href{ http://doi.org/10.15128/r13x816m66s}{doi:10.15128/r13x816m66s}. 

\begin{acknowledgments}
 The authors are grateful to Ian Chaplin and Sophie Edwards (Durham University, Department of Earth Sciences) for the slicing and polishing of the samples used in this work. The authors would like to thank Ifan Hughes and Robert Potvliege for fruitful discussions. This work was supported by the Engineering and Physical Sciences Research Council (EPSRC), United Kingdom, through research grants EP/P011470/1 and EP/P012000/1. The authors also acknowledge seedcorn funding from Durham University. LAPG acknowledges financial support from the UK Defence and Scientific Technology Laboratory via an EPSRC Industrial Case award. VW acknowledges support by the NSF through a grant for the Institute for Theoretical Atomic, Molecular and Optical Physics at Harvard University and the Smithsonian Astrophysical Observatory. 
 
\end{acknowledgments}
\appendix
\section{Frequency response of the microwave antennae}\label{app:antenna}

\begin{figure*}
    \centering
    \includegraphics[width=\linewidth]{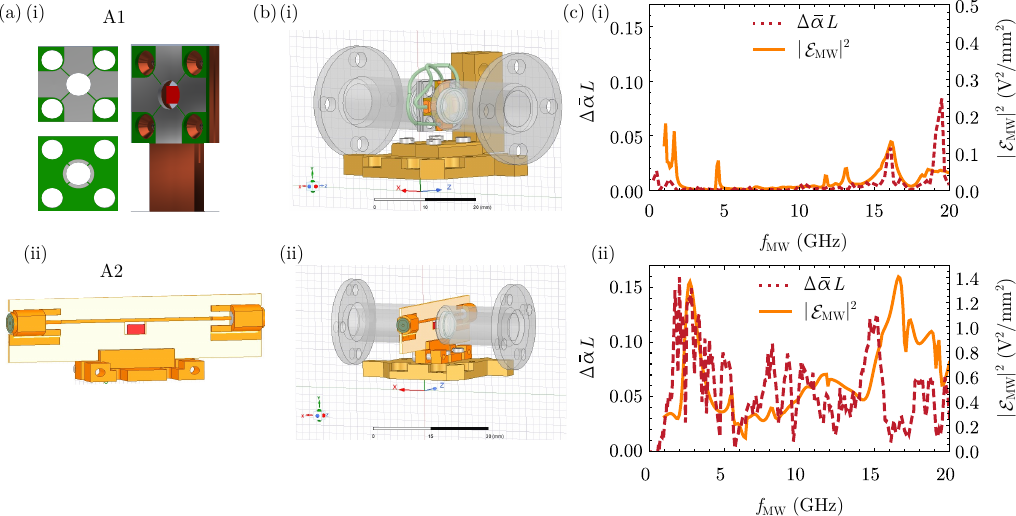}
    \caption{Microwave frequency dependence of antennae. (a) Renderings of antenna A1 (i) and A2 (ii). (b) Rendering of components included in microwave field simulations of antenna A1 (i) and A2 (ii). The  antenna is positioned in the centre between two stainless steel lens tubes (translucent grey) and support by a copper mounting plate (yellow). (c) Measured microwave frequency dependence of the change in absorption at excitation energy $E=E_\mathrm{9D}$ (dashed red, left axis), and the corresponding variation of the modulus squared of the simulated microwave field strength inside the sample (solid orange, right axis) for antenna A1 (i) and A2 (ii). }
    \label{fig:antenna_f_dep}
\end{figure*}

Both antennae (A1 and A2) used in this work were conceived as broadband near-field devices. From one-photon transmission spectroscopy measurements it became clear that the two antennae have radically different frequency responses as shown in Fig.~\ref{fig:antenna_f_dep}. In both cases strong resonances were observed at specific frequencies.

To understand these results, we performed detailed modelling of the microwave field produced inside the sample using commercial finite-element electromagnetic design software. The frequency response of both antennae was found to be strongly modified by the presence of other metal components in the cryostat, in particular the mounts for the aspheric lenses. The components included in the simulations are shown in Fig.~\ref{fig:antenna_f_dep}(b). These parasitic couplings were responsible for the strongly resonant behaviour of antenna A1, and the fine structure in the frequency response of antenna A2.

A comparison of the simulation results with the microwave response obtained from the experimental transmission spectra is shown in~\ref{fig:antenna_f_dep}(c). The agreement is very good for antenna A1 (Fig.~\ref{fig:antenna_f_dep}(c)(i)) predicting major peaks at 16 and 19~GHz as observed in the experimental data. For the stripline antenna A2 (Fig.~\ref{fig:antenna_f_dep}(c)(ii)), agreement is reasonable, with the broad feature at low frequency predicted. The simulation does not predict the fine structure present in the data. However, the details of the simulated result are sensitive to the exact locations and sizes of all components inside the cryostat, and we believe the level of the agreement is reasonable considering the limitations of the CAD model used to construct the simulations. 

The simulation also showed that  measurements of the frequency dependence of the transmitted and reflected microwave power ($S$ parameters) were not directly correlated to the local electric field strength with in the sample, due to the strong coupling with the electromagnetic environment. As a result we were unable to independently determine the local electric field strength from experiment. In light of the agreement observed in Fig.~\ref{fig:antenna_f_dep} we consider that the simulations are a useful tool to estimate the local electric field strength at the sample.

\section{Exciton-polariton model of microwave-optical coupling}\label{app:theory}

Here we provide a detailed derivation of the expressions for the susceptibilities and intensities in section~\ref{sec:theory}. Our model is based on the exciton-polariton model of light-matter coupling. Bosonic operators $\hat X_{nl}$ describe the annihilation of excitons with quantum numbers $n$ and $l$ (denoted S,P,D etc). The light fields are represented by  classical amplitudes $\mathcal E_\text{IN},\mathcal E_\text{OUT},\mathcal E_\text{MW},\mathcal E_\text{SB}, {\mathcal E}_\text{SHG}$, for the laser-input and output, microwave-, sideband-, and SHG-fields, respectively.

\subsection{Single-photon absorption}
We begin by approximating the incident light field  (here produced by an LED) as a monochromatic field with well-defined energy $E$ and field strength $\mathcal{E_\text{IN}}$.  (we note for absorption experiments performed with a laser~\cite{Kazimierczuk2014} this would be exact).
 A dipole-active yellow P-exciton, described by $\hat X_{n\text{P}}$ and of resonance energy $E+\hbar\delta_{n\text{P}}$, is excited and decays with rate $\Gamma_{n\text{P}}$. Detailed analysis of the phonon-induced couplings to the excitons and their decay channels implies Fano-type resonance lines~\cite{TOYOZAWA1964} on top of a broad $\sqrt{E}$- shaped background~\cite{Elliott1957,Schone2017}. Here we neglect the asymmetry of the Fano lineshape and assume a symmetric Lorentzian. The background absorption gains its oscillator strength from the $1S$ exciton. Electric dipole transitions from this state are in the mid-infrared; therefore as we observe in experiment the background is unaffected by the far off-resonant microwave field. Therefore the background  cancels exactly in $\Delta T$ and need not be included here.

Under these conditions the equation for the operator $\hat{X}_{n\text{P}}$ is given by
\begin{equation}
	\dot{\hat X}_{n\text{P}}=-(i\delta_{n\text{P}}+\Gamma_{n\text{P}}){\hat X}_{n\text{P}}-ig^{\text{VB}\to n\text{P}}\mathcal E_\text{IN}.
\end{equation}
Herein $g^{\text{VB}\to n\text{P}}$ is the coupling rate between valence band and P-exciton of principle quantum number $n$, and is connected to the dipole coupling strength, $D^{\text{VB}\to n\text{P}}$, defined in the main text. We will analyze the relation after solving this equation. 

To calculate the field propagation inside the medium and obtain its response, we apply the slowly-varying amplitude approximation~\cite{BONIFACIO1993}, and focus on the steady state, as all fields are CW. This connects the change of the field amplitude to the medium emitters, i.e. the excitons, or in classical electrodynamics terms, the medium polarization. The equations can be derived in the form
\begin{equation}
	\frac{c}{\eta}\partial_z\mathcal E_\text{OUT}=-ig^{n\text{P}\to \text{VB}}\langle\hat X_{n\text{P}}\rangle.
\end{equation}
Note that in the simple case of dipole input and output $g^{n\text{P}\to \text{VB}}=(g^{\text{VB}\to n\text{P}})^*$. Solving for the steady-state polarization and inserting it into the solution for the output field, we obtain
\begin{equation}
	\frac{c}{\eta}\partial_z\mathcal E_\text{OUT}=-\frac{|g^{n\text{P}\to \text{VB}}|^2}{i\delta_{n\text{P}}+\Gamma_{n\text{P}}}\mathcal E_\text{IN}.
\end{equation}
To understand the connection between this result and Eq.~(\ref{eq:chi1}) in the main text, we first clarify that the rate $g^{\text{VB}\to n\text{P}}$ effectively describes a Rabi frequency and thus the physical field amplitude $\mathcal E_\text{phys}$ is normalized outside of this rate, $\mathcal E_\text{phys}=\mathcal E_0\mathcal E_\text{IN/OUT}$. The normalization term $\mathcal E_0$ is included in the coupling rate and can be taken as the coupling between elementary field excitations (photons, even if we view the fields as classical) and the full fields. Its value can be found in any quantum optics text book~\cite{VoWel_Ch2} and reads as
\begin{equation}
	\mathcal E_0=\sqrt{\frac{\hbar\omega_\text{IN}}{2\varepsilon_0\varepsilon_\text{b}V}},
\end{equation}
with $\varepsilon_\text b$ being the background permittivity and $V$ the mode volume. Applying the basic formula for the Rabi frequency $g=d\mathcal E_0/\hbar$ and the notion that the dipole strength in the main text is per volume, we arrive at
\begin{equation}
	|g^{n\text{P}\to \text{VB}}|^2=|D^{n\text{P}\to \text{VB}}|^2\frac{\omega_\text{IN}}{2\varepsilon_0\hbar\eta^2}\label{eq.relgD}.
\end{equation}
Here $\eta$ is the (real) non-resonant background refractive index of Cu$_2$O. The resulting formula for the spatial change in the field amplitude is the well-known case of quasi resonant absorption~\cite{HaKo_Ch1} and reads
\begin{equation}
	\begin{split}
		\frac{c}{\eta}\partial_z\mathcal E_\text{OUT}=&-\frac{\omega_\text{IN}|D^{n\text{P}\to \text{VB}}|^2}{2\varepsilon_0\hbar\eta^2(i\delta_{n\text{P}}+\Gamma_{n\text{P}})}\mathcal E_\text{IN}\\
		=&i\omega_\text{IN}\chi^{(1)}_{n\text P}\mathcal E_\text{IN}.
	\end{split}
\end{equation}

\subsection{SHG without Microwave}
Let us now move on to the SHG process without turning on the microwave. In SHG, the input laser is at half the excitation energy  $\omega_\text{IN}=E/(2\hbar)$, and as a result it couples  off-resonantly to all dipole allowed states. The dominant contribution are from S excitons states belonging to the ``blue'' series~\cite{Schmutzler2013,Schone2017} and to a smaller degree the yellow P-excitons. These intermediate states are described by operators $\hat X_{\alpha}$, $\alpha$ being a general index for all these states.  These states are then coupled to the target yellow S- or D-exciton states via a second electric dipole transition. To begin with we consider only a single final S exciton state, with associated operator by $\hat X_{n\text S}$. The coupling constants of the first and second process with intermediate exciton $\hat X_{\alpha}$ are designated $g^{\text{VB}\to\alpha}$ and $g^{\alpha\to n\text{S}}$, respectively. The S-exciton emits light via quadrupole emission with rate $q^{n\text S\to \text{VB}}$. This rate is analogue to the dipole coupling rates $g$ and hence connects to the quadrupole moment $Q^{n\text S\to\text{VB}}$ in the same way as for the dipolar counter part in Eq.~(\ref{eq.relgD}). Splitting off the laser energy $\omega_\text{IN}$ from the kinetic energy ($2\omega_\text{IN}$ for the S-exciton as it is excited via 2 photons), we arrive at an effective Hamiltonian
\begin{equation}
	\begin{split}
		\hat H/\hbar=&\delta_\text S\hat X^\dagger_{n\text S}\hat X_{n\text S}+\sum\limits_\alpha\Delta_\alpha\hat X^\dagger_{\alpha}\hat X_{\alpha}\\
		&+\sum\limits_\alpha\Big[ g^{\text{VB}\to\alpha}(\mathcal E_\text{IN}\hat X^\dagger_{\alpha}+\mathcal E_\text{IN}^*\hat X_{\alpha})\\
		&+(g^{\alpha\to n\text S}\hat X_{n\text S}^\dagger\mathcal E_\text{IN}\hat X_{\alpha}+g^{n\text S\to \alpha}\mathcal E_\text{IN}^*\hat X^\dagger_{\alpha}\hat X_{n\text S})\\
		&+q^{n\text S\to \text{VB}}\hat X^\dagger_{n\text S}{\mathcal E}_\text{SHG}+q^{\text{VB}\to n\text S}\hat X_{n\text S}{\mathcal E}^*_\text{SHG})\Big].
	\end{split}\label{eqn:fullSHG}
\end{equation} 
Here we define the S-exciton resonance frequency $\omega_{n\text S}=\delta_\text S+E/\hbar$; likewise $\omega_\alpha=\Delta_\alpha+E/(2\hbar)$ for the intermediate states. We note that the detuning from the intermediate states $\Delta_\alpha$ is much larger than any other system parameter (coupling strengths, linewidths). 

This effective Hamiltonian~(\ref{eqn:fullSHG}) is unitary, and therefore treats both SHG and its time-reversed process on an equal footing. However, in practice weak excitation of the SHG light, dissipation, and necessary phase matching suppress the reverse process substantially. In~(\ref{eqn:fullSHG}), the different internal momenta at $E$ and $E/2$ lead to a reduction of $g^{n\text S\to \alpha}$ compared to $g^{\alpha\to n\text S}$. The process $q^{\text{VB}\to n\text S}\hat X^\dagger_{n\text S}{\mathcal E}_\text{SHG}$, where the SHG light creates an exciton via quadrupole absorption, is also much less likely than the dominant two-photon excitation process. Therefore the time-reversed terms are suppressed, yielding different equations for $\hat X_{n\text S}$ and $\hat X_{n\text S}^\dagger$. To ensure conservation of energy (i.e. a stationary in flow and outflow of photons) we ensure that these operators are appropriately normalised.


The resulting equations of motion for the two annihilation operators and one creation operator are
\begin{align}
	\dot{\hat X}_{\alpha}=&-(i\Delta_\alpha+\Gamma_{\alpha})\hat X_{\alpha}-ig^{\text{VB}\to\alpha}\mathcal E_\text{IN}\\
	\dot{\hat X}_{n\text S}=&-(i\delta_\text S+\Gamma_{n\text S})\hat X_{n\text S}-i\sum\limits_\alpha g^{\alpha\to n\text S}\mathcal E_\text{IN}\hat X_{\alpha}\\
	\dot{\hat X}_{n\text S}^\dagger=&(i\delta_\text S-\Gamma_{n\text S})\hat X_{n\text S}^\dagger+iq^{n\text S\to g}{\mathcal E}_\text{SHG}^*.
\end{align}
The fast motion of the off-resonant intermediate excitons allows us to eliminate them adiabatically (setting $\dot{\hat X}_{\alpha}=0$) and insert the solution into the equation for the S-exciton
\begin{equation}
	\dot{\hat X}_{n\text S}=-(i\delta_\text S+\Gamma_{n\text S})\hat X_{n\text S}-\sum\limits_\alpha \frac{g^{\text{VB}\to\alpha} g^{\alpha\to n\text S}\mathcal E_\text{IN}^2}{i\Delta_\alpha+\Gamma_{\alpha}}.
\end{equation}
Due to $|\Delta_\alpha|\gg\Gamma_{\alpha}$, we see that these sums are basically imaginary, allowing us to define an SHG driving strength $m^{\text{VB}\to n\text S}$:
\begin{align}
	\dot{\hat X}_{n\text S}=&-\left(i\delta_\text S+\Gamma_{n\text S}\right)\hat X_{n\text S}+im^{\text{VB}\to n\text S}\mathcal E_\text{IN}^2,\label{eq.dotXns}\\
	m^{\text{VB}\to n\text S}=&\sum\limits_\alpha \frac{g^{\text{VB}\to\alpha} g^{\alpha\to n\text S}}{\Delta_\alpha}.
\end{align}
This effective SHG driving strength is again coupled to an effective dipole moment $M^{\text{VB}\to n\text S}$ in the same way as $g$ and $q$, previously.
The dynamical equation for $\hat X_{n\text S}$ is the same as for a driven bosonic emitter, yielding a coherent state. 

Finally, for the SHG field, we again apply the slowly-varying envelope approximation to obtain
\begin{equation}
	\frac{c}{\eta}\partial_z{\mathcal E}_\text{SHG}=-iq^{n\text S\to \text{VB}}\langle\hat X_{n\text S}\rangle,
\end{equation}
evaluated for the steady-state value of $\langle \hat X_{n\text S}\rangle$. Inserting the solutions to obtain the susceptibility, we get
\begin{equation}
\begin{split}
		\frac{c}{\eta}\partial_z{\mathcal E}_\text{SHG}=&\sum\limits_\alpha\frac{q^{n\text S\to \text{VB}}g^{\alpha\to n\text{S}}g^{\text{VB}\to\alpha}}{(\Gamma_{n\text S}+i\delta_\text S)\Delta_\alpha}		\mathcal E_\text{IN}^2\\
		=&\frac{q^{n\text S\to \text{VB}}m^{\text{VB}\to n\text S}}{(\Gamma_{n\text S}+i\delta_{n\text S})}\mathcal E_\text{IN}^2=2i\omega_\text{IN}\chi^{(2)}_{n\text S}\mathcal E_\text{IN}^2.
	\end{split}
\end{equation}
The frequency $2\omega_\text{IN}$ indicates that we look at only the propagation of SHG light.
The full susceptibility for the SHG 
follows by summing $\chi_{nl}$ over all possible S and D final states.

\begin{widetext}
\subsection{Including the microwave field}
Turning on a microwave field of frequency $\omega_\text{MW}=2\pi\times f_\mathrm{MW}$ and amplitude $\mathcal E_\text{MW}$, enables the laser-excited even-parity (S or D)-exciton $\hat X_{n\ell}$ to couple to a nearby odd-parity (P or F)-exciton described by $\hat X_{n'\ell'}$ and kinetic energy $\hbar\omega_{n'\ell'}$.   The microwave coupling strength is denoted by the rate $\Omega_{n\ell n'\ell'}=g^{n\ell\to n'\ell'}\mathcal E_\text{MW}$, which again is coupled to a dipole moment per volume as described in the single-photon absorption scenario. 

Again, we focus on a single exciton state before summing the results over all relevant dipole-coupled states. Direct laser excitation of odd parity states is negligible. We note that the large width of the exciton states compared to their separation means that the usual near resonant rotating-wave approximation (RWA) cannot be made.
Thus, we explicitly retain time dependence of the microwave field and obtain the differential equations
\begin{align}
	\dot{X}_{n'\ell'}=&-(i\delta_{n'\ell'}+\Gamma_{n'\ell'})X_{n'\ell'}-i\Omega_{n\ell n'\ell'}e^{-i\omega_\text{MW}t}X_{n\ell}-i\Omega_{n\ell n'\ell'}^*e^{i\omega_\text{MW}t}X_{n\ell},\\
	\dot{X}_{n\ell}=&-\left(i\delta_{n\ell}+\Gamma_{n\ell}\right)X_{n\ell}+im^{\text{VB}\to n\ell}\mathcal E_\text{IN}^2-i\Omega_{n\ell n'\ell'}^*e^{i\omega_\text{MW}t}X_{n'\ell'}-i\Omega_{n\ell n'\ell'}e^{-i\omega_\text{MW}t}X_{n'\ell'}.
\end{align}
Here the detuning $\delta_{n'\ell'}=\omega_{n'\ell'}-2\omega_\text{IN}$. The hats have been omitted on all operators, as the linear structure of the equations allows us to cast them directly for the expectation values. 
 
In general, this set of equations is difficult to solve due to the different oscillations prohibiting the filtering out of a slowly varying amplitude. One way to approach this problem is via Fourier series. In particular, we assume as solution for the equations the ansatz
\begin{align}
	X_{n'\ell'}=&\sum\limits_{k=-\infty}^\infty P_ke^{ik\omega_\text{MW}t},\\
	X_{n\ell}=&\sum\limits_{k=-\infty}^\infty S_ke^{ik\omega_\text{MW}t}.
\end{align}
The terms $P_k,S_k$ are time-independent, indicating these formulas to represent steady-state solutions. The equations for each coefficient $k$ couple to those for $k\pm1$. From a physical point of view these couplings represent sidebands shifted by $k\cdot \omega_\text{MW}$ from the carrier at $E/\hbar$. In general, these terms constitute an infinite hierarchy. However, in experiments a second-order sideband at $\pm 2\omega_\text{MW}$ was not observed within our experimental sensitivity. Hence, we retain only the first order terms $k=0,\pm1$, corresponding to the resonances in the emission spectrum at $E$ (carrier) and $E\pm\hbar \omega_\text{MW}$ (sidebands) observed in the experiments.

This approach yields 6 linear equations for the coefficients $P_k,S_k$ as follows:
\begin{align}
	0=&-(i\delta_{n'\ell'}+\Gamma_{n'\ell'})P_0-i\Omega_{n\ell n'\ell'}S_1-i\Omega_{n\ell n'\ell'}^*S_{-1}\\
	i\omega_\text{MW}P_1=&-(i\delta_{n'\ell'}+\Gamma_{n'\ell'})P_1-i\Omega_{n\ell n'\ell'}^*S_0\\
	-i\omega_\text{MW}P_{-1}=&-(i\delta_{n'\ell'}+\Gamma_{n'\ell'})P_{-1}-i\Omega_{n\ell n'\ell'}S_0\\
	0=&-(i\delta_{n\ell }+\Gamma_{n\ell })S_0-i\Omega_{n\ell n'\ell'}P_1-i\Omega_{n\ell n'\ell'}^*P_{-1}+im^{\text{VB}\to n\ell }\mathcal E_\text{IN}^2\\
	i\omega_\text{MW}S_1=&-(i\delta_{n\ell }+\Gamma_{n\ell })S_1-i\Omega_{n\ell n'\ell'}^*P_0\\
	-i\omega_\text{MW}S_{-1}=&-(i\delta_{n \ell }+\Gamma_{n\ell })S_{-1}-i\Omega_{n\ell n'\ell'}P_0.
\end{align}
Inserting the last two equations into the first, we easily see that 
\begin{equation}
	P_0=0=S_{\pm1}.
\end{equation}
This shows, as expected, that to first order the emission at the carrier is exclusively due to the even-parity exciton at the carrier, represented by $S_0$, while the two sidebands form due to the odd-parity exciton coupling with strength $P_{\pm1}$.

Solving the remaining three equations we obtain
\begin{align}
	S_0=&\frac{im^{\text{VB}\to n\ell'}\mathcal E_\text{IN}^2}{i\delta_{n\ell}+\Gamma_{n\ell}+|\Omega_{n\ell n'\ell'}|^2\left(\frac{1}{i\delta_{n'\ell'}^-+\Gamma_{n'\ell'}}+\frac{1}{i\delta_{n'\ell'}^++\Gamma_{n'\ell'}}\right)}, \label{eq.Xns(2)}\\
	P_{-1}=&\frac{m^{\text{VB}\to n\ell }\Omega_{n\ell n'\ell'}\mathcal E_\text{IN}^2}{(i\delta_{n\ell}+\Gamma_{n\ell})(i\delta_{n'\ell'}^-+\Gamma_{n'\ell'})+|\Omega_{n\ell n'\ell'}|^2\left(1+\frac{i\delta_{n'\ell'}^-+\Gamma_{n'\ell'}}{i\delta_{n'\ell'}^++\Gamma_{n'\ell'}}\right)},\label{eq.Xnpm}\\
	P_1=&\frac{m^{\text{VB}\to n\ell }\Omega_{n\ell n'\ell'}\mathcal E_\text{IN}^2}{(i\delta_{n\ell}+\Gamma_{n\ell})(i\delta_{n'\ell'}^++\Gamma_{n'\ell'})+|\Omega_{n\ell n'\ell'}|^2\left(1+\frac{i\delta_{n'\ell'}^++\Gamma_{n'\ell'}}{i\delta_{n'\ell'}^-+\Gamma_{n'\ell'}}\right)}.\label{eq.Xnpp}
\end{align}
Herein we defined as effective detunings $\delta_{n'\ell'}^{\pm}=\delta_{n'\ell'}\pm \omega_\text{MW}$.
Each of these terms now represents the polarization of the medium at one of the three frequencies $k\omega_\text{MW}$.  To obtain the susceptibility, we apply one more time the slowly-varying envelope approximation. All output light stems from the SHG, it is just split due to the microwave coupling. Hence, we may, for example write for the $k=1$ sideband
\begin{align}
	\frac{c}{\eta}\partial_z\mathcal E_{1}=&-iq^{n\ell \to \text{VB}}P_1=
	\frac{-iq^{n\ell \to \text{VB}}m^{\text{VB}\to n\ell }\Omega_{n\ell n'\ell'}}{(i\delta_{n\ell }+\Gamma_{n\ell })(i\delta_{n'\ell'}^++\Gamma_{n'\ell'})+|\Omega_{n\ell n'\ell'}|^2\left(1+\frac{i\delta_{n'\ell'}^++\Gamma_{n'\ell'}}{i\delta_{n'\ell'}^-+\Gamma_{n'\ell'}}\right)}\mathcal E_\text{IN}^2=2i\omega_\text{IN}\chi^{(3)}_{n\ell n'\ell'}\mathcal E_\text{IN}^2.
\end{align}
Summing over all relevant states $n'\ell'$ that couple via microwave from the $n\ell$-excitons yields the total polarization, which is proportional to the electric field outside the crystal, thus yielding the expressions for the carrier and sideband intensities provided in the main text. 

We note that the change in the carrier intensity occurs due to the saturation term proportional to the microwave intensity $|\mathcal E_\text{MW}|^2$ in the denominator of $S_0$. For weak microwave fields this yields a change of the amplitude proportional to the intensity, while at larger field strengths it becomes nonlinear. Likewise, the sidebands increase linearly with $|\mathcal E_\text{MW}|^2$ for weak fields before decreasing again at larger field strengths. We also note that when one sideband is close to the $(n\ell)-(n'\ell')$ transition frequency and the other far away such that $2\omega_\text{MW}\gg \Gamma_{n'\ell'}$, the effect of the second sideband becomes a simple resonance shift of the two exciton resonances. This behaviour is equivalent to an alternative approach detailed in~\cite{James2000}, wherein a first-order correction to the RWA yields a small resonance shift on the strong resonance.

\section{Microwave frequency dependence of SHG }\label{app:SHGMW}

In Fig.~\ref{fig:SHG_f_dep}(a) the predicted variation of the blue sideband intensity with microwave frequency is shown for excitation energies $E$ corresponding to three different Rydberg states. The response is broadband over the range of $f_\mathrm{MW}$ considered. As expected, the peak response shifts to lower frequencies as $n$ increases, reflecting the $n^{-3}$ scaling of the separation between neighbouring Rydberg state. 
\begin{figure*}
    \centering
    \includegraphics[width=\linewidth]{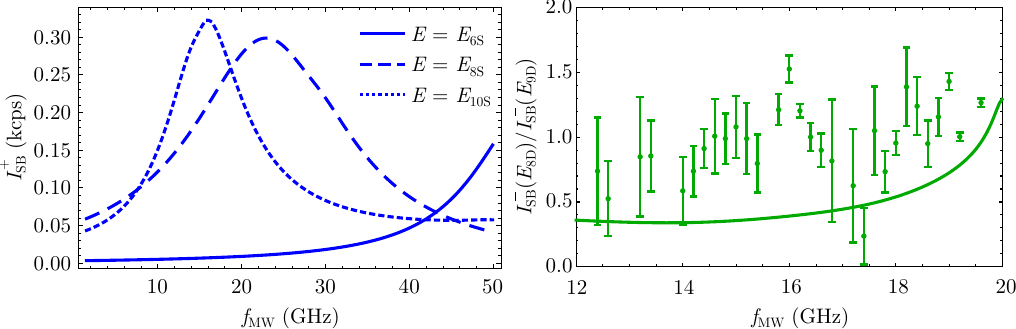}
    \caption{Microwave frequency dependence of sideband intensities. (a) Predicted microwave frequency dependence of blue sideband intensity. Excitation energies resonant with three different states are plotted, $E=E_\mathrm{6S}$ (solid), $E=E_\mathrm{8S}$ (dashed), $E=E_\mathrm{10S}$ (dotted). (b) Ratio of red sideband intensity at two excitation energies ($E=E_\mathrm{8D}$ and $E=E_\mathrm{9D}$) as function of microwave frequency. Solid line shows predictions from equation~\ref{eq:I_sdbnd} with $\mathcal{E}_\mathrm{MW}= 200~\si{\volt\per\meter}$. Points are experimentally measured.
    }
    \label{fig:SHG_f_dep}
\end{figure*}

The experimentally measured microwave frequency dependence is dominated by the antenna used, making comparisons between the theoretical predictions and experimental data difficult. To remove the response of the antenna we take the ratio of the red sideband amplitude $I^-_\mathrm{SB}$, at two different values of $E$. Fig.~\ref{fig:SHG_f_dep}(b) shows the ratio of the red sideband intensities at the 8D and 9D resonances. The model underestimates the ratio but gets the general trend of the ratio increasing with $f_\mathrm{MW}$. We note that the ratio of the red sideband at 8D to 9D is also underestimated by the model in Fig.~\ref{fig:sidebands_I}(d). 
\end{widetext}
\newpage

\bibliography{microwave_paper.bib}

\begin{thebibliography}{93}%
\makeatletter
\providecommand \@ifxundefined [1]{%
 \@ifx{#1\undefined}
}%
\providecommand \@ifnum [1]{%
 \ifnum #1\expandafter \@firstoftwo
 \else \expandafter \@secondoftwo
 \fi
}%
\providecommand \@ifx [1]{%
 \ifx #1\expandafter \@firstoftwo
 \else \expandafter \@secondoftwo
 \fi
}%
\providecommand \natexlab [1]{#1}%
\providecommand \enquote  [1]{``#1''}%
\providecommand \bibnamefont  [1]{#1}%
\providecommand \bibfnamefont [1]{#1}%
\providecommand \citenamefont [1]{#1}%
\providecommand \href@noop [0]{\@secondoftwo}%
\providecommand \href [0]{\begingroup \@sanitize@url \@href}%
\providecommand \@href[1]{\@@startlink{#1}\@@href}%
\providecommand \@@href[1]{\endgroup#1\@@endlink}%
\providecommand \@sanitize@url [0]{\catcode `\\12\catcode `\$12\catcode
  `\&12\catcode `\#12\catcode `\^12\catcode `\_12\catcode `\%12\relax}%
\providecommand \@@startlink[1]{}%
\providecommand \@@endlink[0]{}%
\providecommand \url  [0]{\begingroup\@sanitize@url \@url }%
\providecommand \@url [1]{\endgroup\@href {#1}{\urlprefix }}%
\providecommand \urlprefix  [0]{URL }%
\providecommand \Eprint [0]{\href }%
\providecommand \doibase [0]{http://dx.doi.org/}%
\providecommand \selectlanguage [0]{\@gobble}%
\providecommand \bibinfo  [0]{\@secondoftwo}%
\providecommand \bibfield  [0]{\@secondoftwo}%
\providecommand \translation [1]{[#1]}%
\providecommand \BibitemOpen [0]{}%
\providecommand \bibitemStop [0]{}%
\providecommand \bibitemNoStop [0]{.\EOS\space}%
\providecommand \EOS [0]{\spacefactor3000\relax}%
\providecommand \BibitemShut  [1]{\csname bibitem#1\endcsname}%
\let\auto@bib@innerbib\@empty
\bibitem [{\citenamefont {Devoret}\ and\ \citenamefont
  {Schoelkopf}(2013)}]{Devoret2013}%
  \BibitemOpen
  \bibfield  {author} {\bibinfo {author} {\bibfnamefont {M.~H.}\ \bibnamefont
  {Devoret}}\ and\ \bibinfo {author} {\bibfnamefont {R.~J.}\ \bibnamefont
  {Schoelkopf}},\ }\href {\doibase 10.1126/science.1231930} {\bibfield
  {journal} {\bibinfo  {journal} {Science}\ }\textbf {\bibinfo {volume}
  {339}},\ \bibinfo {pages} {1169} (\bibinfo {year} {2013})}\BibitemShut
  {NoStop}%
\bibitem [{\citenamefont {Kelly}\ \emph {et~al.}(2015)\citenamefont {Kelly},
  \citenamefont {Barends}, \citenamefont {Fowler}, \citenamefont {Megrant},
  \citenamefont {Jeffrey}, \citenamefont {White}, \citenamefont {Sank},
  \citenamefont {Mutus}, \citenamefont {Campbell}, \citenamefont {Chen},
  \citenamefont {Chen}, \citenamefont {Chiaro}, \citenamefont {Dunsworth},
  \citenamefont {Hoi}, \citenamefont {Neill}, \citenamefont {O'Malley},
  \citenamefont {Quintana}, \citenamefont {Roushan}, \citenamefont
  {Vainsencher}, \citenamefont {Wenner}, \citenamefont {Cleland},\ and\
  \citenamefont {Martinis}}]{Kelly2015}%
  \BibitemOpen
  \bibfield  {author} {\bibinfo {author} {\bibfnamefont {J.}~\bibnamefont
  {Kelly}}, \bibinfo {author} {\bibfnamefont {R.}~\bibnamefont {Barends}},
  \bibinfo {author} {\bibfnamefont {A.~G.}\ \bibnamefont {Fowler}}, \bibinfo
  {author} {\bibfnamefont {A.}~\bibnamefont {Megrant}}, \bibinfo {author}
  {\bibfnamefont {E.}~\bibnamefont {Jeffrey}}, \bibinfo {author} {\bibfnamefont
  {T.~C.}\ \bibnamefont {White}}, \bibinfo {author} {\bibfnamefont
  {D.}~\bibnamefont {Sank}}, \bibinfo {author} {\bibfnamefont {J.~Y.}\
  \bibnamefont {Mutus}}, \bibinfo {author} {\bibfnamefont {B.}~\bibnamefont
  {Campbell}}, \bibinfo {author} {\bibfnamefont {Y.}~\bibnamefont {Chen}},
  \bibinfo {author} {\bibfnamefont {Z.}~\bibnamefont {Chen}}, \bibinfo {author}
  {\bibfnamefont {B.}~\bibnamefont {Chiaro}}, \bibinfo {author} {\bibfnamefont
  {A.}~\bibnamefont {Dunsworth}}, \bibinfo {author} {\bibfnamefont {I.-C.}\
  \bibnamefont {Hoi}}, \bibinfo {author} {\bibfnamefont {C.}~\bibnamefont
  {Neill}}, \bibinfo {author} {\bibfnamefont {P.~J.~J.}\ \bibnamefont
  {O'Malley}}, \bibinfo {author} {\bibfnamefont {C.}~\bibnamefont {Quintana}},
  \bibinfo {author} {\bibfnamefont {P.}~\bibnamefont {Roushan}}, \bibinfo
  {author} {\bibfnamefont {A.}~\bibnamefont {Vainsencher}}, \bibinfo {author}
  {\bibfnamefont {J.}~\bibnamefont {Wenner}}, \bibinfo {author} {\bibfnamefont
  {A.~N.}\ \bibnamefont {Cleland}}, \ and\ \bibinfo {author} {\bibfnamefont
  {J.~M.}\ \bibnamefont {Martinis}},\ }\href {\doibase 10.1038/nature14270}
  {\bibfield  {journal} {\bibinfo  {journal} {Nature}\ }\textbf {\bibinfo
  {volume} {519}},\ \bibinfo {pages} {66} (\bibinfo {year} {2015})}\BibitemShut
  {NoStop}%
\bibitem [{\citenamefont {Hofheinz}\ \emph {et~al.}(2008)\citenamefont
  {Hofheinz}, \citenamefont {Weig}, \citenamefont {Ansmann}, \citenamefont
  {Bialczak}, \citenamefont {Lucero}, \citenamefont {Neeley}, \citenamefont
  {O'Connell}, \citenamefont {Wang}, \citenamefont {Martinis},\ and\
  \citenamefont {Cleland}}]{Hofheinz2008}%
  \BibitemOpen
  \bibfield  {author} {\bibinfo {author} {\bibfnamefont {M.}~\bibnamefont
  {Hofheinz}}, \bibinfo {author} {\bibfnamefont {E.~M.}\ \bibnamefont {Weig}},
  \bibinfo {author} {\bibfnamefont {M.}~\bibnamefont {Ansmann}}, \bibinfo
  {author} {\bibfnamefont {R.~C.}\ \bibnamefont {Bialczak}}, \bibinfo {author}
  {\bibfnamefont {E.}~\bibnamefont {Lucero}}, \bibinfo {author} {\bibfnamefont
  {M.}~\bibnamefont {Neeley}}, \bibinfo {author} {\bibfnamefont {A.~D.}\
  \bibnamefont {O'Connell}}, \bibinfo {author} {\bibfnamefont {H.}~\bibnamefont
  {Wang}}, \bibinfo {author} {\bibfnamefont {J.~M.}\ \bibnamefont {Martinis}},
  \ and\ \bibinfo {author} {\bibfnamefont {A.~N.}\ \bibnamefont {Cleland}},\
  }\href {\doibase 10.1038/nature07136} {\bibfield  {journal} {\bibinfo
  {journal} {Nature}\ }\textbf {\bibinfo {volume} {454}},\ \bibinfo {pages}
  {310} (\bibinfo {year} {2008})}\BibitemShut {NoStop}%
\bibitem [{\citenamefont {Watson}\ \emph {et~al.}(2018)\citenamefont {Watson},
  \citenamefont {Philips}, \citenamefont {Kawakami}, \citenamefont {Ward},
  \citenamefont {Scarlino}, \citenamefont {Veldhorst}, \citenamefont {Savage},
  \citenamefont {Lagally}, \citenamefont {Friesen}, \citenamefont
  {Coppersmith}, \citenamefont {Eriksson},\ and\ \citenamefont
  {Vandersypen}}]{Watson2018}%
  \BibitemOpen
  \bibfield  {author} {\bibinfo {author} {\bibfnamefont {T.~F.}\ \bibnamefont
  {Watson}}, \bibinfo {author} {\bibfnamefont {S.~G.~J.}\ \bibnamefont
  {Philips}}, \bibinfo {author} {\bibfnamefont {E.}~\bibnamefont {Kawakami}},
  \bibinfo {author} {\bibfnamefont {D.~R.}\ \bibnamefont {Ward}}, \bibinfo
  {author} {\bibfnamefont {P.}~\bibnamefont {Scarlino}}, \bibinfo {author}
  {\bibfnamefont {M.}~\bibnamefont {Veldhorst}}, \bibinfo {author}
  {\bibfnamefont {D.~E.}\ \bibnamefont {Savage}}, \bibinfo {author}
  {\bibfnamefont {M.~G.}\ \bibnamefont {Lagally}}, \bibinfo {author}
  {\bibfnamefont {M.}~\bibnamefont {Friesen}}, \bibinfo {author} {\bibfnamefont
  {S.~N.}\ \bibnamefont {Coppersmith}}, \bibinfo {author} {\bibfnamefont
  {M.~A.}\ \bibnamefont {Eriksson}}, \ and\ \bibinfo {author} {\bibfnamefont
  {L.~M.~K.}\ \bibnamefont {Vandersypen}},\ }\href {\doibase
  10.1038/nature25766} {\bibfield  {journal} {\bibinfo  {journal} {Nature}\
  }\textbf {\bibinfo {volume} {555}},\ \bibinfo {pages} {633} (\bibinfo {year}
  {2018})}\BibitemShut {NoStop}%
\bibitem [{\citenamefont {Kurpiers}\ \emph {et~al.}(2018)\citenamefont
  {Kurpiers}, \citenamefont {Magnard}, \citenamefont {Walter}, \citenamefont
  {Royer}, \citenamefont {Pechal}, \citenamefont {Heinsoo}, \citenamefont
  {Salath{\'e}}, \citenamefont {Akin}, \citenamefont {Storz}, \citenamefont
  {Besse}, \citenamefont {Gasparinetti}, \citenamefont {Blais},\ and\
  \citenamefont {Wallraff}}]{Kurpiers2018}%
  \BibitemOpen
  \bibfield  {author} {\bibinfo {author} {\bibfnamefont {P.}~\bibnamefont
  {Kurpiers}}, \bibinfo {author} {\bibfnamefont {P.}~\bibnamefont {Magnard}},
  \bibinfo {author} {\bibfnamefont {T.}~\bibnamefont {Walter}}, \bibinfo
  {author} {\bibfnamefont {B.}~\bibnamefont {Royer}}, \bibinfo {author}
  {\bibfnamefont {M.}~\bibnamefont {Pechal}}, \bibinfo {author} {\bibfnamefont
  {J.}~\bibnamefont {Heinsoo}}, \bibinfo {author} {\bibfnamefont
  {Y.}~\bibnamefont {Salath{\'e}}}, \bibinfo {author} {\bibfnamefont
  {A.}~\bibnamefont {Akin}}, \bibinfo {author} {\bibfnamefont {S.}~\bibnamefont
  {Storz}}, \bibinfo {author} {\bibfnamefont {J.-C.}\ \bibnamefont {Besse}},
  \bibinfo {author} {\bibfnamefont {S.}~\bibnamefont {Gasparinetti}}, \bibinfo
  {author} {\bibfnamefont {A.}~\bibnamefont {Blais}}, \ and\ \bibinfo {author}
  {\bibfnamefont {A.}~\bibnamefont {Wallraff}},\ }\href {\doibase
  10.1038/s41586-018-0195-y} {\bibfield  {journal} {\bibinfo  {journal}
  {Nature}\ }\textbf {\bibinfo {volume} {558}},\ \bibinfo {pages} {264}
  (\bibinfo {year} {2018})}\BibitemShut {NoStop}%
\bibitem [{\citenamefont {Zhong}\ \emph {et~al.}(2019)\citenamefont {Zhong},
  \citenamefont {Chang}, \citenamefont {Satzinger}, \citenamefont {Chou},
  \citenamefont {Bienfait}, \citenamefont {Conner}, \citenamefont {Dumur},
  \citenamefont {Grebel}, \citenamefont {Peairs}, \citenamefont {Povey},
  \citenamefont {Schuster},\ and\ \citenamefont {Cleland}}]{Zhong2019}%
  \BibitemOpen
  \bibfield  {author} {\bibinfo {author} {\bibfnamefont {Y.~P.}\ \bibnamefont
  {Zhong}}, \bibinfo {author} {\bibfnamefont {H.-S.}\ \bibnamefont {Chang}},
  \bibinfo {author} {\bibfnamefont {K.~J.}\ \bibnamefont {Satzinger}}, \bibinfo
  {author} {\bibfnamefont {M.-H.}\ \bibnamefont {Chou}}, \bibinfo {author}
  {\bibfnamefont {A.}~\bibnamefont {Bienfait}}, \bibinfo {author}
  {\bibfnamefont {C.~R.}\ \bibnamefont {Conner}}, \bibinfo {author}
  {\bibfnamefont {{\'E}.}~\bibnamefont {Dumur}}, \bibinfo {author}
  {\bibfnamefont {J.}~\bibnamefont {Grebel}}, \bibinfo {author} {\bibfnamefont
  {G.~A.}\ \bibnamefont {Peairs}}, \bibinfo {author} {\bibfnamefont {R.~G.}\
  \bibnamefont {Povey}}, \bibinfo {author} {\bibfnamefont {D.~I.}\ \bibnamefont
  {Schuster}}, \ and\ \bibinfo {author} {\bibfnamefont {A.~N.}\ \bibnamefont
  {Cleland}},\ }\href {\doibase 10.1038/s41567-019-0507-7} {\bibfield
  {journal} {\bibinfo  {journal} {Nature Physics}\ }\textbf {\bibinfo {volume}
  {15}},\ \bibinfo {pages} {741} (\bibinfo {year} {2019})}\BibitemShut
  {NoStop}%
\bibitem [{\citenamefont {Liao}\ \emph {et~al.}(2017)\citenamefont {Liao},
  \citenamefont {Cai}, \citenamefont {Liu}, \citenamefont {Zhang},
  \citenamefont {Li}, \citenamefont {Ren}, \citenamefont {Yin}, \citenamefont
  {Shen}, \citenamefont {Cao}, \citenamefont {Li}, \citenamefont {Li},
  \citenamefont {Chen}, \citenamefont {Sun}, \citenamefont {Jia}, \citenamefont
  {Wu}, \citenamefont {Jiang}, \citenamefont {Wang}, \citenamefont {Huang},
  \citenamefont {Wang}, \citenamefont {Zhou}, \citenamefont {Deng},
  \citenamefont {Xi}, \citenamefont {Ma}, \citenamefont {Hu}, \citenamefont
  {Zhang}, \citenamefont {Chen}, \citenamefont {Liu}, \citenamefont {Wang},
  \citenamefont {Zhu}, \citenamefont {Lu}, \citenamefont {Shu}, \citenamefont
  {Peng}, \citenamefont {Wang},\ and\ \citenamefont {Pan}}]{Liao2017}%
  \BibitemOpen
  \bibfield  {author} {\bibinfo {author} {\bibfnamefont {S.-K.}\ \bibnamefont
  {Liao}}, \bibinfo {author} {\bibfnamefont {W.-Q.}\ \bibnamefont {Cai}},
  \bibinfo {author} {\bibfnamefont {W.-Y.}\ \bibnamefont {Liu}}, \bibinfo
  {author} {\bibfnamefont {L.}~\bibnamefont {Zhang}}, \bibinfo {author}
  {\bibfnamefont {Y.}~\bibnamefont {Li}}, \bibinfo {author} {\bibfnamefont
  {J.-G.}\ \bibnamefont {Ren}}, \bibinfo {author} {\bibfnamefont
  {J.}~\bibnamefont {Yin}}, \bibinfo {author} {\bibfnamefont {Q.}~\bibnamefont
  {Shen}}, \bibinfo {author} {\bibfnamefont {Y.}~\bibnamefont {Cao}}, \bibinfo
  {author} {\bibfnamefont {Z.-P.}\ \bibnamefont {Li}}, \bibinfo {author}
  {\bibfnamefont {F.-Z.}\ \bibnamefont {Li}}, \bibinfo {author} {\bibfnamefont
  {X.-W.}\ \bibnamefont {Chen}}, \bibinfo {author} {\bibfnamefont {L.-H.}\
  \bibnamefont {Sun}}, \bibinfo {author} {\bibfnamefont {J.-J.}\ \bibnamefont
  {Jia}}, \bibinfo {author} {\bibfnamefont {J.-C.}\ \bibnamefont {Wu}},
  \bibinfo {author} {\bibfnamefont {X.-J.}\ \bibnamefont {Jiang}}, \bibinfo
  {author} {\bibfnamefont {J.-F.}\ \bibnamefont {Wang}}, \bibinfo {author}
  {\bibfnamefont {Y.-M.}\ \bibnamefont {Huang}}, \bibinfo {author}
  {\bibfnamefont {Q.}~\bibnamefont {Wang}}, \bibinfo {author} {\bibfnamefont
  {Y.-L.}\ \bibnamefont {Zhou}}, \bibinfo {author} {\bibfnamefont
  {L.}~\bibnamefont {Deng}}, \bibinfo {author} {\bibfnamefont {T.}~\bibnamefont
  {Xi}}, \bibinfo {author} {\bibfnamefont {L.}~\bibnamefont {Ma}}, \bibinfo
  {author} {\bibfnamefont {T.}~\bibnamefont {Hu}}, \bibinfo {author}
  {\bibfnamefont {Q.}~\bibnamefont {Zhang}}, \bibinfo {author} {\bibfnamefont
  {Y.-A.}\ \bibnamefont {Chen}}, \bibinfo {author} {\bibfnamefont {N.-L.}\
  \bibnamefont {Liu}}, \bibinfo {author} {\bibfnamefont {X.-B.}\ \bibnamefont
  {Wang}}, \bibinfo {author} {\bibfnamefont {Z.-C.}\ \bibnamefont {Zhu}},
  \bibinfo {author} {\bibfnamefont {C.-Y.}\ \bibnamefont {Lu}}, \bibinfo
  {author} {\bibfnamefont {R.}~\bibnamefont {Shu}}, \bibinfo {author}
  {\bibfnamefont {C.-Z.}\ \bibnamefont {Peng}}, \bibinfo {author}
  {\bibfnamefont {J.-Y.}\ \bibnamefont {Wang}}, \ and\ \bibinfo {author}
  {\bibfnamefont {J.-W.}\ \bibnamefont {Pan}},\ }\href {\doibase
  10.1038/nature23655} {\bibfield  {journal} {\bibinfo  {journal} {Nature}\
  }\textbf {\bibinfo {volume} {549}},\ \bibinfo {pages} {43} (\bibinfo {year}
  {2017})}\BibitemShut {NoStop}%
\bibitem [{\citenamefont {Boaron}\ \emph {et~al.}(2018)\citenamefont {Boaron},
  \citenamefont {Boso}, \citenamefont {Rusca}, \citenamefont {Vulliez},
  \citenamefont {Autebert}, \citenamefont {Caloz}, \citenamefont {Perrenoud},
  \citenamefont {Gras}, \citenamefont {Bussi\`eres}, \citenamefont {Li},
  \citenamefont {Nolan}, \citenamefont {Martin},\ and\ \citenamefont
  {Zbinden}}]{Boaron2018}%
  \BibitemOpen
  \bibfield  {author} {\bibinfo {author} {\bibfnamefont {A.}~\bibnamefont
  {Boaron}}, \bibinfo {author} {\bibfnamefont {G.}~\bibnamefont {Boso}},
  \bibinfo {author} {\bibfnamefont {D.}~\bibnamefont {Rusca}}, \bibinfo
  {author} {\bibfnamefont {C.}~\bibnamefont {Vulliez}}, \bibinfo {author}
  {\bibfnamefont {C.}~\bibnamefont {Autebert}}, \bibinfo {author}
  {\bibfnamefont {M.}~\bibnamefont {Caloz}}, \bibinfo {author} {\bibfnamefont
  {M.}~\bibnamefont {Perrenoud}}, \bibinfo {author} {\bibfnamefont
  {G.}~\bibnamefont {Gras}}, \bibinfo {author} {\bibfnamefont {F.}~\bibnamefont
  {Bussi\`eres}}, \bibinfo {author} {\bibfnamefont {M.-J.}\ \bibnamefont {Li}},
  \bibinfo {author} {\bibfnamefont {D.}~\bibnamefont {Nolan}}, \bibinfo
  {author} {\bibfnamefont {A.}~\bibnamefont {Martin}}, \ and\ \bibinfo {author}
  {\bibfnamefont {H.}~\bibnamefont {Zbinden}},\ }\href {\doibase
  10.1103/PhysRevLett.121.190502} {\bibfield  {journal} {\bibinfo  {journal}
  {Phys. Rev. Lett.}\ }\textbf {\bibinfo {volume} {121}},\ \bibinfo {pages}
  {190502} (\bibinfo {year} {2018})}\BibitemShut {NoStop}%
\bibitem [{\citenamefont {Yu}\ \emph {et~al.}(2020)\citenamefont {Yu},
  \citenamefont {Ma}, \citenamefont {Luo}, \citenamefont {Jing}, \citenamefont
  {Sun}, \citenamefont {Fang}, \citenamefont {Yang}, \citenamefont {Liu},
  \citenamefont {Zheng}, \citenamefont {Xie}, \citenamefont {Zhang},
  \citenamefont {You}, \citenamefont {Wang}, \citenamefont {Chen},
  \citenamefont {Zhang}, \citenamefont {Bao},\ and\ \citenamefont
  {Pan}}]{Yu2020}%
  \BibitemOpen
  \bibfield  {author} {\bibinfo {author} {\bibfnamefont {Y.}~\bibnamefont
  {Yu}}, \bibinfo {author} {\bibfnamefont {F.}~\bibnamefont {Ma}}, \bibinfo
  {author} {\bibfnamefont {X.-Y.}\ \bibnamefont {Luo}}, \bibinfo {author}
  {\bibfnamefont {B.}~\bibnamefont {Jing}}, \bibinfo {author} {\bibfnamefont
  {P.-F.}\ \bibnamefont {Sun}}, \bibinfo {author} {\bibfnamefont {R.-Z.}\
  \bibnamefont {Fang}}, \bibinfo {author} {\bibfnamefont {C.-W.}\ \bibnamefont
  {Yang}}, \bibinfo {author} {\bibfnamefont {H.}~\bibnamefont {Liu}}, \bibinfo
  {author} {\bibfnamefont {M.-Y.}\ \bibnamefont {Zheng}}, \bibinfo {author}
  {\bibfnamefont {X.-P.}\ \bibnamefont {Xie}}, \bibinfo {author} {\bibfnamefont
  {W.-J.}\ \bibnamefont {Zhang}}, \bibinfo {author} {\bibfnamefont {L.-X.}\
  \bibnamefont {You}}, \bibinfo {author} {\bibfnamefont {Z.}~\bibnamefont
  {Wang}}, \bibinfo {author} {\bibfnamefont {T.-Y.}\ \bibnamefont {Chen}},
  \bibinfo {author} {\bibfnamefont {Q.}~\bibnamefont {Zhang}}, \bibinfo
  {author} {\bibfnamefont {X.-H.}\ \bibnamefont {Bao}}, \ and\ \bibinfo
  {author} {\bibfnamefont {J.-W.}\ \bibnamefont {Pan}},\ }\href {\doibase
  10.1038/s41586-020-1976-7} {\bibfield  {journal} {\bibinfo  {journal}
  {Nature}\ }\textbf {\bibinfo {volume} {578}},\ \bibinfo {pages} {240}
  (\bibinfo {year} {2020})}\BibitemShut {NoStop}%
\bibitem [{\citenamefont {Xiang}\ \emph {et~al.}(2013)\citenamefont {Xiang},
  \citenamefont {Ashhab}, \citenamefont {You},\ and\ \citenamefont
  {Nori}}]{Xiang2013}%
  \BibitemOpen
  \bibfield  {author} {\bibinfo {author} {\bibfnamefont {Z.-L.}\ \bibnamefont
  {Xiang}}, \bibinfo {author} {\bibfnamefont {S.}~\bibnamefont {Ashhab}},
  \bibinfo {author} {\bibfnamefont {J.~Q.}\ \bibnamefont {You}}, \ and\
  \bibinfo {author} {\bibfnamefont {F.}~\bibnamefont {Nori}},\ }\href {\doibase
  10.1103/RevModPhys.85.623} {\bibfield  {journal} {\bibinfo  {journal} {Rev.
  Mod. Phys.}\ }\textbf {\bibinfo {volume} {85}},\ \bibinfo {pages} {623}
  (\bibinfo {year} {2013})}\BibitemShut {NoStop}%
\bibitem [{\citenamefont {Clerk}\ \emph {et~al.}(2020)\citenamefont {Clerk},
  \citenamefont {Lehnert}, \citenamefont {Bertet}, \citenamefont {Petta},\ and\
  \citenamefont {Nakamura}}]{Clerk2020}%
  \BibitemOpen
  \bibfield  {author} {\bibinfo {author} {\bibfnamefont {A.~A.}\ \bibnamefont
  {Clerk}}, \bibinfo {author} {\bibfnamefont {K.~W.}\ \bibnamefont {Lehnert}},
  \bibinfo {author} {\bibfnamefont {P.}~\bibnamefont {Bertet}}, \bibinfo
  {author} {\bibfnamefont {J.~R.}\ \bibnamefont {Petta}}, \ and\ \bibinfo
  {author} {\bibfnamefont {Y.}~\bibnamefont {Nakamura}},\ }\href {\doibase
  10.1038/s41567-020-0797-9} {\bibfield  {journal} {\bibinfo  {journal} {Nature
  Physics}\ }\textbf {\bibinfo {volume} {16}},\ \bibinfo {pages} {257}
  (\bibinfo {year} {2020})}\BibitemShut {NoStop}%
\bibitem [{\citenamefont {Stannigel}\ \emph {et~al.}(2010)\citenamefont
  {Stannigel}, \citenamefont {Rabl}, \citenamefont {S\o{}rensen}, \citenamefont
  {Zoller},\ and\ \citenamefont {Lukin}}]{Stannigel2010}%
  \BibitemOpen
  \bibfield  {author} {\bibinfo {author} {\bibfnamefont {K.}~\bibnamefont
  {Stannigel}}, \bibinfo {author} {\bibfnamefont {P.}~\bibnamefont {Rabl}},
  \bibinfo {author} {\bibfnamefont {A.~S.}\ \bibnamefont {S\o{}rensen}},
  \bibinfo {author} {\bibfnamefont {P.}~\bibnamefont {Zoller}}, \ and\ \bibinfo
  {author} {\bibfnamefont {M.~D.}\ \bibnamefont {Lukin}},\ }\href {\doibase
  10.1103/PhysRevLett.105.220501} {\bibfield  {journal} {\bibinfo  {journal}
  {Phys. Rev. Lett.}\ }\textbf {\bibinfo {volume} {105}},\ \bibinfo {pages}
  {220501} (\bibinfo {year} {2010})}\BibitemShut {NoStop}%
\bibitem [{\citenamefont {Forsch}\ \emph {et~al.}(2020)\citenamefont {Forsch},
  \citenamefont {Stockill}, \citenamefont {Wallucks}, \citenamefont
  {Marinkovic}, \citenamefont {G{\"a}rtner}, \citenamefont {Norte},
  \citenamefont {van Otten}, \citenamefont {Fiore}, \citenamefont
  {Srinivasan},\ and\ \citenamefont {Gr{\"o}blacher}}]{Forsch2020}%
  \BibitemOpen
  \bibfield  {author} {\bibinfo {author} {\bibfnamefont {M.}~\bibnamefont
  {Forsch}}, \bibinfo {author} {\bibfnamefont {R.}~\bibnamefont {Stockill}},
  \bibinfo {author} {\bibfnamefont {A.}~\bibnamefont {Wallucks}}, \bibinfo
  {author} {\bibfnamefont {I.}~\bibnamefont {Marinkovic}}, \bibinfo {author}
  {\bibfnamefont {C.}~\bibnamefont {G{\"a}rtner}}, \bibinfo {author}
  {\bibfnamefont {R.~A.}\ \bibnamefont {Norte}}, \bibinfo {author}
  {\bibfnamefont {F.}~\bibnamefont {van Otten}}, \bibinfo {author}
  {\bibfnamefont {A.}~\bibnamefont {Fiore}}, \bibinfo {author} {\bibfnamefont
  {K.}~\bibnamefont {Srinivasan}}, \ and\ \bibinfo {author} {\bibfnamefont
  {S.}~\bibnamefont {Gr{\"o}blacher}},\ }\href {\doibase
  10.1038/s41567-019-0673-7} {\bibfield  {journal} {\bibinfo  {journal} {Nature
  Physics}\ }\textbf {\bibinfo {volume} {16}},\ \bibinfo {pages} {69} (\bibinfo
  {year} {2020})}\BibitemShut {NoStop}%
\bibitem [{\citenamefont {Fan}\ \emph {et~al.}(2018)\citenamefont {Fan},
  \citenamefont {Zou}, \citenamefont {Cheng}, \citenamefont {Guo},
  \citenamefont {Han}, \citenamefont {Gong}, \citenamefont {Wang},\ and\
  \citenamefont {Tang}}]{Fan2018}%
  \BibitemOpen
  \bibfield  {author} {\bibinfo {author} {\bibfnamefont {L.}~\bibnamefont
  {Fan}}, \bibinfo {author} {\bibfnamefont {C.-L.}\ \bibnamefont {Zou}},
  \bibinfo {author} {\bibfnamefont {R.}~\bibnamefont {Cheng}}, \bibinfo
  {author} {\bibfnamefont {X.}~\bibnamefont {Guo}}, \bibinfo {author}
  {\bibfnamefont {X.}~\bibnamefont {Han}}, \bibinfo {author} {\bibfnamefont
  {Z.}~\bibnamefont {Gong}}, \bibinfo {author} {\bibfnamefont {S.}~\bibnamefont
  {Wang}}, \ and\ \bibinfo {author} {\bibfnamefont {H.~X.}\ \bibnamefont
  {Tang}},\ }\href {\doibase 10.1126/sciadv.aar4994} {\bibfield  {journal}
  {\bibinfo  {journal} {Science Advances}\ }\textbf {\bibinfo {volume} {4}}
  (\bibinfo {year} {2018}),\ 10.1126/sciadv.aar4994}\BibitemShut {NoStop}%
\bibitem [{\citenamefont {Fan}\ \emph {et~al.}(2015)\citenamefont {Fan},
  \citenamefont {Kumar}, \citenamefont {Sedlacek}, \citenamefont {Kübler},
  \citenamefont {Karimkashi},\ and\ \citenamefont {Shaffer}}]{Fan_2015}%
  \BibitemOpen
  \bibfield  {author} {\bibinfo {author} {\bibfnamefont {H.}~\bibnamefont
  {Fan}}, \bibinfo {author} {\bibfnamefont {S.}~\bibnamefont {Kumar}}, \bibinfo
  {author} {\bibfnamefont {J.}~\bibnamefont {Sedlacek}}, \bibinfo {author}
  {\bibfnamefont {H.}~\bibnamefont {Kübler}}, \bibinfo {author} {\bibfnamefont
  {S.}~\bibnamefont {Karimkashi}}, \ and\ \bibinfo {author} {\bibfnamefont
  {J.~P.}\ \bibnamefont {Shaffer}},\ }\href {\doibase
  10.1088/0953-4075/48/20/202001} {\bibfield  {journal} {\bibinfo  {journal}
  {Journal of Physics B: Atomic, Molecular and Optical Physics}\ }\textbf
  {\bibinfo {volume} {48}},\ \bibinfo {pages} {202001} (\bibinfo {year}
  {2015})}\BibitemShut {NoStop}%
\bibitem [{\citenamefont {Gard}\ \emph {et~al.}(2017)\citenamefont {Gard},
  \citenamefont {Jacobs}, \citenamefont {McDermott},\ and\ \citenamefont
  {Saffman}}]{Gard2017}%
  \BibitemOpen
  \bibfield  {author} {\bibinfo {author} {\bibfnamefont {B.~T.}\ \bibnamefont
  {Gard}}, \bibinfo {author} {\bibfnamefont {K.}~\bibnamefont {Jacobs}},
  \bibinfo {author} {\bibfnamefont {R.}~\bibnamefont {McDermott}}, \ and\
  \bibinfo {author} {\bibfnamefont {M.}~\bibnamefont {Saffman}},\ }\href
  {\doibase 10.1103/PhysRevA.96.013833} {\bibfield  {journal} {\bibinfo
  {journal} {Phys. Rev. A}\ }\textbf {\bibinfo {volume} {96}},\ \bibinfo
  {pages} {013833} (\bibinfo {year} {2017})}\BibitemShut {NoStop}%
\bibitem [{\citenamefont {Han}\ \emph {et~al.}(2018)\citenamefont {Han},
  \citenamefont {Vogt}, \citenamefont {Gross}, \citenamefont {Jaksch},
  \citenamefont {Kiffner},\ and\ \citenamefont {Li}}]{Han2018}%
  \BibitemOpen
  \bibfield  {author} {\bibinfo {author} {\bibfnamefont {J.}~\bibnamefont
  {Han}}, \bibinfo {author} {\bibfnamefont {T.}~\bibnamefont {Vogt}}, \bibinfo
  {author} {\bibfnamefont {C.}~\bibnamefont {Gross}}, \bibinfo {author}
  {\bibfnamefont {D.}~\bibnamefont {Jaksch}}, \bibinfo {author} {\bibfnamefont
  {M.}~\bibnamefont {Kiffner}}, \ and\ \bibinfo {author} {\bibfnamefont
  {W.}~\bibnamefont {Li}},\ }\href {\doibase 10.1103/PhysRevLett.120.093201}
  {\bibfield  {journal} {\bibinfo  {journal} {Phys. Rev. Lett.}\ }\textbf
  {\bibinfo {volume} {120}},\ \bibinfo {pages} {093201} (\bibinfo {year}
  {2018})}\BibitemShut {NoStop}%
\bibitem [{\citenamefont {Petrosyan}\ \emph {et~al.}(2019)\citenamefont
  {Petrosyan}, \citenamefont {M{\o}lmer}, \citenamefont {Fort{\'{a}}gh},\ and\
  \citenamefont {Saffman}}]{Petrosyan_2019}%
  \BibitemOpen
  \bibfield  {author} {\bibinfo {author} {\bibfnamefont {D.}~\bibnamefont
  {Petrosyan}}, \bibinfo {author} {\bibfnamefont {K.}~\bibnamefont
  {M{\o}lmer}}, \bibinfo {author} {\bibfnamefont {J.}~\bibnamefont
  {Fort{\'{a}}gh}}, \ and\ \bibinfo {author} {\bibfnamefont {M.}~\bibnamefont
  {Saffman}},\ }\href {\doibase 10.1088/1367-2630/ab307c} {\bibfield  {journal}
  {\bibinfo  {journal} {New Journal of Physics}\ }\textbf {\bibinfo {volume}
  {21}},\ \bibinfo {pages} {073033} (\bibinfo {year} {2019})}\BibitemShut
  {NoStop}%
\bibitem [{\citenamefont {Chopinaud}\ and\ \citenamefont
  {Pritchard}(2021)}]{chopinaud2021}%
  \BibitemOpen
  \bibfield  {author} {\bibinfo {author} {\bibfnamefont {A.}~\bibnamefont
  {Chopinaud}}\ and\ \bibinfo {author} {\bibfnamefont {J.~D.}\ \bibnamefont
  {Pritchard}},\ }\href {\doibase 10.1103/PhysRevApplied.16.024008} {\bibfield
  {journal} {\bibinfo  {journal} {Phys. Rev. Applied}\ }\textbf {\bibinfo
  {volume} {16}},\ \bibinfo {pages} {024008} (\bibinfo {year}
  {2021})}\BibitemShut {NoStop}%
\bibitem [{\citenamefont {Hattermann}\ \emph {et~al.}(2017)\citenamefont
  {Hattermann}, \citenamefont {Bothner}, \citenamefont {Ley}, \citenamefont
  {Ferdinand}, \citenamefont {Wiedmaier}, \citenamefont {S{\'a}rk{\'a}ny},
  \citenamefont {Kleiner}, \citenamefont {Koelle},\ and\ \citenamefont
  {Fort{\'a}gh}}]{Hattermann2017}%
  \BibitemOpen
  \bibfield  {author} {\bibinfo {author} {\bibfnamefont {H.}~\bibnamefont
  {Hattermann}}, \bibinfo {author} {\bibfnamefont {D.}~\bibnamefont {Bothner}},
  \bibinfo {author} {\bibfnamefont {L.~Y.}\ \bibnamefont {Ley}}, \bibinfo
  {author} {\bibfnamefont {B.}~\bibnamefont {Ferdinand}}, \bibinfo {author}
  {\bibfnamefont {D.}~\bibnamefont {Wiedmaier}}, \bibinfo {author}
  {\bibfnamefont {L.}~\bibnamefont {S{\'a}rk{\'a}ny}}, \bibinfo {author}
  {\bibfnamefont {R.}~\bibnamefont {Kleiner}}, \bibinfo {author} {\bibfnamefont
  {D.}~\bibnamefont {Koelle}}, \ and\ \bibinfo {author} {\bibfnamefont
  {J.}~\bibnamefont {Fort{\'a}gh}},\ }\href {\doibase
  10.1038/s41467-017-02439-7} {\bibfield  {journal} {\bibinfo  {journal}
  {Nature Communications}\ }\textbf {\bibinfo {volume} {8}},\ \bibinfo {pages}
  {2254} (\bibinfo {year} {2017})}\BibitemShut {NoStop}%
\bibitem [{\citenamefont {Sch\"one}\ \emph
  {et~al.}(2017{\natexlab{a}})\citenamefont {Sch\"one}, \citenamefont {Stolz},\
  and\ \citenamefont {Naka}}]{morgan2019}%
  \BibitemOpen
  \bibfield  {author} {\bibinfo {author} {\bibfnamefont {F.}~\bibnamefont
  {Sch\"one}}, \bibinfo {author} {\bibfnamefont {H.}~\bibnamefont {Stolz}}, \
  and\ \bibinfo {author} {\bibfnamefont {N.}~\bibnamefont {Naka}},\ }\href
  {\doibase 10.1103/PhysRevB.96.115207} {\bibfield  {journal} {\bibinfo
  {journal} {Phys. Rev. B}\ }\textbf {\bibinfo {volume} {96}},\ \bibinfo
  {pages} {115207} (\bibinfo {year} {2017}{\natexlab{a}})}\BibitemShut
  {NoStop}%
\bibitem [{\citenamefont {Petrosyan}\ \emph {et~al.}(2009)\citenamefont
  {Petrosyan}, \citenamefont {Bensky}, \citenamefont {Kurizki}, \citenamefont
  {Mazets}, \citenamefont {Majer},\ and\ \citenamefont
  {Schmiedmayer}}]{Petrosyan2009}%
  \BibitemOpen
  \bibfield  {author} {\bibinfo {author} {\bibfnamefont {D.}~\bibnamefont
  {Petrosyan}}, \bibinfo {author} {\bibfnamefont {G.}~\bibnamefont {Bensky}},
  \bibinfo {author} {\bibfnamefont {G.}~\bibnamefont {Kurizki}}, \bibinfo
  {author} {\bibfnamefont {I.}~\bibnamefont {Mazets}}, \bibinfo {author}
  {\bibfnamefont {J.}~\bibnamefont {Majer}}, \ and\ \bibinfo {author}
  {\bibfnamefont {J.}~\bibnamefont {Schmiedmayer}},\ }\href {\doibase
  10.1103/PhysRevA.79.040304} {\bibfield  {journal} {\bibinfo  {journal} {Phys.
  Rev. A}\ }\textbf {\bibinfo {volume} {79}},\ \bibinfo {pages} {040304}
  (\bibinfo {year} {2009})}\BibitemShut {NoStop}%
\bibitem [{\citenamefont {Kaiser}\ \emph {et~al.}(2021)\citenamefont {Kaiser},
  \citenamefont {Glaser}, \citenamefont {Ley}, \citenamefont {Grimmel},
  \citenamefont {Hattermann}, \citenamefont {Bothner}, \citenamefont {Koelle},
  \citenamefont {Kleiner}, \citenamefont {Petrosyan}, \citenamefont
  {Günther},\ and\ \citenamefont {Fortágh}}]{kaiser2021}%
  \BibitemOpen
  \bibfield  {author} {\bibinfo {author} {\bibfnamefont {M.}~\bibnamefont
  {Kaiser}}, \bibinfo {author} {\bibfnamefont {C.}~\bibnamefont {Glaser}},
  \bibinfo {author} {\bibfnamefont {L.~Y.}\ \bibnamefont {Ley}}, \bibinfo
  {author} {\bibfnamefont {J.}~\bibnamefont {Grimmel}}, \bibinfo {author}
  {\bibfnamefont {H.}~\bibnamefont {Hattermann}}, \bibinfo {author}
  {\bibfnamefont {D.}~\bibnamefont {Bothner}}, \bibinfo {author} {\bibfnamefont
  {D.}~\bibnamefont {Koelle}}, \bibinfo {author} {\bibfnamefont
  {R.}~\bibnamefont {Kleiner}}, \bibinfo {author} {\bibfnamefont
  {D.}~\bibnamefont {Petrosyan}}, \bibinfo {author} {\bibfnamefont
  {A.}~\bibnamefont {Günther}}, \ and\ \bibinfo {author} {\bibfnamefont
  {J.}~\bibnamefont {Fortágh}},\ }\href@noop {} {\enquote {\bibinfo {title}
  {Cavity driven \text{Rabi} oscillations between \text{Rydberg} states of
  atoms trapped on a superconducting atom chip},}\ } (\bibinfo {year} {2021}),\
  \Eprint {http://arxiv.org/abs/2105.05188} {arXiv:2105.05188 [quant-ph]}
  \BibitemShut {NoStop}%
\bibitem [{\citenamefont {Hayashi}\ and\ \citenamefont
  {Katsuki}(1952)}]{hayashi1952}%
  \BibitemOpen
  \bibfield  {author} {\bibinfo {author} {\bibfnamefont {M.}~\bibnamefont
  {Hayashi}}\ and\ \bibinfo {author} {\bibfnamefont {K.}~\bibnamefont
  {Katsuki}},\ }\href {\doibase 10.1143/JPSJ.7.599} {\bibfield  {journal}
  {\bibinfo  {journal} {Journal of the Physical Society of Japan}\ }\textbf
  {\bibinfo {volume} {7}},\ \bibinfo {pages} {599} (\bibinfo {year}
  {1952})}\BibitemShut {NoStop}%
\bibitem [{\citenamefont {Gross}(1956)}]{Gross1956}%
  \BibitemOpen
  \bibfield  {author} {\bibinfo {author} {\bibfnamefont {E.~F.}\ \bibnamefont
  {Gross}},\ }\href {\doibase 10.1007/BF02746069} {\bibfield  {journal}
  {\bibinfo  {journal} {Il Nuovo Cimento}\ }\textbf {\bibinfo {volume} {3}},\
  \bibinfo {pages} {672} (\bibinfo {year} {1956})}\BibitemShut {NoStop}%
\bibitem [{\citenamefont {Agekyan}(1977)}]{Agekyan1977}%
  \BibitemOpen
  \bibfield  {author} {\bibinfo {author} {\bibfnamefont {V.~T.}\ \bibnamefont
  {Agekyan}},\ }\href {\doibase 10.1002/pssa.2210430102} {\bibfield  {journal}
  {\bibinfo  {journal} {physica status solidi (a)}\ }\textbf {\bibinfo {volume}
  {43}},\ \bibinfo {pages} {11} (\bibinfo {year} {1977})}\BibitemShut {NoStop}%
\bibitem [{\citenamefont {Washington}\ \emph {et~al.}(1977)\citenamefont
  {Washington}, \citenamefont {Genack}, \citenamefont {Cummins}, \citenamefont
  {Bruce}, \citenamefont {Compaan},\ and\ \citenamefont
  {Forman}}]{Washington1977}%
  \BibitemOpen
  \bibfield  {author} {\bibinfo {author} {\bibfnamefont {M.~A.}\ \bibnamefont
  {Washington}}, \bibinfo {author} {\bibfnamefont {A.~Z.}\ \bibnamefont
  {Genack}}, \bibinfo {author} {\bibfnamefont {H.~Z.}\ \bibnamefont {Cummins}},
  \bibinfo {author} {\bibfnamefont {R.~H.}\ \bibnamefont {Bruce}}, \bibinfo
  {author} {\bibfnamefont {A.}~\bibnamefont {Compaan}}, \ and\ \bibinfo
  {author} {\bibfnamefont {R.~A.}\ \bibnamefont {Forman}},\ }\href {\doibase
  10.1103/PhysRevB.15.2145} {\bibfield  {journal} {\bibinfo  {journal} {Phys.
  Rev. B}\ }\textbf {\bibinfo {volume} {15}},\ \bibinfo {pages} {2145}
  (\bibinfo {year} {1977})}\BibitemShut {NoStop}%
\bibitem [{\citenamefont {Kavoulakis}\ \emph {et~al.}(1997)\citenamefont
  {Kavoulakis}, \citenamefont {Chang},\ and\ \citenamefont
  {Baym}}]{Kavoulakis1997}%
  \BibitemOpen
  \bibfield  {author} {\bibinfo {author} {\bibfnamefont {G.~M.}\ \bibnamefont
  {Kavoulakis}}, \bibinfo {author} {\bibfnamefont {Y.-C.}\ \bibnamefont
  {Chang}}, \ and\ \bibinfo {author} {\bibfnamefont {G.}~\bibnamefont {Baym}},\
  }\href {\doibase 10.1103/PhysRevB.55.7593} {\bibfield  {journal} {\bibinfo
  {journal} {Phys. Rev. B}\ }\textbf {\bibinfo {volume} {55}},\ \bibinfo
  {pages} {7593} (\bibinfo {year} {1997})}\BibitemShut {NoStop}%
\bibitem [{\citenamefont {Kazimierczuk}\ \emph {et~al.}(2014)\citenamefont
  {Kazimierczuk}, \citenamefont {Fr{\"{o}}hlich}, \citenamefont {Scheel},
  \citenamefont {Stolz},\ and\ \citenamefont {Bayer}}]{Kazimierczuk2014}%
  \BibitemOpen
  \bibfield  {author} {\bibinfo {author} {\bibfnamefont {T.}~\bibnamefont
  {Kazimierczuk}}, \bibinfo {author} {\bibfnamefont {D.}~\bibnamefont
  {Fr{\"{o}}hlich}}, \bibinfo {author} {\bibfnamefont {S.}~\bibnamefont
  {Scheel}}, \bibinfo {author} {\bibfnamefont {H.}~\bibnamefont {Stolz}}, \
  and\ \bibinfo {author} {\bibfnamefont {M.}~\bibnamefont {Bayer}},\ }\href
  {\doibase 10.1038/nature13832} {\bibfield  {journal} {\bibinfo  {journal}
  {Nature}\ }\textbf {\bibinfo {volume} {514}},\ \bibinfo {pages} {343}
  (\bibinfo {year} {2014})}\BibitemShut {NoStop}%
\bibitem [{\citenamefont {Versteegh}\ \emph {et~al.}(2021)\citenamefont
  {Versteegh}, \citenamefont {Steinhauer}, \citenamefont {Bajo}, \citenamefont
  {Lettner}, \citenamefont {Soro}, \citenamefont {Romanova}, \citenamefont
  {Gyger}, \citenamefont {Schweickert}, \citenamefont {Mysyrowicz},\ and\
  \citenamefont {Zwiller}}]{versteegh2021}%
  \BibitemOpen
  \bibfield  {author} {\bibinfo {author} {\bibfnamefont {M.~A.~M.}\
  \bibnamefont {Versteegh}}, \bibinfo {author} {\bibfnamefont {S.}~\bibnamefont
  {Steinhauer}}, \bibinfo {author} {\bibfnamefont {J.}~\bibnamefont {Bajo}},
  \bibinfo {author} {\bibfnamefont {T.}~\bibnamefont {Lettner}}, \bibinfo
  {author} {\bibfnamefont {A.}~\bibnamefont {Soro}}, \bibinfo {author}
  {\bibfnamefont {A.}~\bibnamefont {Romanova}}, \bibinfo {author}
  {\bibfnamefont {S.}~\bibnamefont {Gyger}}, \bibinfo {author} {\bibfnamefont
  {L.}~\bibnamefont {Schweickert}}, \bibinfo {author} {\bibfnamefont
  {A.}~\bibnamefont {Mysyrowicz}}, \ and\ \bibinfo {author} {\bibfnamefont
  {V.}~\bibnamefont {Zwiller}},\ }\href@noop {} {\enquote {\bibinfo {title}
  {Giant \text{Rydberg} excitons in $\mathrm{Cu}_{2}\mathrm{O}$ probed by
  photoluminescence excitation spectroscopy},}\ } (\bibinfo {year} {2021}),\
  \Eprint {http://arxiv.org/abs/2105.07942} {arXiv:2105.07942
  [cond-mat.mes-hall]} \BibitemShut {NoStop}%
\bibitem [{\citenamefont {Snoke}(2002)}]{Snoke2002}%
  \BibitemOpen
  \bibfield  {author} {\bibinfo {author} {\bibfnamefont {D.}~\bibnamefont
  {Snoke}},\ }\href {\doibase 10.1126/science.1078082} {\bibfield  {journal}
  {\bibinfo  {journal} {Science}\ }\textbf {\bibinfo {volume} {298}},\ \bibinfo
  {pages} {1368} (\bibinfo {year} {2002})}\BibitemShut {NoStop}%
\bibitem [{\citenamefont {Heck\"otter}\ \emph {et~al.}(2020)\citenamefont
  {Heck\"otter}, \citenamefont {Janas}, \citenamefont {Schwartz}, \citenamefont
  {A\ss{}mann},\ and\ \citenamefont {Bayer}}]{Heckotter2020}%
  \BibitemOpen
  \bibfield  {author} {\bibinfo {author} {\bibfnamefont {J.}~\bibnamefont
  {Heck\"otter}}, \bibinfo {author} {\bibfnamefont {D.}~\bibnamefont {Janas}},
  \bibinfo {author} {\bibfnamefont {R.}~\bibnamefont {Schwartz}}, \bibinfo
  {author} {\bibfnamefont {M.}~\bibnamefont {A\ss{}mann}}, \ and\ \bibinfo
  {author} {\bibfnamefont {M.}~\bibnamefont {Bayer}},\ }\href {\doibase
  10.1103/PhysRevB.101.235207} {\bibfield  {journal} {\bibinfo  {journal}
  {Phys. Rev. B}\ }\textbf {\bibinfo {volume} {101}},\ \bibinfo {pages}
  {235207} (\bibinfo {year} {2020})}\BibitemShut {NoStop}%
\bibitem [{\citenamefont {Fr\"ohlich}\ \emph {et~al.}(1985)\citenamefont
  {Fr\"ohlich}, \citenamefont {N\"othe},\ and\ \citenamefont
  {Reimann}}]{Frohlich1985}%
  \BibitemOpen
  \bibfield  {author} {\bibinfo {author} {\bibfnamefont {D.}~\bibnamefont
  {Fr\"ohlich}}, \bibinfo {author} {\bibfnamefont {A.}~\bibnamefont {N\"othe}},
  \ and\ \bibinfo {author} {\bibfnamefont {K.}~\bibnamefont {Reimann}},\ }\href
  {\doibase 10.1103/PhysRevLett.55.1335} {\bibfield  {journal} {\bibinfo
  {journal} {Phys. Rev. Lett.}\ }\textbf {\bibinfo {volume} {55}},\ \bibinfo
  {pages} {1335} (\bibinfo {year} {1985})}\BibitemShut {NoStop}%
\bibitem [{\citenamefont {Fröhlich}\ \emph {et~al.}(1987)\citenamefont
  {Fröhlich}, \citenamefont {Reimann},\ and\ \citenamefont
  {Wille}}]{FROHLICH1987}%
  \BibitemOpen
  \bibfield  {author} {\bibinfo {author} {\bibfnamefont {D.}~\bibnamefont
  {Fröhlich}}, \bibinfo {author} {\bibfnamefont {K.}~\bibnamefont {Reimann}},
  \ and\ \bibinfo {author} {\bibfnamefont {R.}~\bibnamefont {Wille}},\ }\href
  {\doibase https://doi.org/10.1016/0022-2313(87)90114-1} {\bibfield  {journal}
  {\bibinfo  {journal} {Journal of Luminescence}\ }\textbf {\bibinfo {volume}
  {38}},\ \bibinfo {pages} {235} (\bibinfo {year} {1987})}\BibitemShut
  {NoStop}%
\bibitem [{\citenamefont {Froehlich}\ \emph {et~al.}(1990)\citenamefont
  {Froehlich}, \citenamefont {Neumann}, \citenamefont {Uebbing},\ and\
  \citenamefont {Wille}}]{Frohlich1990}%
  \BibitemOpen
  \bibfield  {author} {\bibinfo {author} {\bibfnamefont {D.~H.}\ \bibnamefont
  {Froehlich}}, \bibinfo {author} {\bibfnamefont {C.}~\bibnamefont {Neumann}},
  \bibinfo {author} {\bibfnamefont {B.}~\bibnamefont {Uebbing}}, \ and\
  \bibinfo {author} {\bibfnamefont {R.~H.}\ \bibnamefont {Wille}},\ }in\ \href
  {\doibase 10.1117/12.18123} {\emph {\bibinfo {booktitle} {Nonlinear Optical
  Materials and Devices for Photonic Switching}}},\ Vol.\ \bibinfo {volume}
  {1216},\ \bibinfo {editor} {edited by\ \bibinfo {editor} {\bibfnamefont
  {N.}~\bibnamefont {Peyghambarian}}},\ \bibinfo {organization} {International
  Society for Optics and Photonics}\ (\bibinfo  {publisher} {SPIE},\ \bibinfo
  {year} {1990})\ pp.\ \bibinfo {pages} {198 -- 205}\BibitemShut {NoStop}%
\bibitem [{\citenamefont {Fröhlich}\ \emph {et~al.}(1990)\citenamefont
  {Fröhlich}, \citenamefont {Neumann}, \citenamefont {Uebbing},\ and\
  \citenamefont {Wille}}]{Frohlich1990-2}%
  \BibitemOpen
  \bibfield  {author} {\bibinfo {author} {\bibfnamefont {D.}~\bibnamefont
  {Fröhlich}}, \bibinfo {author} {\bibfnamefont {C.}~\bibnamefont {Neumann}},
  \bibinfo {author} {\bibfnamefont {B.}~\bibnamefont {Uebbing}}, \ and\
  \bibinfo {author} {\bibfnamefont {R.}~\bibnamefont {Wille}},\ }\href
  {\doibase https://doi.org/10.1002/pssb.2221590135} {\bibfield  {journal}
  {\bibinfo  {journal} {physica status solidi (b)}\ }\textbf {\bibinfo {volume}
  {159}},\ \bibinfo {pages} {297} (\bibinfo {year} {1990})}\BibitemShut
  {NoStop}%
\bibitem [{\citenamefont {Jörger}\ \emph {et~al.}(2003)\citenamefont
  {Jörger}, \citenamefont {Tsitsishvili}, \citenamefont {Fleck},\ and\
  \citenamefont {Klingshirn}}]{Jorger2003}%
  \BibitemOpen
  \bibfield  {author} {\bibinfo {author} {\bibfnamefont {M.}~\bibnamefont
  {Jörger}}, \bibinfo {author} {\bibfnamefont {E.}~\bibnamefont
  {Tsitsishvili}}, \bibinfo {author} {\bibfnamefont {T.}~\bibnamefont {Fleck}},
  \ and\ \bibinfo {author} {\bibfnamefont {C.}~\bibnamefont {Klingshirn}},\
  }\href {\doibase 10.1002/pssb.200303164} {\bibfield  {journal} {\bibinfo
  {journal} {physica status solidi (b)}\ }\textbf {\bibinfo {volume} {238}},\
  \bibinfo {pages} {470} (\bibinfo {year} {2003})}\BibitemShut {NoStop}%
\bibitem [{\citenamefont {Bassani}\ \emph {et~al.}(2004)\citenamefont
  {Bassani}, \citenamefont {{La Rocca}},\ and\ \citenamefont
  {Artoni}}]{BASSANI2004}%
  \BibitemOpen
  \bibfield  {author} {\bibinfo {author} {\bibfnamefont {F.}~\bibnamefont
  {Bassani}}, \bibinfo {author} {\bibfnamefont {G.}~\bibnamefont {{La Rocca}}},
  \ and\ \bibinfo {author} {\bibfnamefont {M.}~\bibnamefont {Artoni}},\ }\href
  {\doibase https://doi.org/10.1016/j.jlumin.2004.08.005} {\bibfield  {journal}
  {\bibinfo  {journal} {Journal of Luminescence}\ }\textbf {\bibinfo {volume}
  {110}},\ \bibinfo {pages} {174} (\bibinfo {year} {2004})},\ \bibinfo {note}
  {325th Wilhelm and Else Heraeus Workshop. Organic Molecular Solids : Excited
  Electronic States and Optical Properties}\BibitemShut {NoStop}%
\bibitem [{\citenamefont {Kuwata-Gonokami}\ \emph {et~al.}(2004)\citenamefont
  {Kuwata-Gonokami}, \citenamefont {Kubouchi}, \citenamefont {Shimano},\ and\
  \citenamefont {Mysyrowicz}}]{Kuwata2004}%
  \BibitemOpen
  \bibfield  {author} {\bibinfo {author} {\bibfnamefont {M.}~\bibnamefont
  {Kuwata-Gonokami}}, \bibinfo {author} {\bibfnamefont {M.}~\bibnamefont
  {Kubouchi}}, \bibinfo {author} {\bibfnamefont {R.}~\bibnamefont {Shimano}}, \
  and\ \bibinfo {author} {\bibfnamefont {A.}~\bibnamefont {Mysyrowicz}},\
  }\href {\doibase 10.1143/JPSJ.73.1065} {\bibfield  {journal} {\bibinfo
  {journal} {Journal of the Physical Society of Japan}\ }\textbf {\bibinfo
  {volume} {73}},\ \bibinfo {pages} {1065} (\bibinfo {year}
  {2004})}\BibitemShut {NoStop}%
\bibitem [{\citenamefont {Kubouchi}\ \emph {et~al.}(2005)\citenamefont
  {Kubouchi}, \citenamefont {Yoshioka}, \citenamefont {Shimano}, \citenamefont
  {Mysyrowicz},\ and\ \citenamefont {Kuwata-Gonokami}}]{Kubouchi2005}%
  \BibitemOpen
  \bibfield  {author} {\bibinfo {author} {\bibfnamefont {M.}~\bibnamefont
  {Kubouchi}}, \bibinfo {author} {\bibfnamefont {K.}~\bibnamefont {Yoshioka}},
  \bibinfo {author} {\bibfnamefont {R.}~\bibnamefont {Shimano}}, \bibinfo
  {author} {\bibfnamefont {A.}~\bibnamefont {Mysyrowicz}}, \ and\ \bibinfo
  {author} {\bibfnamefont {M.}~\bibnamefont {Kuwata-Gonokami}},\ }\href
  {\doibase 10.1103/PhysRevLett.94.016403} {\bibfield  {journal} {\bibinfo
  {journal} {Phys. Rev. Lett.}\ }\textbf {\bibinfo {volume} {94}},\ \bibinfo
  {pages} {016403} (\bibinfo {year} {2005})}\BibitemShut {NoStop}%
\bibitem [{\citenamefont {Tayagaki}\ \emph {et~al.}(2005)\citenamefont
  {Tayagaki}, \citenamefont {Mysyrowicz},\ and\ \citenamefont
  {Kuwata-Gonokami}}]{Tayagaki2005}%
  \BibitemOpen
  \bibfield  {author} {\bibinfo {author} {\bibfnamefont {T.}~\bibnamefont
  {Tayagaki}}, \bibinfo {author} {\bibfnamefont {A.}~\bibnamefont
  {Mysyrowicz}}, \ and\ \bibinfo {author} {\bibfnamefont {M.}~\bibnamefont
  {Kuwata-Gonokami}},\ }\href {\doibase 10.1143/JPSJ.74.1423} {\bibfield
  {journal} {\bibinfo  {journal} {Journal of the Physical Society of Japan}\
  }\textbf {\bibinfo {volume} {74}},\ \bibinfo {pages} {1423} (\bibinfo {year}
  {2005})}\BibitemShut {NoStop}%
\bibitem [{\citenamefont {Huber}\ \emph {et~al.}(2006)\citenamefont {Huber},
  \citenamefont {Schmid}, \citenamefont {Shen}, \citenamefont {Chemla},\ and\
  \citenamefont {Kaindl}}]{Huber2006}%
  \BibitemOpen
  \bibfield  {author} {\bibinfo {author} {\bibfnamefont {R.}~\bibnamefont
  {Huber}}, \bibinfo {author} {\bibfnamefont {B.~A.}\ \bibnamefont {Schmid}},
  \bibinfo {author} {\bibfnamefont {Y.~R.}\ \bibnamefont {Shen}}, \bibinfo
  {author} {\bibfnamefont {D.~S.}\ \bibnamefont {Chemla}}, \ and\ \bibinfo
  {author} {\bibfnamefont {R.~A.}\ \bibnamefont {Kaindl}},\ }\href {\doibase
  10.1103/PhysRevLett.96.017402} {\bibfield  {journal} {\bibinfo  {journal}
  {Phys. Rev. Lett.}\ }\textbf {\bibinfo {volume} {96}},\ \bibinfo {pages}
  {017402} (\bibinfo {year} {2006})}\BibitemShut {NoStop}%
\bibitem [{\citenamefont {Yoshioka}\ \emph {et~al.}(2007)\citenamefont
  {Yoshioka}, \citenamefont {Ideguchi},\ and\ \citenamefont
  {Kuwata-Gonokami}}]{Yoshiaka2007}%
  \BibitemOpen
  \bibfield  {author} {\bibinfo {author} {\bibfnamefont {K.}~\bibnamefont
  {Yoshioka}}, \bibinfo {author} {\bibfnamefont {T.}~\bibnamefont {Ideguchi}},
  \ and\ \bibinfo {author} {\bibfnamefont {M.}~\bibnamefont
  {Kuwata-Gonokami}},\ }\href {\doibase 10.1103/PhysRevB.76.033204} {\bibfield
  {journal} {\bibinfo  {journal} {Phys. Rev. B}\ }\textbf {\bibinfo {volume}
  {76}},\ \bibinfo {pages} {033204} (\bibinfo {year} {2007})}\BibitemShut
  {NoStop}%
\bibitem [{\citenamefont {Huber}\ \emph {et~al.}(2008)\citenamefont {Huber},
  \citenamefont {Schmid}, \citenamefont {Kaindl},\ and\ \citenamefont
  {Chemla}}]{Huber2008}%
  \BibitemOpen
  \bibfield  {author} {\bibinfo {author} {\bibfnamefont {R.}~\bibnamefont
  {Huber}}, \bibinfo {author} {\bibfnamefont {B.~A.}\ \bibnamefont {Schmid}},
  \bibinfo {author} {\bibfnamefont {R.~A.}\ \bibnamefont {Kaindl}}, \ and\
  \bibinfo {author} {\bibfnamefont {D.~S.}\ \bibnamefont {Chemla}},\ }\href
  {\doibase 10.1002/pssb.200777603} {\bibfield  {journal} {\bibinfo  {journal}
  {physica status solidi (b)}\ }\textbf {\bibinfo {volume} {245}},\ \bibinfo
  {pages} {1041} (\bibinfo {year} {2008})}\BibitemShut {NoStop}%
\bibitem [{\citenamefont {Yoshioka}\ \emph {et~al.}(2010)\citenamefont
  {Yoshioka}, \citenamefont {Ideguchi}, \citenamefont {Mysyrowicz},\ and\
  \citenamefont {Kuwata-Gonokami}}]{Yoshiaka2010}%
  \BibitemOpen
  \bibfield  {author} {\bibinfo {author} {\bibfnamefont {K.}~\bibnamefont
  {Yoshioka}}, \bibinfo {author} {\bibfnamefont {T.}~\bibnamefont {Ideguchi}},
  \bibinfo {author} {\bibfnamefont {A.}~\bibnamefont {Mysyrowicz}}, \ and\
  \bibinfo {author} {\bibfnamefont {M.}~\bibnamefont {Kuwata-Gonokami}},\
  }\href {\doibase 10.1103/PhysRevB.82.041201} {\bibfield  {journal} {\bibinfo
  {journal} {Phys. Rev. B}\ }\textbf {\bibinfo {volume} {82}},\ \bibinfo
  {pages} {041201} (\bibinfo {year} {2010})}\BibitemShut {NoStop}%
\bibitem [{\citenamefont {Yoshioka}\ and\ \citenamefont
  {Kuwata-Gonokami}(2012)}]{Yoshioka2012}%
  \BibitemOpen
  \bibfield  {author} {\bibinfo {author} {\bibfnamefont {K.}~\bibnamefont
  {Yoshioka}}\ and\ \bibinfo {author} {\bibfnamefont {M.}~\bibnamefont
  {Kuwata-Gonokami}},\ }\href {\doibase 10.1088/1367-2630/14/5/055024}
  {\bibfield  {journal} {\bibinfo  {journal} {New Journal of Physics}\ }\textbf
  {\bibinfo {volume} {14}} (\bibinfo {year} {2012}),\
  10.1088/1367-2630/14/5/055024}\BibitemShut {NoStop}%
\bibitem [{\citenamefont {Heck\"otter}\ \emph {et~al.}(2018)\citenamefont
  {Heck\"otter}, \citenamefont {Freitag}, \citenamefont {Fr\"ohlich},
  \citenamefont {A\ss{}mann}, \citenamefont {Bayer}, \citenamefont
  {Gr\"unwald}, \citenamefont {Sch\"one}, \citenamefont {Semkat}, \citenamefont
  {Stolz},\ and\ \citenamefont {Scheel}}]{Heckotter2018}%
  \BibitemOpen
  \bibfield  {author} {\bibinfo {author} {\bibfnamefont {J.}~\bibnamefont
  {Heck\"otter}}, \bibinfo {author} {\bibfnamefont {M.}~\bibnamefont
  {Freitag}}, \bibinfo {author} {\bibfnamefont {D.}~\bibnamefont {Fr\"ohlich}},
  \bibinfo {author} {\bibfnamefont {M.}~\bibnamefont {A\ss{}mann}}, \bibinfo
  {author} {\bibfnamefont {M.}~\bibnamefont {Bayer}}, \bibinfo {author}
  {\bibfnamefont {P.}~\bibnamefont {Gr\"unwald}}, \bibinfo {author}
  {\bibfnamefont {F.}~\bibnamefont {Sch\"one}}, \bibinfo {author}
  {\bibfnamefont {D.}~\bibnamefont {Semkat}}, \bibinfo {author} {\bibfnamefont
  {H.}~\bibnamefont {Stolz}}, \ and\ \bibinfo {author} {\bibfnamefont
  {S.}~\bibnamefont {Scheel}},\ }\href {\doibase
  10.1103/PhysRevLett.121.097401} {\bibfield  {journal} {\bibinfo  {journal}
  {Phys. Rev. Lett.}\ }\textbf {\bibinfo {volume} {121}},\ \bibinfo {pages}
  {097401} (\bibinfo {year} {2018})}\BibitemShut {NoStop}%
\bibitem [{\citenamefont {Walther}\ \emph
  {et~al.}(2018{\natexlab{a}})\citenamefont {Walther}, \citenamefont
  {Kr\"uger}, \citenamefont {Scheel},\ and\ \citenamefont
  {Pohl}}]{Walther2018}%
  \BibitemOpen
  \bibfield  {author} {\bibinfo {author} {\bibfnamefont {V.}~\bibnamefont
  {Walther}}, \bibinfo {author} {\bibfnamefont {S.~O.}\ \bibnamefont
  {Kr\"uger}}, \bibinfo {author} {\bibfnamefont {S.}~\bibnamefont {Scheel}}, \
  and\ \bibinfo {author} {\bibfnamefont {T.}~\bibnamefont {Pohl}},\ }\href
  {\doibase 10.1103/PhysRevB.98.165201} {\bibfield  {journal} {\bibinfo
  {journal} {Phys. Rev. B}\ }\textbf {\bibinfo {volume} {98}},\ \bibinfo
  {pages} {165201} (\bibinfo {year} {2018}{\natexlab{a}})}\BibitemShut
  {NoStop}%
\bibitem [{\citenamefont {Gr\"unwald}\ \emph {et~al.}(2016)\citenamefont
  {Gr\"unwald}, \citenamefont {A\ss{}mann}, \citenamefont {Heck\"otter},
  \citenamefont {Fr\"ohlich}, \citenamefont {Bayer}, \citenamefont {Stolz},\
  and\ \citenamefont {Scheel}}]{Grunwald2016}%
  \BibitemOpen
  \bibfield  {author} {\bibinfo {author} {\bibfnamefont {P.}~\bibnamefont
  {Gr\"unwald}}, \bibinfo {author} {\bibfnamefont {M.}~\bibnamefont
  {A\ss{}mann}}, \bibinfo {author} {\bibfnamefont {J.}~\bibnamefont
  {Heck\"otter}}, \bibinfo {author} {\bibfnamefont {D.}~\bibnamefont
  {Fr\"ohlich}}, \bibinfo {author} {\bibfnamefont {M.}~\bibnamefont {Bayer}},
  \bibinfo {author} {\bibfnamefont {H.}~\bibnamefont {Stolz}}, \ and\ \bibinfo
  {author} {\bibfnamefont {S.}~\bibnamefont {Scheel}},\ }\href {\doibase
  10.1103/PhysRevLett.117.133003} {\bibfield  {journal} {\bibinfo  {journal}
  {Phys. Rev. Lett.}\ }\textbf {\bibinfo {volume} {117}},\ \bibinfo {pages}
  {133003} (\bibinfo {year} {2016})}\BibitemShut {NoStop}%
\bibitem [{\citenamefont {Walther}\ \emph
  {et~al.}(2018{\natexlab{b}})\citenamefont {Walther}, \citenamefont {Johne},\
  and\ \citenamefont {Pohl}}]{Walther2018nl}%
  \BibitemOpen
  \bibfield  {author} {\bibinfo {author} {\bibfnamefont {V.}~\bibnamefont
  {Walther}}, \bibinfo {author} {\bibfnamefont {R.}~\bibnamefont {Johne}}, \
  and\ \bibinfo {author} {\bibfnamefont {T.}~\bibnamefont {Pohl}},\ }\href
  {\doibase 10.1038/s41467-018-03742-7} {\bibfield  {journal} {\bibinfo
  {journal} {Nature Communications}\ }\textbf {\bibinfo {volume} {9}},\
  \bibinfo {pages} {1309} (\bibinfo {year} {2018}{\natexlab{b}})}\BibitemShut
  {NoStop}%
\bibitem [{\citenamefont {Khazali}\ \emph {et~al.}(2017)\citenamefont
  {Khazali}, \citenamefont {Heshami},\ and\ \citenamefont
  {Simon}}]{Khazali2017}%
  \BibitemOpen
  \bibfield  {author} {\bibinfo {author} {\bibfnamefont {M.}~\bibnamefont
  {Khazali}}, \bibinfo {author} {\bibfnamefont {K.}~\bibnamefont {Heshami}}, \
  and\ \bibinfo {author} {\bibfnamefont {C.}~\bibnamefont {Simon}},\ }\href
  {\doibase 10.1088/1361-6455/aa8d7c} {\bibfield  {journal} {\bibinfo
  {journal} {Journal of Physics B: Atomic, Molecular and Optical Physics}\
  }\textbf {\bibinfo {volume} {50}},\ \bibinfo {pages} {215301} (\bibinfo
  {year} {2017})}\BibitemShut {NoStop}%
\bibitem [{\citenamefont {Ziemkiewicz}\ and\ \citenamefont
  {Zieli\'{n}ska-Raczy\'{n}ska}(2018)}]{Ziemkiewicz2018}%
  \BibitemOpen
  \bibfield  {author} {\bibinfo {author} {\bibfnamefont {D.}~\bibnamefont
  {Ziemkiewicz}}\ and\ \bibinfo {author} {\bibfnamefont {S.}~\bibnamefont
  {Zieli\'{n}ska-Raczy\'{n}ska}},\ }\href {\doibase 10.1364/OL.43.003742}
  {\bibfield  {journal} {\bibinfo  {journal} {Opt. Lett.}\ }\textbf {\bibinfo
  {volume} {43}},\ \bibinfo {pages} {3742} (\bibinfo {year}
  {2018})}\BibitemShut {NoStop}%
\bibitem [{\citenamefont {Matsumoto}\ \emph {et~al.}(1996)\citenamefont
  {Matsumoto}, \citenamefont {Saito}, \citenamefont {Hasuo}, \citenamefont
  {Kono},\ and\ \citenamefont {Nagasawa}}]{MATSUMOTO1996}%
  \BibitemOpen
  \bibfield  {author} {\bibinfo {author} {\bibfnamefont {H.}~\bibnamefont
  {Matsumoto}}, \bibinfo {author} {\bibfnamefont {K.}~\bibnamefont {Saito}},
  \bibinfo {author} {\bibfnamefont {M.}~\bibnamefont {Hasuo}}, \bibinfo
  {author} {\bibfnamefont {S.}~\bibnamefont {Kono}}, \ and\ \bibinfo {author}
  {\bibfnamefont {N.}~\bibnamefont {Nagasawa}},\ }\href {\doibase
  https://doi.org/10.1016/0038-1098(95)00601-X} {\bibfield  {journal} {\bibinfo
   {journal} {Solid State Communications}\ }\textbf {\bibinfo {volume} {97}},\
  \bibinfo {pages} {125 } (\bibinfo {year} {1996})}\BibitemShut {NoStop}%
\bibitem [{\citenamefont {Sun}\ \emph {et~al.}(2001)\citenamefont {Sun},
  \citenamefont {Wong},\ and\ \citenamefont {Ketterson}}]{Ketterson2001}%
  \BibitemOpen
  \bibfield  {author} {\bibinfo {author} {\bibfnamefont {Y.}~\bibnamefont
  {Sun}}, \bibinfo {author} {\bibfnamefont {G.~K.~L.}\ \bibnamefont {Wong}}, \
  and\ \bibinfo {author} {\bibfnamefont {J.~B.}\ \bibnamefont {Ketterson}},\
  }\href {\doibase 10.1103/PhysRevB.63.125323} {\bibfield  {journal} {\bibinfo
  {journal} {Phys. Rev. B}\ }\textbf {\bibinfo {volume} {63}},\ \bibinfo
  {pages} {125323} (\bibinfo {year} {2001})}\BibitemShut {NoStop}%
\bibitem [{\citenamefont {Naka}\ \emph {et~al.}(2013)\citenamefont {Naka},
  \citenamefont {Akimoto},\ and\ \citenamefont {Shirai}}]{Naka2013}%
  \BibitemOpen
  \bibfield  {author} {\bibinfo {author} {\bibfnamefont {N.}~\bibnamefont
  {Naka}}, \bibinfo {author} {\bibfnamefont {I.}~\bibnamefont {Akimoto}}, \
  and\ \bibinfo {author} {\bibfnamefont {M.}~\bibnamefont {Shirai}},\ }\href
  {\doibase 10.1002/pssb.201200713} {\bibfield  {journal} {\bibinfo  {journal}
  {physica status solidi (b)}\ }\textbf {\bibinfo {volume} {250}},\ \bibinfo
  {pages} {1773} (\bibinfo {year} {2013})}\BibitemShut {NoStop}%
\bibitem [{\citenamefont {Mund}\ \emph {et~al.}(2018)\citenamefont {Mund},
  \citenamefont {Fr\"ohlich}, \citenamefont {Yakovlev},\ and\ \citenamefont
  {Bayer}}]{Mund2018}%
  \BibitemOpen
  \bibfield  {author} {\bibinfo {author} {\bibfnamefont {J.}~\bibnamefont
  {Mund}}, \bibinfo {author} {\bibfnamefont {D.}~\bibnamefont {Fr\"ohlich}},
  \bibinfo {author} {\bibfnamefont {D.~R.}\ \bibnamefont {Yakovlev}}, \ and\
  \bibinfo {author} {\bibfnamefont {M.}~\bibnamefont {Bayer}},\ }\href
  {\doibase 10.1103/PhysRevB.98.085203} {\bibfield  {journal} {\bibinfo
  {journal} {Phys. Rev. B}\ }\textbf {\bibinfo {volume} {98}},\ \bibinfo
  {pages} {085203} (\bibinfo {year} {2018})}\BibitemShut {NoStop}%
\bibitem [{\citenamefont {Mund}\ \emph {et~al.}(2019)\citenamefont {Mund},
  \citenamefont {Uihlein}, \citenamefont {Fr\"ohlich}, \citenamefont
  {Yakovlev},\ and\ \citenamefont {Bayer}}]{Mund2019}%
  \BibitemOpen
  \bibfield  {author} {\bibinfo {author} {\bibfnamefont {J.}~\bibnamefont
  {Mund}}, \bibinfo {author} {\bibfnamefont {C.}~\bibnamefont {Uihlein}},
  \bibinfo {author} {\bibfnamefont {D.}~\bibnamefont {Fr\"ohlich}}, \bibinfo
  {author} {\bibfnamefont {D.~R.}\ \bibnamefont {Yakovlev}}, \ and\ \bibinfo
  {author} {\bibfnamefont {M.}~\bibnamefont {Bayer}},\ }\href {\doibase
  10.1103/PhysRevB.99.195204} {\bibfield  {journal} {\bibinfo  {journal} {Phys.
  Rev. B}\ }\textbf {\bibinfo {volume} {99}},\ \bibinfo {pages} {195204}
  (\bibinfo {year} {2019})}\BibitemShut {NoStop}%
\bibitem [{\citenamefont {Farenbruch}\ \emph
  {et~al.}(2020{\natexlab{a}})\citenamefont {Farenbruch}, \citenamefont {Mund},
  \citenamefont {Fr\"ohlich}, \citenamefont {Yakovlev}, \citenamefont {Bayer},
  \citenamefont {Semina},\ and\ \citenamefont {Glazov}}]{Farenbruch2020}%
  \BibitemOpen
  \bibfield  {author} {\bibinfo {author} {\bibfnamefont {A.}~\bibnamefont
  {Farenbruch}}, \bibinfo {author} {\bibfnamefont {J.}~\bibnamefont {Mund}},
  \bibinfo {author} {\bibfnamefont {D.}~\bibnamefont {Fr\"ohlich}}, \bibinfo
  {author} {\bibfnamefont {D.~R.}\ \bibnamefont {Yakovlev}}, \bibinfo {author}
  {\bibfnamefont {M.}~\bibnamefont {Bayer}}, \bibinfo {author} {\bibfnamefont
  {M.~A.}\ \bibnamefont {Semina}}, \ and\ \bibinfo {author} {\bibfnamefont
  {M.~M.}\ \bibnamefont {Glazov}},\ }\href {\doibase
  10.1103/PhysRevB.101.115201} {\bibfield  {journal} {\bibinfo  {journal}
  {Phys. Rev. B}\ }\textbf {\bibinfo {volume} {101}},\ \bibinfo {pages}
  {115201} (\bibinfo {year} {2020}{\natexlab{a}})}\BibitemShut {NoStop}%
\bibitem [{\citenamefont {Rommel}\ \emph {et~al.}(2020)\citenamefont {Rommel},
  \citenamefont {Main}, \citenamefont {Farenbruch}, \citenamefont {Mund},
  \citenamefont {Fr\"ohlich}, \citenamefont {Yakovlev}, \citenamefont {Bayer},\
  and\ \citenamefont {Uihlein}}]{Rommel2020}%
  \BibitemOpen
  \bibfield  {author} {\bibinfo {author} {\bibfnamefont {P.}~\bibnamefont
  {Rommel}}, \bibinfo {author} {\bibfnamefont {J.}~\bibnamefont {Main}},
  \bibinfo {author} {\bibfnamefont {A.}~\bibnamefont {Farenbruch}}, \bibinfo
  {author} {\bibfnamefont {J.}~\bibnamefont {Mund}}, \bibinfo {author}
  {\bibfnamefont {D.}~\bibnamefont {Fr\"ohlich}}, \bibinfo {author}
  {\bibfnamefont {D.~R.}\ \bibnamefont {Yakovlev}}, \bibinfo {author}
  {\bibfnamefont {M.}~\bibnamefont {Bayer}}, \ and\ \bibinfo {author}
  {\bibfnamefont {C.}~\bibnamefont {Uihlein}},\ }\href {\doibase
  10.1103/PhysRevB.101.115202} {\bibfield  {journal} {\bibinfo  {journal}
  {Phys. Rev. B}\ }\textbf {\bibinfo {volume} {101}},\ \bibinfo {pages}
  {115202} (\bibinfo {year} {2020})}\BibitemShut {NoStop}%
\bibitem [{\citenamefont {Dyubko}\ \emph {et~al.}(1995)\citenamefont {Dyubko},
  \citenamefont {Efimenko}, \citenamefont {Efremov},\ and\ \citenamefont
  {Podnos}}]{Dyubko1995}%
  \BibitemOpen
  \bibfield  {author} {\bibinfo {author} {\bibfnamefont {S.}~\bibnamefont
  {Dyubko}}, \bibinfo {author} {\bibfnamefont {M.}~\bibnamefont {Efimenko}},
  \bibinfo {author} {\bibfnamefont {V.}~\bibnamefont {Efremov}}, \ and\
  \bibinfo {author} {\bibfnamefont {S.}~\bibnamefont {Podnos}},\ }\href
  {\doibase 10.1103/PhysRevA.52.514} {\bibfield  {journal} {\bibinfo  {journal}
  {Phys. Rev. A}\ }\textbf {\bibinfo {volume} {52}},\ \bibinfo {pages} {514}
  (\bibinfo {year} {1995})}\BibitemShut {NoStop}%
\bibitem [{\citenamefont {Ryabtsev}\ and\ \citenamefont
  {Tret'yakov}(2001)}]{Ryabtsev2001}%
  \BibitemOpen
  \bibfield  {author} {\bibinfo {author} {\bibfnamefont {I.~I.}\ \bibnamefont
  {Ryabtsev}}\ and\ \bibinfo {author} {\bibfnamefont {D.~B.}\ \bibnamefont
  {Tret'yakov}},\ }\href {\doibase 10.1134/1.1351567} {\bibfield  {journal}
  {\bibinfo  {journal} {Optics and Spectroscopy}\ }\textbf {\bibinfo {volume}
  {90}},\ \bibinfo {pages} {145} (\bibinfo {year} {2001})}\BibitemShut
  {NoStop}%
\bibitem [{\citenamefont {Lynch}\ \emph {et~al.}(2021)\citenamefont {Lynch},
  \citenamefont {Hodges}, \citenamefont {Mandal}, \citenamefont {Langbein},
  \citenamefont {Singh}, \citenamefont {Gallagher}, \citenamefont {Pritchett},
  \citenamefont {Pizzey}, \citenamefont {Rogers}, \citenamefont {Adams},\ and\
  \citenamefont {Jones}}]{Lynch2021}%
  \BibitemOpen
  \bibfield  {author} {\bibinfo {author} {\bibfnamefont {S.~A.}\ \bibnamefont
  {Lynch}}, \bibinfo {author} {\bibfnamefont {C.}~\bibnamefont {Hodges}},
  \bibinfo {author} {\bibfnamefont {S.}~\bibnamefont {Mandal}}, \bibinfo
  {author} {\bibfnamefont {W.}~\bibnamefont {Langbein}}, \bibinfo {author}
  {\bibfnamefont {R.~P.}\ \bibnamefont {Singh}}, \bibinfo {author}
  {\bibfnamefont {L.~A.~P.}\ \bibnamefont {Gallagher}}, \bibinfo {author}
  {\bibfnamefont {J.~D.}\ \bibnamefont {Pritchett}}, \bibinfo {author}
  {\bibfnamefont {D.}~\bibnamefont {Pizzey}}, \bibinfo {author} {\bibfnamefont
  {J.~P.}\ \bibnamefont {Rogers}}, \bibinfo {author} {\bibfnamefont {C.~S.}\
  \bibnamefont {Adams}}, \ and\ \bibinfo {author} {\bibfnamefont {M.~P.~A.}\
  \bibnamefont {Jones}},\ }\href {\doibase 10.1103/PhysRevMaterials.5.084602}
  {\bibfield  {journal} {\bibinfo  {journal} {Phys. Rev. Materials}\ }\textbf
  {\bibinfo {volume} {5}},\ \bibinfo {pages} {084602} (\bibinfo {year}
  {2021})}\BibitemShut {NoStop}%
\bibitem [{\citenamefont {Baumeister}(1961)}]{Baumeister1961}%
  \BibitemOpen
  \bibfield  {author} {\bibinfo {author} {\bibfnamefont {P.~W.}\ \bibnamefont
  {Baumeister}},\ }\href {\doibase 10.1103/PhysRev.121.359} {\bibfield
  {journal} {\bibinfo  {journal} {Phys. Rev.}\ }\textbf {\bibinfo {volume}
  {121}},\ \bibinfo {pages} {359} (\bibinfo {year} {1961})}\BibitemShut
  {NoStop}%
\bibitem [{\citenamefont {Sch\"one}\ \emph
  {et~al.}(2017{\natexlab{b}})\citenamefont {Sch\"one}, \citenamefont {Stolz},\
  and\ \citenamefont {Naka}}]{Schone2017}%
  \BibitemOpen
  \bibfield  {author} {\bibinfo {author} {\bibfnamefont {F.}~\bibnamefont
  {Sch\"one}}, \bibinfo {author} {\bibfnamefont {H.}~\bibnamefont {Stolz}}, \
  and\ \bibinfo {author} {\bibfnamefont {N.}~\bibnamefont {Naka}},\ }\href
  {\doibase 10.1103/PhysRevB.96.115207} {\bibfield  {journal} {\bibinfo
  {journal} {Phys. Rev. B}\ }\textbf {\bibinfo {volume} {96}},\ \bibinfo
  {pages} {115207} (\bibinfo {year} {2017}{\natexlab{b}})}\BibitemShut
  {NoStop}%
\bibitem [{\citenamefont {Takahata}\ and\ \citenamefont
  {Naka}(2018)}]{Naka2018}%
  \BibitemOpen
  \bibfield  {author} {\bibinfo {author} {\bibfnamefont {M.}~\bibnamefont
  {Takahata}}\ and\ \bibinfo {author} {\bibfnamefont {N.}~\bibnamefont
  {Naka}},\ }\href {\doibase 10.1103/PhysRevB.98.195205} {\bibfield  {journal}
  {\bibinfo  {journal} {Phys. Rev. B}\ }\textbf {\bibinfo {volume} {98}},\
  \bibinfo {pages} {195205} (\bibinfo {year} {2018})}\BibitemShut {NoStop}%
\bibitem [{\citenamefont {Saikan}\ \emph {et~al.}(1985)\citenamefont {Saikan},
  \citenamefont {Hashimoto}, \citenamefont {Kushida},\ and\ \citenamefont
  {Namba}}]{Saikan1985}%
  \BibitemOpen
  \bibfield  {author} {\bibinfo {author} {\bibfnamefont {S.}~\bibnamefont
  {Saikan}}, \bibinfo {author} {\bibfnamefont {N.}~\bibnamefont {Hashimoto}},
  \bibinfo {author} {\bibfnamefont {T.}~\bibnamefont {Kushida}}, \ and\
  \bibinfo {author} {\bibfnamefont {K.}~\bibnamefont {Namba}},\ }\href
  {\doibase 10.1063/1.448624} {\bibfield  {journal} {\bibinfo  {journal} {The
  Journal of Chemical Physics}\ }\textbf {\bibinfo {volume} {82}},\ \bibinfo
  {pages} {5409} (\bibinfo {year} {1985})}\BibitemShut {NoStop}%
\bibitem [{\citenamefont {Adams}\ and\ \citenamefont
  {Hughes}(2018)}]{f2f_lightmatter}%
  \BibitemOpen
  \bibfield  {author} {\bibinfo {author} {\bibfnamefont {C.~S.}\ \bibnamefont
  {Adams}}\ and\ \bibinfo {author} {\bibfnamefont {I.~G.}\ \bibnamefont
  {Hughes}},\ }\enquote {\bibinfo {title} {Optics f2f: From \text{Fourier} to
  \text{Fresnel}},}\ \ (\bibinfo  {publisher} {Oxford University Press},\
  \bibinfo {address} {Oxford},\ \bibinfo {year} {2018})\ Chap.~\bibinfo
  {chapter} {13}\BibitemShut {NoStop}%
\bibitem [{\citenamefont {Sch\"one}\ \emph {et~al.}(2016)\citenamefont
  {Sch\"one}, \citenamefont {Kr\"uger}, \citenamefont {Gr\"unwald},
  \citenamefont {Stolz}, \citenamefont {Scheel}, \citenamefont {A\ss{}mann},
  \citenamefont {Heck\"otter}, \citenamefont {Thewes}, \citenamefont
  {Fr\"ohlich},\ and\ \citenamefont {Bayer}}]{Schone2016-2}%
  \BibitemOpen
  \bibfield  {author} {\bibinfo {author} {\bibfnamefont {F.}~\bibnamefont
  {Sch\"one}}, \bibinfo {author} {\bibfnamefont {S.-O.}\ \bibnamefont
  {Kr\"uger}}, \bibinfo {author} {\bibfnamefont {P.}~\bibnamefont
  {Gr\"unwald}}, \bibinfo {author} {\bibfnamefont {H.}~\bibnamefont {Stolz}},
  \bibinfo {author} {\bibfnamefont {S.}~\bibnamefont {Scheel}}, \bibinfo
  {author} {\bibfnamefont {M.}~\bibnamefont {A\ss{}mann}}, \bibinfo {author}
  {\bibfnamefont {J.}~\bibnamefont {Heck\"otter}}, \bibinfo {author}
  {\bibfnamefont {J.}~\bibnamefont {Thewes}}, \bibinfo {author} {\bibfnamefont
  {D.}~\bibnamefont {Fr\"ohlich}}, \ and\ \bibinfo {author} {\bibfnamefont
  {M.}~\bibnamefont {Bayer}},\ }\href {\doibase 10.1103/PhysRevB.93.075203}
  {\bibfield  {journal} {\bibinfo  {journal} {Phys. Rev. B}\ }\textbf {\bibinfo
  {volume} {93}},\ \bibinfo {pages} {075203} (\bibinfo {year}
  {2016})}\BibitemShut {NoStop}%
\bibitem [{\citenamefont {Fröhlich}\ \emph {et~al.}(2005)\citenamefont
  {Fröhlich}, \citenamefont {Dasbach}, \citenamefont {{Baldassarri Höger von
  Högersthal}}, \citenamefont {Bayer}, \citenamefont {Klieber}, \citenamefont
  {Suter},\ and\ \citenamefont {Stolz}}]{FROHLICH2005}%
  \BibitemOpen
  \bibfield  {author} {\bibinfo {author} {\bibfnamefont {D.}~\bibnamefont
  {Fröhlich}}, \bibinfo {author} {\bibfnamefont {G.}~\bibnamefont {Dasbach}},
  \bibinfo {author} {\bibfnamefont {G.}~\bibnamefont {{Baldassarri Höger von
  Högersthal}}}, \bibinfo {author} {\bibfnamefont {M.}~\bibnamefont {Bayer}},
  \bibinfo {author} {\bibfnamefont {R.}~\bibnamefont {Klieber}}, \bibinfo
  {author} {\bibfnamefont {D.}~\bibnamefont {Suter}}, \ and\ \bibinfo {author}
  {\bibfnamefont {H.}~\bibnamefont {Stolz}},\ }\href {\doibase
  https://doi.org/10.1016/j.ssc.2004.06.044} {\bibfield  {journal} {\bibinfo
  {journal} {Solid State Communications}\ }\textbf {\bibinfo {volume} {134}},\
  \bibinfo {pages} {139} (\bibinfo {year} {2005})},\ \bibinfo {note}
  {spontaneous Coherence in Excitonic Systems}\BibitemShut {NoStop}%
\bibitem [{\citenamefont {Farenbruch}\ \emph {et~al.}(2021)\citenamefont
  {Farenbruch}, \citenamefont {Fr\"ohlich}, \citenamefont {Stolz},
  \citenamefont {Yakovlev},\ and\ \citenamefont {Bayer}}]{Farenbruch2021}%
  \BibitemOpen
  \bibfield  {author} {\bibinfo {author} {\bibfnamefont {A.}~\bibnamefont
  {Farenbruch}}, \bibinfo {author} {\bibfnamefont {D.}~\bibnamefont
  {Fr\"ohlich}}, \bibinfo {author} {\bibfnamefont {H.}~\bibnamefont {Stolz}},
  \bibinfo {author} {\bibfnamefont {D.~R.}\ \bibnamefont {Yakovlev}}, \ and\
  \bibinfo {author} {\bibfnamefont {M.}~\bibnamefont {Bayer}},\ }\href
  {\doibase 10.1103/PhysRevB.104.075203} {\bibfield  {journal} {\bibinfo
  {journal} {Phys. Rev. B}\ }\textbf {\bibinfo {volume} {104}},\ \bibinfo
  {pages} {075203} (\bibinfo {year} {2021})}\BibitemShut {NoStop}%
\bibitem [{\citenamefont {Busson}\ and\ \citenamefont
  {Tadjeddine}(2009)}]{Busson2009}%
  \BibitemOpen
  \bibfield  {author} {\bibinfo {author} {\bibfnamefont {B.}~\bibnamefont
  {Busson}}\ and\ \bibinfo {author} {\bibfnamefont {A.}~\bibnamefont
  {Tadjeddine}},\ }\href {\doibase 10.1021/jp908240d} {\bibfield  {journal}
  {\bibinfo  {journal} {The Journal of Physical Chemistry C}\ }\textbf
  {\bibinfo {volume} {113}},\ \bibinfo {pages} {21895} (\bibinfo {year}
  {2009})}\BibitemShut {NoStop}%
\bibitem [{\citenamefont {Rogers}\ \emph {et~al.}()\citenamefont {Rogers},
  \citenamefont {Gallagher}, \citenamefont {Pizzey}, \citenamefont {Pritchett},
  \citenamefont {Adams}, \citenamefont {Jones}, \citenamefont {Hodges},
  \citenamefont {Langbein},\ and\ \citenamefont {Lynch}}]{Rogers2021}%
  \BibitemOpen
  \bibfield  {author} {\bibinfo {author} {\bibfnamefont {J.}~\bibnamefont
  {Rogers}}, \bibinfo {author} {\bibfnamefont {L.~A.~P.}\ \bibnamefont
  {Gallagher}}, \bibinfo {author} {\bibfnamefont {D.}~\bibnamefont {Pizzey}},
  \bibinfo {author} {\bibfnamefont {J.~P.}\ \bibnamefont {Pritchett}}, \bibinfo
  {author} {\bibfnamefont {C.~S.}\ \bibnamefont {Adams}}, \bibinfo {author}
  {\bibfnamefont {M.~P. A.~J.}\ \bibnamefont {Jones}}, \bibinfo {author}
  {\bibfnamefont {C.}~\bibnamefont {Hodges}}, \bibinfo {author} {\bibfnamefont
  {W.}~\bibnamefont {Langbein}}, \ and\ \bibinfo {author} {\bibfnamefont
  {S.~A.}\ \bibnamefont {Lynch}},\ }\href@noop {} {\enquote {\bibinfo {title}
  {High resolution nanosecond spectroscopy of even-parity \text{Rydberg}
  excitons in \text{Cu}$_{2}$\text{O}},}\ }\Eprint {http://arxiv.org/abs/in
  preparation} {in preparation} \BibitemShut {NoStop}%
\bibitem [{\citenamefont {Heck\"otter}\ \emph {et~al.}(2017)\citenamefont
  {Heck\"otter}, \citenamefont {Freitag}, \citenamefont {Fr\"ohlich},
  \citenamefont {A\ss{}mann}, \citenamefont {Bayer}, \citenamefont {Semina},\
  and\ \citenamefont {Glazov}}]{Heckotter2017-2}%
  \BibitemOpen
  \bibfield  {author} {\bibinfo {author} {\bibfnamefont {J.}~\bibnamefont
  {Heck\"otter}}, \bibinfo {author} {\bibfnamefont {M.}~\bibnamefont
  {Freitag}}, \bibinfo {author} {\bibfnamefont {D.}~\bibnamefont {Fr\"ohlich}},
  \bibinfo {author} {\bibfnamefont {M.}~\bibnamefont {A\ss{}mann}}, \bibinfo
  {author} {\bibfnamefont {M.}~\bibnamefont {Bayer}}, \bibinfo {author}
  {\bibfnamefont {M.~A.}\ \bibnamefont {Semina}}, \ and\ \bibinfo {author}
  {\bibfnamefont {M.~M.}\ \bibnamefont {Glazov}},\ }\href {\doibase
  10.1103/PhysRevB.95.035210} {\bibfield  {journal} {\bibinfo  {journal} {Phys.
  Rev. B}\ }\textbf {\bibinfo {volume} {95}},\ \bibinfo {pages} {035210}
  (\bibinfo {year} {2017})}\BibitemShut {NoStop}%
\bibitem [{\citenamefont {Uihlein}\ \emph {et~al.}(1981)\citenamefont
  {Uihlein}, \citenamefont {Fr\"ohlich},\ and\ \citenamefont
  {Kenklies}}]{Uihlein1981}%
  \BibitemOpen
  \bibfield  {author} {\bibinfo {author} {\bibfnamefont {C.}~\bibnamefont
  {Uihlein}}, \bibinfo {author} {\bibfnamefont {D.}~\bibnamefont {Fr\"ohlich}},
  \ and\ \bibinfo {author} {\bibfnamefont {R.}~\bibnamefont {Kenklies}},\
  }\href {\doibase 10.1103/PhysRevB.23.2731} {\bibfield  {journal} {\bibinfo
  {journal} {Phys. Rev. B}\ }\textbf {\bibinfo {volume} {23}},\ \bibinfo
  {pages} {2731} (\bibinfo {year} {1981})}\BibitemShut {NoStop}%
\bibitem [{\citenamefont {Semina}(2018)}]{Semina2018}%
  \BibitemOpen
  \bibfield  {author} {\bibinfo {author} {\bibfnamefont {M.~A.}\ \bibnamefont
  {Semina}},\ }\href {\doibase 10.1134/S1063783418080218} {\bibfield  {journal}
  {\bibinfo  {journal} {Physics of the Solid State}\ }\textbf {\bibinfo
  {volume} {60}},\ \bibinfo {pages} {1527} (\bibinfo {year}
  {2018})}\BibitemShut {NoStop}%
\bibitem [{\citenamefont {Heckötter}\ \emph {et~al.}(2021)\citenamefont
  {Heckötter}, \citenamefont {Rommel}, \citenamefont {Main}, \citenamefont
  {Aßmann},\ and\ \citenamefont {Bayer}}]{Heckotter2021}%
  \BibitemOpen
  \bibfield  {author} {\bibinfo {author} {\bibfnamefont {J.}~\bibnamefont
  {Heckötter}}, \bibinfo {author} {\bibfnamefont {P.}~\bibnamefont {Rommel}},
  \bibinfo {author} {\bibfnamefont {J.}~\bibnamefont {Main}}, \bibinfo {author}
  {\bibfnamefont {M.}~\bibnamefont {Aßmann}}, \ and\ \bibinfo {author}
  {\bibfnamefont {M.}~\bibnamefont {Bayer}},\ }\href {\doibase
  https://doi.org/10.1002/pssr.202100335} {\bibfield  {journal} {\bibinfo
  {journal} {physica status solidi (RRL) – Rapid Research Letters}\ ,\
  \bibinfo {pages} {2100335}} (\bibinfo {year} {2021})}\BibitemShut {NoStop}%
\bibitem [{\citenamefont {Hogan}\ \emph {et~al.}(2012)\citenamefont {Hogan},
  \citenamefont {Agner}, \citenamefont {Merkt}, \citenamefont {Thiele},
  \citenamefont {Filipp},\ and\ \citenamefont {Wallraff}}]{Hogan2012}%
  \BibitemOpen
  \bibfield  {author} {\bibinfo {author} {\bibfnamefont {S.~D.}\ \bibnamefont
  {Hogan}}, \bibinfo {author} {\bibfnamefont {J.~A.}\ \bibnamefont {Agner}},
  \bibinfo {author} {\bibfnamefont {F.}~\bibnamefont {Merkt}}, \bibinfo
  {author} {\bibfnamefont {T.}~\bibnamefont {Thiele}}, \bibinfo {author}
  {\bibfnamefont {S.}~\bibnamefont {Filipp}}, \ and\ \bibinfo {author}
  {\bibfnamefont {A.}~\bibnamefont {Wallraff}},\ }\href {\doibase
  10.1103/PhysRevLett.108.063004} {\bibfield  {journal} {\bibinfo  {journal}
  {Phys. Rev. Lett.}\ }\textbf {\bibinfo {volume} {108}},\ \bibinfo {pages}
  {063004} (\bibinfo {year} {2012})}\BibitemShut {NoStop}%
\bibitem [{\citenamefont {Park}\ \emph {et~al.}(2011)\citenamefont {Park},
  \citenamefont {Tanner}, \citenamefont {Claessens}, \citenamefont {Shuman},\
  and\ \citenamefont {Gallagher}}]{Park2011}%
  \BibitemOpen
  \bibfield  {author} {\bibinfo {author} {\bibfnamefont {H.}~\bibnamefont
  {Park}}, \bibinfo {author} {\bibfnamefont {P.~J.}\ \bibnamefont {Tanner}},
  \bibinfo {author} {\bibfnamefont {B.~J.}\ \bibnamefont {Claessens}}, \bibinfo
  {author} {\bibfnamefont {E.~S.}\ \bibnamefont {Shuman}}, \ and\ \bibinfo
  {author} {\bibfnamefont {T.~F.}\ \bibnamefont {Gallagher}},\ }\href {\doibase
  10.1103/PhysRevA.84.022704} {\bibfield  {journal} {\bibinfo  {journal} {Phys.
  Rev. A}\ }\textbf {\bibinfo {volume} {84}},\ \bibinfo {pages} {022704}
  (\bibinfo {year} {2011})}\BibitemShut {NoStop}%
\bibitem [{\citenamefont {Tanasittikosol}\ \emph {et~al.}(2011)\citenamefont
  {Tanasittikosol}, \citenamefont {Pritchard}, \citenamefont {Maxwell},
  \citenamefont {Gauguet}, \citenamefont {Weatherill}, \citenamefont
  {Potvliege},\ and\ \citenamefont {Adams}}]{Tanasittikosol2011}%
  \BibitemOpen
  \bibfield  {author} {\bibinfo {author} {\bibfnamefont {M.}~\bibnamefont
  {Tanasittikosol}}, \bibinfo {author} {\bibfnamefont {J.~D.}\ \bibnamefont
  {Pritchard}}, \bibinfo {author} {\bibfnamefont {D.}~\bibnamefont {Maxwell}},
  \bibinfo {author} {\bibfnamefont {A.}~\bibnamefont {Gauguet}}, \bibinfo
  {author} {\bibfnamefont {K.~J.}\ \bibnamefont {Weatherill}}, \bibinfo
  {author} {\bibfnamefont {R.~M.}\ \bibnamefont {Potvliege}}, \ and\ \bibinfo
  {author} {\bibfnamefont {C.~S.}\ \bibnamefont {Adams}},\ }\href {\doibase
  10.1088/0953-4075/44/18/184020} {\bibfield  {journal} {\bibinfo  {journal}
  {Journal of Physics B: Atomic, Molecular and Optical Physics}\ }\textbf
  {\bibinfo {volume} {44}},\ \bibinfo {pages} {184020} (\bibinfo {year}
  {2011})}\BibitemShut {NoStop}%
\bibitem [{\citenamefont {Sevin{\c{c}}li}\ and\ \citenamefont
  {Pohl}(2014)}]{Sevincli2014}%
  \BibitemOpen
  \bibfield  {author} {\bibinfo {author} {\bibfnamefont {S.}~\bibnamefont
  {Sevin{\c{c}}li}}\ and\ \bibinfo {author} {\bibfnamefont {T.}~\bibnamefont
  {Pohl}},\ }\href {\doibase 10.1088/1367-2630/16/12/123036} {\bibfield
  {journal} {\bibinfo  {journal} {New Journal of Physics}\ }\textbf {\bibinfo
  {volume} {16}},\ \bibinfo {pages} {123036} (\bibinfo {year}
  {2014})}\BibitemShut {NoStop}%
\bibitem [{\citenamefont {Barredo}\ \emph {et~al.}(2015)\citenamefont
  {Barredo}, \citenamefont {Labuhn}, \citenamefont {Ravets}, \citenamefont
  {Lahaye}, \citenamefont {Browaeys},\ and\ \citenamefont
  {Adams}}]{Barredeo2015}%
  \BibitemOpen
  \bibfield  {author} {\bibinfo {author} {\bibfnamefont {D.}~\bibnamefont
  {Barredo}}, \bibinfo {author} {\bibfnamefont {H.}~\bibnamefont {Labuhn}},
  \bibinfo {author} {\bibfnamefont {S.}~\bibnamefont {Ravets}}, \bibinfo
  {author} {\bibfnamefont {T.}~\bibnamefont {Lahaye}}, \bibinfo {author}
  {\bibfnamefont {A.}~\bibnamefont {Browaeys}}, \ and\ \bibinfo {author}
  {\bibfnamefont {C.~S.}\ \bibnamefont {Adams}},\ }\href {\doibase
  10.1103/PhysRevLett.114.113002} {\bibfield  {journal} {\bibinfo  {journal}
  {Phys. Rev. Lett.}\ }\textbf {\bibinfo {volume} {114}},\ \bibinfo {pages}
  {113002} (\bibinfo {year} {2015})}\BibitemShut {NoStop}%
\bibitem [{\citenamefont {Tommey}\ and\ \citenamefont
  {Hogan}(2019)}]{Tommey2020}%
  \BibitemOpen
  \bibfield  {author} {\bibinfo {author} {\bibfnamefont {J.~D.~R.}\
  \bibnamefont {Tommey}}\ and\ \bibinfo {author} {\bibfnamefont {S.~D.}\
  \bibnamefont {Hogan}},\ }\href {\doibase 10.1103/PhysRevA.100.053417}
  {\bibfield  {journal} {\bibinfo  {journal} {Phys. Rev. A}\ }\textbf {\bibinfo
  {volume} {100}},\ \bibinfo {pages} {053417} (\bibinfo {year}
  {2019})}\BibitemShut {NoStop}%
\bibitem [{\citenamefont {Farenbruch}\ \emph
  {et~al.}(2020{\natexlab{b}})\citenamefont {Farenbruch}, \citenamefont
  {Fr\"ohlich}, \citenamefont {Yakovlev},\ and\ \citenamefont
  {Bayer}}]{Farenbruch2020-2}%
  \BibitemOpen
  \bibfield  {author} {\bibinfo {author} {\bibfnamefont {A.}~\bibnamefont
  {Farenbruch}}, \bibinfo {author} {\bibfnamefont {D.}~\bibnamefont
  {Fr\"ohlich}}, \bibinfo {author} {\bibfnamefont {D.~R.}\ \bibnamefont
  {Yakovlev}}, \ and\ \bibinfo {author} {\bibfnamefont {M.}~\bibnamefont
  {Bayer}},\ }\href {\doibase 10.1103/PhysRevLett.125.207402} {\bibfield
  {journal} {\bibinfo  {journal} {Phys. Rev. Lett.}\ }\textbf {\bibinfo
  {volume} {125}},\ \bibinfo {pages} {207402} (\bibinfo {year}
  {2020}{\natexlab{b}})}\BibitemShut {NoStop}%
\bibitem [{\citenamefont {Weisbuch}\ \emph {et~al.}(1992)\citenamefont
  {Weisbuch}, \citenamefont {Nishioka}, \citenamefont {Ishikawa},\ and\
  \citenamefont {Arakawa}}]{Weisbuch1992}%
  \BibitemOpen
  \bibfield  {author} {\bibinfo {author} {\bibfnamefont {C.}~\bibnamefont
  {Weisbuch}}, \bibinfo {author} {\bibfnamefont {M.}~\bibnamefont {Nishioka}},
  \bibinfo {author} {\bibfnamefont {A.}~\bibnamefont {Ishikawa}}, \ and\
  \bibinfo {author} {\bibfnamefont {Y.}~\bibnamefont {Arakawa}},\ }\href
  {\doibase 10.1103/PhysRevLett.69.3314} {\bibfield  {journal} {\bibinfo
  {journal} {Phys. Rev. Lett.}\ }\textbf {\bibinfo {volume} {69}},\ \bibinfo
  {pages} {3314} (\bibinfo {year} {1992})}\BibitemShut {NoStop}%
\bibitem [{\citenamefont {Bernardot}\ \emph {et~al.}(1992)\citenamefont
  {Bernardot}, \citenamefont {Nussenzveig}, \citenamefont {Brune},
  \citenamefont {Raimond},\ and\ \citenamefont {Haroche}}]{Bernardot1992}%
  \BibitemOpen
  \bibfield  {author} {\bibinfo {author} {\bibfnamefont {F.}~\bibnamefont
  {Bernardot}}, \bibinfo {author} {\bibfnamefont {P.}~\bibnamefont
  {Nussenzveig}}, \bibinfo {author} {\bibfnamefont {M.}~\bibnamefont {Brune}},
  \bibinfo {author} {\bibfnamefont {J.~M.}\ \bibnamefont {Raimond}}, \ and\
  \bibinfo {author} {\bibfnamefont {S.}~\bibnamefont {Haroche}},\ }\href
  {\doibase 10.1209/0295-5075/17/1/007} {\bibfield  {journal} {\bibinfo
  {journal} {Europhysics Letters ({EPL})}\ }\textbf {\bibinfo {volume} {17}},\
  \bibinfo {pages} {33} (\bibinfo {year} {1992})}\BibitemShut {NoStop}%
\bibitem [{\citenamefont {Walther}\ \emph {et~al.}(2020)\citenamefont
  {Walther}, \citenamefont {Gr\"unwald},\ and\ \citenamefont
  {Pohl}}]{Grunwald2020}%
  \BibitemOpen
  \bibfield  {author} {\bibinfo {author} {\bibfnamefont {V.}~\bibnamefont
  {Walther}}, \bibinfo {author} {\bibfnamefont {P.}~\bibnamefont {Gr\"unwald}},
  \ and\ \bibinfo {author} {\bibfnamefont {T.}~\bibnamefont {Pohl}},\ }\href
  {\doibase 10.1103/PhysRevLett.125.173601} {\bibfield  {journal} {\bibinfo
  {journal} {Phys. Rev. Lett.}\ }\textbf {\bibinfo {volume} {125}},\ \bibinfo
  {pages} {173601} (\bibinfo {year} {2020})}\BibitemShut {NoStop}%
\bibitem [{\citenamefont {Toyozawa}(1964)}]{TOYOZAWA1964}%
  \BibitemOpen
  \bibfield  {author} {\bibinfo {author} {\bibfnamefont {Y.}~\bibnamefont
  {Toyozawa}},\ }\href {\doibase https://doi.org/10.1016/0022-3697(64)90162-3}
  {\bibfield  {journal} {\bibinfo  {journal} {Journal of Physics and Chemistry
  of Solids}\ }\textbf {\bibinfo {volume} {25}},\ \bibinfo {pages} {59 }
  (\bibinfo {year} {1964})}\BibitemShut {NoStop}%
\bibitem [{\citenamefont {Elliott}(1957)}]{Elliott1957}%
  \BibitemOpen
  \bibfield  {author} {\bibinfo {author} {\bibfnamefont {R.~J.}\ \bibnamefont
  {Elliott}},\ }\href {\doibase 10.1103/PhysRev.108.1384} {\bibfield  {journal}
  {\bibinfo  {journal} {Phys. Rev.}\ }\textbf {\bibinfo {volume} {108}},\
  \bibinfo {pages} {1384} (\bibinfo {year} {1957})}\BibitemShut {NoStop}%
\bibitem [{\citenamefont {Bonifacio}\ \emph {et~al.}(1993)\citenamefont
  {Bonifacio}, \citenamefont {Caloi},\ and\ \citenamefont
  {Maroli}}]{BONIFACIO1993}%
  \BibitemOpen
  \bibfield  {author} {\bibinfo {author} {\bibfnamefont {R.}~\bibnamefont
  {Bonifacio}}, \bibinfo {author} {\bibfnamefont {R.}~\bibnamefont {Caloi}}, \
  and\ \bibinfo {author} {\bibfnamefont {C.}~\bibnamefont {Maroli}},\ }\href
  {\doibase https://doi.org/10.1016/0030-4018(93)90363-A} {\bibfield  {journal}
  {\bibinfo  {journal} {Optics Communications}\ }\textbf {\bibinfo {volume}
  {101}},\ \bibinfo {pages} {185} (\bibinfo {year} {1993})}\BibitemShut
  {NoStop}%
\bibitem [{\citenamefont {Vogel}\ and\ \citenamefont
  {Welsch}(2006)}]{VoWel_Ch2}%
  \BibitemOpen
  \bibfield  {author} {\bibinfo {author} {\bibfnamefont {W.}~\bibnamefont
  {Vogel}}\ and\ \bibinfo {author} {\bibfnamefont {D.-G.}\ \bibnamefont
  {Welsch}},\ }\enquote {\bibinfo {title} {Elements of quantum
  electrodynamics},}\ in\ \href {\doibase
  https://doi.org/10.1002/3527608524.ch2} {\emph {\bibinfo {booktitle} {Quantum
  Optics}}}\ (\bibinfo  {publisher} {Wiley},\ \bibinfo {address} {Weinheim},\
  \bibinfo {year} {2006})\ Chap.~\bibinfo {chapter} {2}, pp.\ \bibinfo {pages}
  {15--72}\BibitemShut {NoStop}%
\bibitem [{\citenamefont {Haug}\ and\ \citenamefont {Koch}(1994)}]{HaKo_Ch1}%
  \BibitemOpen
  \bibfield  {author} {\bibinfo {author} {\bibfnamefont {H.}~\bibnamefont
  {Haug}}\ and\ \bibinfo {author} {\bibfnamefont {S.}~\bibnamefont {Koch}},\
  }\enquote {\bibinfo {title} {Oscillator model},}\ in\ \href@noop {} {\emph
  {\bibinfo {booktitle} {Quantum theory of the optical and electronic
  properties of semiconductors}}}\ (\bibinfo  {publisher} {World Scientific},\
  \bibinfo {address} {Singapore},\ \bibinfo {year} {1994})\ Chap.~\bibinfo
  {chapter} {1}, pp.\ \bibinfo {pages} {1--15}\BibitemShut {NoStop}%
\bibitem [{\citenamefont {Schmutzler}\ \emph {et~al.}(2013)\citenamefont
  {Schmutzler}, \citenamefont {Fr\"ohlich},\ and\ \citenamefont
  {Bayer}}]{Schmutzler2013}%
  \BibitemOpen
  \bibfield  {author} {\bibinfo {author} {\bibfnamefont {J.}~\bibnamefont
  {Schmutzler}}, \bibinfo {author} {\bibfnamefont {D.}~\bibnamefont
  {Fr\"ohlich}}, \ and\ \bibinfo {author} {\bibfnamefont {M.}~\bibnamefont
  {Bayer}},\ }\href {\doibase 10.1103/PhysRevB.87.245202} {\bibfield  {journal}
  {\bibinfo  {journal} {Phys. Rev. B}\ }\textbf {\bibinfo {volume} {87}},\
  \bibinfo {pages} {245202} (\bibinfo {year} {2013})}\BibitemShut {NoStop}%
\bibitem [{\citenamefont {James}(2000)}]{James2000}%
  \BibitemOpen
  \bibfield  {author} {\bibinfo {author} {\bibfnamefont {D.}~\bibnamefont
  {James}},\ }\href {\doibase
  https://doi.org/10.1002/1521-3978(200009)48:9/11<823::AID-PROP823>3.0.CO;2-M}
  {\bibfield  {journal} {\bibinfo  {journal} {Fortschritte der Physik}\
  }\textbf {\bibinfo {volume} {48}},\ \bibinfo {pages} {823} (\bibinfo {year}
  {2000})}\BibitemShut {NoStop}%
\end{thebibliography}%

\end{document}